\RequirePackage{ifpdf}
\ifpdf 
\documentclass[pdftex]{sigma}
\else
\documentclass{sigma}
\fi

\numberwithin{equation}{section}

\begin{document}

\allowdisplaybreaks

\renewcommand{\PaperNumber}{055}

\FirstPageHeading

\ShortArticleName{Eigenfunction Expansions of Functions Describing
Systems with Symmetries}

\ArticleName{Eigenfunction Expansions of Functions Describing\\
Systems with Symmetries}

\Author{Ivan KACHURYK~$^\dag$ and Anatoliy KLIMYK~$^\ddag$}

\AuthorNameForHeading{I. Kachuryk and A. Klimyk}

\Address{$^\dag$~Khmel'nyts'kyy National University,
Khmel'nyts'kyy, Ukraine}
\EmailD{\href{mailto:kachuryk@ief.tup.km.ua}{kachuryk@ief.tup.km.ua}}

\Address{$^\ddag$~Bogolyubov Institute for Theoretical Physics,
      14-b   Metrologichna Str., Kyiv-143, 03143 Ukraine}
\EmailD{\href{mailto:aklimyk@bitp.kiev.ua}{aklimyk@bitp.kiev.ua}}

\ArticleDates{Received March 02, 2007; Published online March 28,
2007}

\Abstract{Physical systems with symmetries are described by
functions containing kinematical and dynamical parts. We consider
the case when kinematical symmetries are described by a noncompact
semisimple real Lie group $G$. Then  separation of kinematical
parts in the functions is fulf\/illed by means of harmonic
analysis related to the group $G$. This separation depends on
choice of  a coordinate system on the space where a physical
system exists. In the paper we review how coordinate systems can
be chosen and how the corresponding harmonic analysis can be done.
In the f\/irst part we consider in detail the case when $G$ is the
de Sitter group $SO_0(1,4)$. In the second part we show how the
corresponding theory can be developed for any noncompact
semisimple real Lie group.}


\Classification{22E43; 22E46; 33C80; 42C10; 45C05; 81Q10}

\section{Introduction}

A symmetry is mathematically described by some group $G$. If a
system (a system of particles in quantum mechanics, a system of
dif\/ferential equations, a system of particles in macrophysics
etc) admits a symmetry that is given by a group $G$, then a
function (functions) that describes this system (wave functions,
scattering amplitudes, solutions of a system of dif\/ferential
equations, functions describing a motion in macrophysics etc)
contains a part (parts), that is determined by the symmetry (and
is independent of a concrete system), and a part (parts) that
characterizes a concrete system.

For example, if a system $A$ of dif\/ferential equations admits a
symmetry group $G$, then, using the symmetry, very often one can
reduces this system $A$ to simpler system $B$ that does not have a
symmetry described by the group $G$. In fact, the system $B$ is
obtained from the system~$A$ by excluding  the symmetry (see, for
example, \cite{1}).

As another example, we cite scattering theory (see, for example,
\cite{2}). A scattering amplitude decomposes into series in
spherical functions $Y^l_m(\theta,\varphi)$ that characterize a
symmetry with respect to the rotation group $SO(3)$.
Coef\/f\/icients of this decomposition are called partial
amplitudes. Partial amplitudes depend on smaller number of
variables and they characterize the scattering under
consideration. There are dif\/ferent collections of partial
amplitudes corresponding to dif\/ferent types of scatterings,
whereas the functions $Y^l_m(\theta,\varphi)$ are the same for all
types of scatterings.

Separation in the functions, characterizing a system, of a part
(parts) depending on symmetry and of a part (parts) characterizing
a concrete system is in fact separation of kinematic and dynamical
parts. Essential part for studying of a concrete system is a
dynamical one. For this reason, separation of a dynamical part in
a function (in functions), characterizing a given physical system,
is a very important procedure. For example, a dynamical part in
scattering theory is represented by partial amplitudes which can
be directly compared with experimental data.

In a simplest case, separation of kinematic and dynamical parts is
a representation of functions, characterizing a system, as a
product of kinematic and dynamical parts. It is a degenerate case.

In a more general case (when a symmetry group is compact),
functions describing a system are represented as a sum of products
of kinematic and dynamical parts (it is a case in the scattering
theory).

A main aim of this paper is separation of kinematic parts in
functions, describing a system, and studying these parts. The
f\/irst step in separation of parts, depending on symmetries, is
choosing variables in such a way that a part of these variables
corresponds to the symmetries (kinematical variables) and
remaining ones correspond to dynamics of the system. Further on,
one studies harmonic analysis of the kinematical part. Under such
analysis, functions describing a physical system are considered as
functions depending only on kinematical variables, that is, one
pays no attention to dynamical variables.

In the framework of the harmonic analysis of the kinematical part,
the functions are expanded into basis functions which are common
eigenfunctions of a collection of self-adjoint operators that are
determined by the symmetry group. It is clear that
coef\/f\/icients of such expansion depend on dynamical variables.

Such collection of self-adjoint operators is not determined
uniquely. This collection depends on kinematical variables.
Kinematical variables are not determined uniquely too. As a~rule,
choice of kinematical variables depends on a dynamical problem
that has to be solved.

There exists a one-to-one correspondence between the following
collections:
\begin{enumerate}\itemsep=0pt

\item[(a)] a collection of kinematical variables;

\item[(b)] a chain of subgroups of the symmetry group $G$;

\item[(c)] a collection of self-adjoint dif\/ferential operators
that are, as a rule, Casimir operators of the group $G$ and of
members of the chain of subgroups. These Casimir operators are
very often Laplace operators expressed in the corresponding
coordinate systems.
\end{enumerate}

A description of such triples for an example of the sphere
$S^{n-1}$ in the $n$-dimensional Euclidean space with the symmetry
group $SO(n)$ is given in \cite{VKII,NSU}. Coordinates
(kinematical variables) in this case are polyspherical. There are
dif\/ferent types of polyspherical coordinates. Laplace operators
on the corresponding manifolds of $S^{n-1}$ expressed in the
corresponding polyspherical coordinates serve as a collection of
self-adjoint operators. Subgroups of the rotation group~$SO(n)$
(corresponding to the chosen type of polyspherical coordinates)
with successive inclusions serve as a chain of subgroups of
$SO(n)$. Note that harmonic analysis of functions given on
quotient spaces of the group $SO(n)$ is used extensively in
nuclear physics (see \cite{Sm,FOS} and references therein).

In this paper we review the case when a symmetry group $G$ is a
simple noncompact Lie group (for example, the Lorentz group, the
de Sitter group, the conformal group etc). The case when $G$
coincides with the motion group $SO_0(1,2)$ of the upper sheet of
the hyperboloid in the 3-dimensional Minkowski space-time is
simple and well-known (see, for example, \cite[Chapter 7]{VKI} and
\cite{BW}). The case when $G=SO_0(1,3)$ is also well-studied (see,
for example, \cite{VSm, WSS,Wint}). In the relativistic physics,
the groups $SO(3)$, $SO_0(1,2)$, $ISO(3)$ and $SO_0(1,3)$ appear
as little groups for the Poincar\'e group $ISO_0(1,3)$ (see
\cite{KPSW,Win,WLS}).

As an example, in the f\/irst part of the paper we consider in
detail the case when $G$ is the de Sitter group $SO_0(1,4)$ that
is a motion group of the 5-dimensional Minkowski space-time. This
space is a base space for the Caluza--Klein theory.

In the second part of the paper we consider the case when $G$ is a
generic linear simple noncompact real Lie group. Then $G$ is a
group of $n\times n$ matrices, where $n$ is some positive integer.
We def\/ine a hyperboloid and a cone with motion group $G$ and
consider how dif\/ferent coordinate systems can be determined on
them. Then we perform harmonic analysis of functions given on
these general hyperboloids and cones in these coordinate systems.
For this purpose, the general harmonic analysis on semisimple
noncompact Lie groups \cite{War} and on the corresponding
homogeneous spaces \cite{Hel} is used. Dif\/ferent coordinate
systems on a hyperboloid and on a cone with a motion group $G$ are
connected to dif\/ferent decompositions of the group $G$ into
products of its subgroups (the Iwasawa decomposition, the Cartan
decomposition, the generalized Cartan decomposition etc).

The basis for this review is our books \cite{VKII,KK86}. Our
consideration is closely related to special functions and
orthogonal polynomials, especially  with those special functions
which admit orthogonality relations. In \cite{VKII,KK86} and also
in \cite{VKI,VKIII,VKRA} one can f\/ind a detailed description of
the relation between the group representation and special
functions. The basic information on special functions and
orthogonal polynomials may be found in \cite{BEI,BEII}.

\section[The de Sitter group $SO_0(1,4)$ and its representations]{The de Sitter group
$\boldsymbol{SO_0(1,4)}$ and its representations} \label{de
Sitter}

The de Sitter group $SO_0(1,4)$ consists of all real $5\times 5$
matrices $g$ with $\det g=1$ which leave invariant the quadratic
form
\[
x_0^2-x_1^2-x_2^2-x_3^2-x_4^2.
\]
The Lie algebra ${\rm so}(1,4)$ of $SO_0(1,4)$ consists of all
real matrices
\begin{equation}\label{I-1}
\left[\begin{array} {ccccc} 0& a_{01}& a_{02}& a_{03}& a_{04}\\
a_{01}& 0& -a_{12}& -a_{13}& -a_{14}\\ a_{02}& a_{12}& 0& -a_{23}&
-a_{24}\\ a_{03}& a_{13}& a_{23}& 0& -a_{34}\\ a_{04}& a_{14}&
a_{24}& a_{34}& 0\end{array}\right].
\end{equation}
Thus, the Lie algebra ${\rm so}(1,4)$ is spanned upon the basis
elements
\begin{gather}\label{I-2}
L_{rs}=-e_{rs}+e_{sr},\qquad s,r=1,2,3,4,\qquad s<r,
\\
\label{I-3} L_{0r}=e_{0r}+e_{r0},\qquad r=1,2,3,4,
\end{gather}
where $e_{rs}$ is a matrix with matrix elements $(e_{rs})_{pq}=
\delta_{rp}\delta_{sq}$. The basis elements \eqref{I-2} and
\eqref{I-3} satisfy the commutation relations
\begin{equation}\label{I-4}
[L_{\mu\nu},L_{\rho\delta}]=g_{\nu\rho}L_{\mu\delta}+g_{\mu\delta}L_{\nu\rho}-
g_{\mu\rho}L_{\nu\delta}-g_{\nu\delta}L_{\mu\rho},\qquad
\rho,\mu,\nu,\delta=0,1,2,3,4,
\end{equation}
where $g_{k0}=g_{0k}=\delta_{0k}$, $g_{ks}=-\delta_{ks}$,
$k,s=1,2,3,4$. The maximal compact subgroup $K$ of $SO_0(1,4)$ is
isomorphic to the group $SO(4)$ and consists of matrices
\[
\left(\begin{array} {cc} 1&
0\\
 0& k
\end{array}\right), \qquad k\in SO(4).
\]

In construction of representations of the group $SO_0(1,4)$ one
uses the Cartan decomposition of the Lie algebra ${\rm so(1,4)}$
and the Iwasawa decomposition of $SO_0(1,4)$. In the Cartan
decomposition ${\rm so}(1,4)={\rm so}(4)+\mathfrak{p}$ the
subspace $\mathfrak{p}$ is spanned by basis elements \eqref{I-3}.
Let $\mathfrak{a}$ be a maximal commutative subalgebra in
$\mathfrak{p}$. This subalgebra is one-dimensional. The matrix
$L_{04}$ can be taken as a basis element of $\mathfrak{a}$. The
subgroup $A=\exp \mathfrak{a}$ is important in the representation
theory of the group $SO_0(1,4)$. This subgroup consists of
matrices
\begin{equation}\label{I-5}
\left[
\begin{array}{ccccc}
  \cosh\alpha & 0 & 0 & 0 & \sinh\alpha \\
  0 & 1 & 0 & 0 & 0 \\
  0 & 0 & 1 & 0 & 0 \\
  0 & 0 & 0 & 1 & 0 \\
  \sinh\alpha & 0 & 0 & 0 & \cosh\alpha
\end{array}\right], \qquad 0\leqslant\alpha<\infty.
\end{equation}
Using the commutation relations \eqref{I-4}, one directly checks
that the Lie subalgebra $\mathfrak{n}$ of ${\rm so}(1,4)$ with the
basis $L_{01}+L_{14}$, $L_{02}+L_{24}$, $L_{03}+L_{34}$ is
nilpotent and, moreover, commutative. The subgroup $N=\exp
\mathfrak{n}$ of $SO_0(1,4)$ consists of the matrices
\begin{equation}\label{I-6}
\left[
\begin{array}{ccccc}
  1+(r^2+s^2+t^2)/2 & t & r & s & -(r^2+s^2+t^2)/2 \\
  t & 1 & 0 & 0 & -t \\
  r & 0 & 1 & 0 & -r \\
  s & 0 & 0 & 1 & -s \\
  (r^2+s^2+t^2)/2 & t & r & s & 1-(r^2+s^2+t^2)/2
\end{array}   \right]
\end{equation}
with $t,r,s\in \mathbb{R}$. The subgroups $K\equiv SO(4)$, $A$ and
$N$ determine the Iwasawa decomposition
\[
SO_0(1,4)=SO(4)\cdot{NA}
\]
of $SO_0(1,4)$. The subgroup $M$ of $SO(4)\subset SO_0(1,4)$,
whose elements commute with elements of $A$, is isomorphic to the
group $SO(3)$. The subgroup
\[
P=SO(3)\cdot NA
\]
is called a {\it parabolic subgroup} of $SO_0(1,4)$. This subgroup
is used for construction of irreducible unitary representations of
$SO_0(1,4)$.

Now we construct representations $\pi_{\delta\lambda}$ of the
principal nonunitary series (and, consequently, of the principal
unitary series) of the group $SO_0(1,4)$. These representations
are constructed by means of irreducible unitary representations
$\delta$ of the subgroup $SO(3)$ and of complex li\-near
forms~$\lambda$ on the subalgebra $\mathfrak{a}$. Since
irreducible unitary representations $\delta$ of the
subgroup~$SO(3)$ are given by a non-negative integer or
half-integer $l$ and a linear form $\lambda$ is given by a~complex
number $\sigma=\lambda (L_{04})$, then the representations
$\pi_{\delta\lambda}$ are determined by the numbers~$l$
and~$\sigma$. For this reason, we use the notation $\pi_{l\sigma}$
for the representations $\pi_{\delta\lambda}$. We shall use only
those representations $\pi_{l\sigma}$ for which $l=0$. These
representations will be denoted by $\pi^\sigma$. The
representations $\pi^\sigma$ are characterized by the property
that the restriction $\pi^\sigma {\downarrow}_{SO(4)}$ contains
the trivial (one-dimensional) representation of the subgroup
$SO(4)$. Moreover, they exhaust all principal nonunitary series
representations of $SO_0(1,4)$ with this property. These
representations are called {\it representations of class $1$ with
respect to $SO(4)$}.

For $\sigma={\rm i}\rho-\frac32$, $\rho\in \mathbb{R}$, the
representations $\pi^\sigma$ are unitary and belong to the {\it
principal unitary series}.

The group $SO_0(1,4)$ has two independent Casimir operators
\begin{gather}\label{I-7}
F=L_{12}^2+L_{13}^2+L_{14}^2+L_{23}^2+L_{24}^2
+L_{34}^2-L_{01}^2-L_{02}^2-L_{03}^2-L_{04}^2,
\\
W=(L_{12}L_{24}-L_{13}L_{24}+L_{14}L_{23})^2\nonumber\\
\phantom{W=}{}
-(L_{02}L_{34}-L_{03}L_{24}+L_{04}L_{23})^2-(L_{01}L_{34}-L_{03}L_{14}+L_{04}L_{13})^2\nonumber\\
\label{I-8}
\phantom{W=}{}-(L_{01}L_{24}-L_{02}L_{14}+L_{04}L_{12})^2-(L_{01}L_{23}-L_{02}L_{13}+L_{03}L_{12})^2.
\end{gather}
It is known (see, for example, \cite{Boy}) that the Casimir
operator $W$ vanishes on the representations~$\pi^\sigma$
of~$SO_0(1,4)$. The Casimir operator $F$ takes on the
representations $\pi^\sigma$ the values $-\sigma(\sigma+3)$.

Under restriction to the subgroup $K=SO(4)$, the representation
$\pi^\sigma$ decomposes into a direct sum of irreducible
representations of $SO(4)$ with highest weights $(m,0)$,
$m=0,1,2,\dots$. The group $SO(4)$ is locally isomorphic to the
group $SO(3)\times SO(3)$. The irreducible representation of
$SO(4)$ with the highest weight $(m,0)$, as a representation of
$SO(3)\times SO(3)$, is given by the numbers $\frac m2$ and $\frac
m2$.

Let us consider representations of $SO_0(1,4)$ realized on
functions on the upper sheet of the two-sheeted hyperboloid and on
functions on the upper sheet of the cone. The upper sheet $H^4_+$
of the two-sheeted hyperboloid $H^4$ can be obtained as a quotient
space $SO_0(1,4)/SO(4)$. In order to show this, we consider the
upper sheet $H^4_+$,
\begin{equation}\label{I-9}
H^4_+:\; x_0^2-x_1^2-x_2^2-x_3^2-x_4^2=1,\qquad x_0>0,
\end{equation}
and the point $x^0=(1,0,0,0,0)$ on it. Elements of $SO_0(1,4)$
transform the hyperboloid $H^4_+$ into $H^4_+$. Besides, for any
two points $x'$ and $x''$ of $H^4_+$ there exists an element $g\in
SO_0(1,4)$ such that $gx'=x''$, that is, the group $SO_0(1,4)$
acts transitively on $H^4_+$. A set of points of $SO_0(1,4)$,
leaving the point $x^0$ invariant, coincides with the subgroup
$SO(4)$. Therefore, $H^4_+$ can be identif\/ied with
$SO_0(1,4)/SO(4)$. Note that on $H^4_+$ the {\it 4-dimensional
Lobachevsky space}~$\mathcal{L}^4$ is realized, which is also
called the {\it de Sitter space}. As to the spherical functions on
this space, see in~\cite{Var}.

One constructs on functions $f(x)$ given on $H^4_+$ the {\it
quasi-regular representation} of the group $SO_0(1,4)$. Let
$L^2(H^4_+)$ be the Hilbert space of functions on $H^4_+$ with the
scalar product
\begin{equation}\label{I-10}
\langle f_1,f_2 \rangle = \int_{H_{+}^4} f_1(x)\overline{f_2(x)}\,
d\mu(x),
\end{equation}
where $d\mu(x)$ is an invariant (with respect to $SO_0(1,4)$)
measure on $H^4_+$. This measure is determined by the formula
\[
d\mu(x)=d^4x/x_0=dx_1dx_2dx_3dx_4/x_0.
\]
The quasi-regular representation $\pi$ of $SO_0(1,4)$ is given on
$L^2(H^4_+)$ by the formula
\begin{equation}\label{I-11}
\pi(g)f(x)=f(g^{-1}x),\qquad x\in H_{+}^4.
\end{equation}
It is easy to show that this representation is unitary. However,
this representation is reducible and decomposes \cite{GG} into a
direct integral of irreducible unitary representations
$\pi^\sigma$ ($\sigma=-\frac{3}{2}+i\rho$, $0\le \rho<\infty$).

The quasi-regular representation of $SO_0(1,4)$ on the Hilbert
space $L^2(C^4_+)$ of functions on the upper sheet $C^4_+$ of the
cone $C^4$,
\[
C^4_+:\  x_0^2-x_1^2-x_2^2-x_3^2-x_4^2=0,\qquad x_0>0,
\]
is constructed analogously:
\begin{equation}\label{I-11a}
\pi(g)f(x)=f(g^{-1}x),\qquad x\in C_{+}^4.
\end{equation}

The cone $C^4_+$ can be identif\/ied with the homogeneous space
$SO_0(1,4)/(SO(3)\times N)$. In order to check this we have to
take the point $x^0=(1,0,0,0,1)\in C^4_+$ and to verify that
$SO(3)\times N$ is a~subgroup of $SO_0(1,4)$ whose elements leave
$x^0$ invariant. The subgroup $SO(3)\times N$ is isomorphic to the
group $ISO(3)$ of motions of the 3-dimensional Euclidean space.

The representation \eqref{I-11a} is unitary with respect to the
scalar product
\begin{equation}\label{I-10a}
\langle f_1,f_2 \rangle = \int_{C_{+}^4} f_1(x)\overline{f_2(x)}\,
d\mu(x)
\end{equation}
on $L^2(C^4_+)$. Here $d\mu(x)=d^4x/x_0$ is an invariant (with
respect to $SO_0(1,4)$) measure on $C^4_+$. This representation is
also reducible. Irreducible unitary representations of $SO_0(1,4)$
can be constructed on spaces of homogeneous functions on the cone
(see, for example, \cite{Vil1}). Representations of the group
$SO_0(1,4)$ (as well as of the group $SO_0(1,p)$) are well
investigated (see \cite{Thom, Newt, Dix, Tak, Hir}).

Under exposition of harmonic analysis of functions on the
homogeneous space $H^4_+$ of the group $SO_0(1,4)$ we shall use
the method developed by Vilenkin and Smorodinsky \cite{VSm} for
the Lorentz group $SO_0(1,3)$ and the results of the paper
\cite{GG} related to the integral geometry.

The de Sitter group $SO_0(1,4)$ is used in dif\/ferent branches of
contemporary physics \cite{Gur}. This is a group of motions of a
symmetric Riemannian space-time, which generalizes the Poincar\'e
group $ISO_0(1,3)$ \cite{Frons}. The de Sitter group $SO_0(1,4)$
is known as a group of invariance of non-relativistic hydrogen
atom since it contains the dynamical group $SO_0(1,3)$ (continuous
spectrum) and the dynamical group $SO(4)$ (discrete spectrum)
\cite{Mal, Barut}. There exist also other directions of
applications of the de Sitter group \cite{Kach1}.

\section[Subgroups of the group $SO_0(1,4)$]{Subgroups of the group $\boldsymbol{SO_0(1,4)}$}
\label{Subgroups}

Below we shall construct dif\/ferent bases of the space
$L^2(H^4_+)$ and shall derive formulas for expansion of functions
of~$L^2(H^4_+)$ in these bases. However, we shall consider not
arbitrary bases of~$L^2(H^4_+)$ but only those of them which
correspond to chains of subgroups of the group~$SO_0(1,4)$. These
bases can be constructed also by means of collections of commuting
self-adjoint operators. Basis functions on~$H^4_+$ consist of
common eigenfunctions of these collections of operators. Bases of
the space~$L^2(C^4_+)$ are constructed analogously.

A standard way for construction of a collection of self-adjoint
operators is to construct the corresponding chain of subgroups
$G'\supset G''\supset G'''\supset \cdots$ of $SO_0(1,4)$. Then one
creates commuting self-adjoint operators, consisting of Casimir
operators of the subgroups of the chain. To each such chain
there corresponds a collection of self-adjoint
operators.

We consider these collections of operators as dif\/ferential
operators on homogeneous spaces of the group $SO_0(1,4)$ (on the
hyperboloid $H^4_+$ or on the cone $C^4_+$). For obtaining these
dif\/ferential operators, we construct  coordinate systems such
that the corresponding variables can be separated in the
dif\/ferential equations for eigenvalues and eigenfunctions of the
collection of operators. To each chain  of subgroups  (and, therefore, to each
collection of self-adjoint operators) there corresponds such coordinate system.
  
In this section we determine subgroups of the group $SO_0(1,4)$
that will be used for construction of chains of subgroups.

In order to deal with self-adjoint operators, we shall use the
elements $J_{\mu\nu}={\rm i}L_{\mu\nu}$ of ${\rm so}(1,4)$ instead
of elements $L_{\mu\nu}$. For 10 generators $J_{\mu\nu}$ of the
group $SO_0(1,4)$ we introduce the notations
\begin{gather}\label{I-12a}
{\bf M} =(M_{1}\equiv J_{23},\;M_{2}\equiv J_{31},\;M_{3}\equiv
J_{12}),
\\
\label{I-12b} {\bf P}=(P_{1}\equiv J_{14},\;P_{2}\equiv
J_{24},\;P_{3}\equiv J_{34}),
\\
\label{I-12c} {\bf N}=(N_{1}\equiv J_{01},\;
 N_{2}\equiv J_{02},\;N_{3}\equiv J_{03}),\quad
\\
\label{I-12d} P_{0}=J_{04}.
\end{gather}
In these notations we have the following expressions for Casimir
operators:
\begin{gather*}
F=(P_{0}^2+{\bf N}^2)-({\bf P}^2+{\bf M}^2),
\\
W=({\bf M}\cdot{\bf P})^2-(P_{0}{\bf M}-{\bf P}\times {\bf
N})^2-({\bf M}\cdot{\bf N})^2.
\end{gather*}
Commutation relations for the generators ${\bf M}$, ${\bf P}$,
${\bf N}$ and $P_0$ are of the form
\begin{gather}
[M_{k},M_{l}]={\rm i}\varepsilon_{klm}M_{m},\qquad
[N_{k},N_{l}]=-{\rm i}\varepsilon_{klm}M_{m},\qquad
[P_{k},P_{l}]={\rm i}\varepsilon_{klm}M_{m},\nonumber\\
[M_{k},N_{l}]={\rm i}\varepsilon_{klm}N_{m},\qquad [M_{k},
P_{l}]={\rm i}\varepsilon_{klm}P_{M},\nonumber\\
[M_{k},N_{k}]=[M_{k},P_{k}]=[M_{k},P_{0}]=0,\nonumber\\
\label{I-13} [P_{0},N_{k}]={\rm i}P_{k},\qquad [P_{0},P_{k}]={\rm
i}N_{k},\qquad [P_{k},N_{l}]={\rm i}\delta_{kl}P_{0},
\end{gather}
where $\varepsilon_{klm}$, $k,l,m=1,2,3$, is the antisymmetric
tensor of the third order equal to 0 or $\pm 1$. Using the
generators ${\bf M}$, ${\bf P}$, ${\bf N}$ and $P_0$ we construct
subgroups of the group $SO_0(1,4)$.

{\bf Subgroup $SO(3)$.} This subgroup corresponds to the
generators ${\bf M}=(M_1,M_2,M_3)$ of the Lie algebra ${\rm
so}(1,4)$:
\begin{equation}\label{I-14}
[M_k,M_l]={\rm i}\varepsilon_{klm}M_m.
\end{equation}

{\bf Subgroup $SO(4)$.} There are the generators ${\bf
M}=(M_1,M_2,M_3)$ and ${\bf P}=(P_1,P_2,P_3)$ of the Lie algebra
${\rm so}(1,4)$ corresponding to this subgroup:
\begin{equation}\label{I-15}
[M_k,M_l]={\rm i}\varepsilon_{klm}M_m,\qquad[M_k,P_l]={\rm
i}\varepsilon_{klm}P_m,\qquad [P_k,P_l]={\rm
i}\varepsilon_{klm}M_m.
\end{equation}

Taking the linear combinations ${\bf V}=({\bf M}+{\bf P})/2$,
${\bf V}'=({\bf M}-{\bf P})/2$, we obtain instead of \eqref{I-15}
the relations
\begin{equation}\label{I-16}
[V_k,V_l]={\rm i}\varepsilon_{klm}V_m,\qquad [V_k',V_l']={\rm
i}\varepsilon_{klm}V_m',\qquad [V_k,V'_l]=0,
\end{equation}
that is, triples of the operators ${\bf V}$ and ${\bf V}'$
constitute bases of two independent Lie algebras ${\rm so}(3)$.
This means that the group $SO(4)$ is locally isomorphic to
$SO(3)\otimes SO(3)$.

{\bf Subgroup $SO_0(1,3)$.} It is the Lorentz group which is
generated by ${\bf M}=(M_1,M_2,M_3)$ and ${\bf N}=(N_1,N_2,N_3)$.
We have the commutation relations
\begin{equation}\label{I-17}
[M_k,M_l]={\rm i}\varepsilon_{klm}M_m,\qquad [M_k,N_l]={\rm
i}\varepsilon_{klm}N_m, \qquad [N_k,N_l]=-{\rm
i}\varepsilon_{klm}M_m.
\end{equation}
If we introduce the generators ${\bf L}^{(1)}=({\bf M}+{\rm i}{\bf
N})/2$ and ${\bf L}^{(2)}= ({\bf M}-{\rm i}{\bf N})/2$, then we
obtain the commutation relations for the generators of the Lie
algebra ${\rm so}(3)$:
\begin{equation}\label{I-18}
[L_k^{(\tau)},L_l^{(\tau')}]={\rm i}\varepsilon_{klm}L_m^{(\tau)}
\delta_{\tau\tau'},\qquad \tau,\tau'=1,2.
\end{equation}

{\bf Subgroup $SO_0(1,1)\otimes SO(3)$.} This subgroup is
generated by the generators ${\bf M}=(M_1,M_2$, $M_3)$, $P_0$. We
have
\begin{equation}\label{I-19}
[M_k,M_l]={\rm i}\varepsilon_{klm}M_m,\qquad [M_k,P_0]=0.
\end{equation}

{\bf Subgroup $SO_0(1,2)\otimes SO'(2)$}. This subgroup is
generated by the generators $N_1$, $N_2$, $M_3$,~$P_3$. They
satisfy the commutation relations
\begin{gather}
[N_1,N_2]=-{\rm i}M_3,\qquad [M_3,N_1]={\rm i}N_2,\qquad
[M_3,N_2]=-{\rm i}N_1,\nonumber\\
\label{I-20} [M_3,P_3]=[N_1,P_3]=[N_2,P_3]=0.
\end{gather}
The subgroup $SO'(2)$ of the group $SO_{0}(1,2)\otimes SO'(2)$ is
generated by the generator $P_{3}$.

{\bf Subgroup $ISO(3)$.} We create from the generators $P_1$,
$P_2$, $P_3$ and $N_1$, $N_2$, $N_3$ of the Lie algebra ${\rm
so}(1,4)$ the following linear combinations:
\begin{equation}\label{I-21}
E_{1}=P_{1}+N_{1},\qquad E_{2}=P_{2}+N_{2}, \qquad
E_{3}=P_{3}+N_{3}.
\end{equation}
It is easy to show that two triples of generators ${\bf
E}=(E_{1},E_{2},E_{3})$ and ${\bf M}=(M_{1},M_{2},M_{3})$ satisfy
the commutation relations for the basis generators of the Lie
algebra $ISO(3)$, which is the Lie algebra of the group of motions
of the 3-dimensional Euclidean space $\mathbb{R}^3$. These
relations are
\begin{equation}\label{I-22}
[M_{k},M_{l}]={\rm i}\varepsilon _{klm}M_{m},\qquad
[M_{k},E_{l}]={\rm i}\varepsilon _{klm}E_{m},\qquad
[E_{k},E_{l}]=0.
\end{equation}
This group is a semidirect product of the group $SO(3)$ and the
group $T(3)$ of shifts in $\mathbb{R}^3$, $ISO(3)=SO(3)\times
T(3)$. Note that $T(3)$ coincides with the subgroup $N$.

{\bf Subgroup $ISO(2)$.} Three operators
\begin{equation}\label{I-23}
\mathcal{E}_{1}=E_{1},\qquad \mathcal{E}_{2}=E_{2},\qquad
\mathcal{E}_{3}=E_{3}+M_{3}
\end{equation}
generate the Lie algebra of the group $ISO(2)$ such that
$ISO(2)\subset ISO(3)$. We have
\begin{equation}\label{I-24}
[\mathcal{E}_{1},\mathcal{E}_2]=0,\qquad
[\mathcal{E}_{1},\mathcal{E}_3]=-{\rm i}\mathcal{E}_2,\qquad
[\mathcal{E}_{2},\mathcal{E}_3]={\rm i} \mathcal{E}_{1}.
\end{equation}

{\bf Subgroup $ISO(2)\otimes T_\perp$.} The set of generators
$E_1$, $E_2$, $M_3$, $E_3$ satisfying the commutation relations
\begin{equation}\label{I-25}
[E_k,E_j]=0,\qquad [M_3,E_3]=0,\qquad [M_3,E_2]=-{\rm i}E_1,\qquad
[M_3,E_1]={\rm i}E_2
\end{equation}
generate a Lie algebra of the group $ISO(2)\otimes T_\perp$
belonging to the group $ISO(3)$. The group $T_\perp$ is a group of
shifts along of the direct line, which is perpendicular to the
plane with the group of motion $ISO(2)$. The generators $M_3$ and
$E_3$ generate the subgroup $SO(2)\otimes T_\perp$ of the group
$ISO(2)\otimes T_\perp$ and we have $[M_3,E_3]=0$.

{\bf Subgroup $T(3)$.} A basis of the Lie algebra of the group
$T(3)\subset ISO(3)$ consists of the generators $E_1$, $E_2$,
$E_3$. Each of the generators $E_k$ generates a one-parameter
subgroup $T_k\subset T(3)$ of shifts along the corresponding
coordinate axes.

\section[Coordinate systems on hyperboloid $H^4_+$ and generators
of the group $SO_0(1,4)$]{Coordinate systems on hyperboloid
$\boldsymbol{H^4_+}$\\ and generators of the group
$\boldsymbol{SO_0(1,4)}$} \label{Coord-sys-H^4}

In this section we consider coordinate systems on the hyperboloid
$H^4_+$ and f\/ind dif\/ferential form of generators of the de
Sitter group $SO_0(1,4)$ in an explicit form. Points of the
hyperboloid $H^4_+$ are characterized by 5 orthogonal coordinates
$x_\mu$, $\mu=0,1,2,3,4$, such that
\[
H^4_+:\; [x,x]:=x_0^2-x_1^2-x_2^2-x_3^2-x_4^2=1,\qquad x_0>0.
\]
These coordinates are called {\it homogeneous}. Since $[x,x]=1$,
then the numbers $x_\mu$, giving a point $x$ of the space
$H^4_+\equiv \mathcal{L}^4$, are in fact its projective
coordinates.

New coordinates on $H^4_+$ will be given by means of relations
connecting them with the coordinates $x_\mu$, $\mu=0,1,2,3,4$.
 \bigskip

{\bf Spherical coordinate system $S$} (coordinates $a$, $\beta$,
$\theta$, $\varphi$):
\begin{gather}
x_0=\cosh a,\qquad x_1=\sinh
a\sin\beta\sin\theta\cos\varphi,\qquad x_2=\sinh
a\sin\beta\sin\theta\sin\varphi,
\nonumber\\
\label{I-26} x_3=\sinh a\sin\beta\cos\theta,\qquad x_4=\sinh
a\cos\beta,
\\
0\leqslant a<\infty,\qquad 0\leqslant \beta,\theta<\pi,\qquad
0\leqslant \phi<2\pi.\nonumber
\end{gather}

{\bf Hyperbolic coordinate system $H$} (coordinates $a$, $b$,
$\theta$, $\varphi$):
\begin{gather}
x_0=\cosh a\cosh b,\qquad x_1=\cosh a\sinh b\sin\theta\cos\varphi,
\nonumber\\
\label{I-27} x_2=\cosh a\sinh b\sin\theta\sin\varphi,\qquad
x_3=\cosh a\sinh b\cos\theta,\qquad x_4=\sinh a,
\\
{-}\infty<a<\infty,\qquad  0\leqslant b<\infty,\qquad 0\leqslant
\theta<\pi,\qquad 0\leqslant \varphi<2\pi.\nonumber
\end{gather}

{\bf Orispherical coordinate system $O$} (coordinates $a$, $r$,
$\theta$, $\varphi$):
\begin{gather}
x_0-x_4=e^a,\qquad x_0+x_4=e^{-a}+r^2e^a,
\nonumber\\
\label{I-28} x_1=e^ar\sin\theta\cos\varphi,\qquad
x_2=e^ar\sin\theta\sin\varphi,\qquad x_3=e^ar\cos\theta,
\\
{-}\infty<a<\infty,\qquad 0\leqslant r<\infty,\qquad 0\leqslant
\theta<\pi,\qquad 0\leqslant \varphi<2\pi.\nonumber
\end{gather}

{\bf Orispherically-cylindric coordinate system $OC$} (coordinates
$a$, $\xi$, $z$, $\varphi$):
\begin{gather}
x_0-x_4=e^a,\qquad x_0+x_4=e^{-a}+(\xi^2+ z^2)e^a,\nonumber
\\
\label{I-29} x_1=e^a\xi\cos\varphi,\qquad
x_2=e^a\xi\sin\varphi,\qquad x_3=e^a\xi,
\\
{-}\infty<a<\infty,\qquad 0\leqslant \xi<\infty,\qquad
{-}\infty<z<\infty,\qquad 0\leqslant \varphi<2\pi.\nonumber
\end{gather}

{\bf Orispherically-translational coordinate system $OT$}
(coordinates $a$, $y_1$, $y_2$, $y_3$):
\begin{gather}
x_0-x_4=e^a,\qquad x_0+x_4=e^{-a}{+}y^2e^a,\nonumber
\\
\label{I-30} x_1=e^ay_1,\qquad x_2=e^ay_2,\qquad p_3=e^ay_3,
\\
y^2=y_1^2+y_2^2+y_3^2,\qquad {-}\infty<a<\infty,\qquad
{-}\infty<y_i<\infty.\nonumber
\end{gather}

{\bf Cylindric coordinate system $C$} (coordinates $a$, $b$,
$\theta$, $\varphi$):
\begin{gather}
x_0=\cosh a\cosh b,\qquad x_1=\sinh a\sin\theta\cos\varphi,\qquad
x_2=\sinh a\sin\theta\sin\varphi,\nonumber
\\
\label{I-31} x_3=\sinh a\cos\theta,\qquad x_4=\cosh a\sinh b,
\\
0\leqslant a<\infty,\qquad {-}\infty<b<\infty,\qquad 0\leqslant
\theta<\pi,\qquad 0\leqslant \varphi<2\pi.\nonumber
\end{gather}

{\bf Spherically-hyperbolic coordinate system $SH$} (coordinates
$a$, $b$, $\varphi$, $\Phi$):
\begin{gather}
x_0=\cosh a\cosh b,\qquad x_1=\cosh a\sinh b\cos\varphi,\qquad
x_2=\cosh a\sinh b\sin\varphi,
\nonumber\\
\label{I-32} x_3=\sinh a\cos\Phi,\qquad x_4=\sinh a\sin\Phi,
\\
0\leqslant a,b<\infty,\qquad 0\leqslant \varphi,\Phi<2\pi.
\nonumber
\end{gather}

Now we go to the representation \eqref{I-11} of the group
$SO_0(1,4)$ realized on the space $L^2(H^4_+)$. This
representation gives in fact a realization of this group on the
hyperboloid $H^4_+$. Functions of the space $L^2(H^4_+)$ can be
considered as functions of parameters of any of the coordinate
systems on $H^4_+$. We shall give a dif\/ferential form of
inf\/initesimal operators of the representation \eqref{I-11} or,
equivalently, a dif\/ferential form of generators of the group
$SO_0(1,4)$, realized on $H^4_+$.

Let $I$ be an inf\/initesimal generator of the group $SO_0(1,4)$,
realized on $H^4_+$. Then $\exp tI$ is a~one-parameter subgroup of
$SO_0(1,4)$. The operator $\pi(\exp tI)$ acts on $L^2(H^4_+)$ as
\[
\pi(\exp tI) f(x)=f((\exp tI)^{-1}x).
\]
Since $\pi(I)=\frac d{dt} \pi(\exp tI)_{t=0}$, then
\begin{equation}\label{I-33}
\pi(I)f(x)=\lim_{t\to 0} \frac{f((\exp tI)^{-1}x)-f(x)}{t}.
\end{equation}
Thus, in the homogeneous coordinates $x_\mu$ we have
\begin{equation}\label{I-34}
J_{rs}={-}{\rm i}\left( x_r\frac{\partial}{\partial
x_s}-x_s\frac{\partial}{\partial x_r} \right),\qquad J_{0s}=-{\rm
i}\left( x_0 \frac{\partial}{\partial x_s}+x_s
\frac{\partial}{\partial x_0} \right).
\end{equation}
(Here and below we write down the operators $\pi(I)$ as $I$.)
Substituting into \eqref{I-34} the expressions for $x_\mu$,
$\mu=0,1,2,3,4$, in terms of the corresponding coordinates we
f\/ind a dif\/ferential form of the generators $I$ in these
coordinates. Let us give the result of such calculation.

{\bf $S$-system:}
\begin{gather*}
M_1=-{\rm i}\left(-\sin\varphi\frac{\partial}{\partial
\theta}-\cot\theta\cos\varphi\frac{\partial}{\partial
\varphi}\right), \qquad M_2=-{\rm
i}\left(\cos\varphi\frac{\partial}{\partial
\theta}-\cot\theta\sin\varphi\frac{\partial}{\partial
\varphi}\right),
\\
 M_3=-{\rm i}\frac{\partial}{\partial \varphi},
\qquad P_0=-{\rm i}\left(\cos\beta\frac{\partial}{\partial
a}-\coth a\sin\beta\frac{\partial}{\partial \beta}\right),
\\
P_1=-{\rm i}\left(-\sin\theta\cos\varphi\frac{\partial}{\partial
\beta}-\cot\beta\cos\theta\cos\varphi\frac{\partial}{\partial\theta}+
\cot\beta\frac{\sin\varphi}{\sin\theta}\frac{\partial}{\partial\varphi}
\right),
\\
P_2=-{\rm i}\left(-\sin\theta\sin\varphi\frac{\partial}{\partial
\beta}-\cot\beta\cos\theta\sin\varphi\frac{\partial}{\partial\theta}-
\cot\beta\frac{\cos\varphi}{\sin\theta}\frac{\partial}{\partial\varphi}
\right),
\\
P_3=-{\rm i}\left(-\cos\theta\frac{\partial}{\partial
\beta}+\cot\beta\sin\theta\frac{\partial}{\partial\theta}\right),
\\
N_1=-{\rm
i}\left(\sin\beta\sin\theta\cos\varphi\frac{\partial}{\partial
a}+\coth a
\cos\beta\sin\theta\cos\varphi\frac{\partial}{\partial\beta}
\right.
\\
\left.\phantom{N_1=}{} +\coth
a\frac{\cos\theta\cos\varphi}{\sin\beta}
\frac{\partial}{\partial\theta}-\coth
a\frac{\sin\varphi}{\sin\beta\sin\theta}\frac{\partial}{\partial\varphi}
\right),
\\
N_2=-{\rm
i}\left(\sin\beta\sin\theta\sin\varphi\frac{\partial}{\partial
a}+\coth a
\cos\beta\sin\theta\sin\varphi\frac{\partial}{\partial\beta}
\right.
\\
\left.\phantom{N_2=}{}+\coth a
\frac{\cos\theta\sin\varphi}{\sin\beta}\frac{\partial}{\partial\theta}
+ \coth
a\frac{\cos\varphi}{\sin\beta\sin\theta}\frac{\partial}{\partial\varphi}
\right),
\\
 N_3=-{\rm i}\left(\sin\beta\cos\theta\frac{\partial}{\partial
a}+\coth a \cos\beta\cos\theta\frac{\partial}{\partial\beta}-\coth
a \frac{\sin\theta}{\sin\beta} \frac{\partial}{\partial\theta}
\right).
\end{gather*}

{\bf $H$-system:}
\begin{gather*}
M_1=- {\rm i}\left( {-}\sin\varphi\frac{\partial}{\partial
\theta}-\cot\theta\cos\varphi\frac{\partial}{\partial \varphi}
\right), \qquad
 M_2=-{\rm i}\left( \cos\varphi\frac{\partial}{\partial
\theta}{-}\cot\theta\sin\varphi\frac{\partial}{\partial \varphi}
\right),\\ M_3=-{\rm i}\frac{\partial}{\partial \varphi}, \qquad
P_0=-{\rm i}\left( \cosh b\frac{\partial}{\partial a}-\tanh a\sinh
b\frac{\partial}{\partial b} \right),
\\
P_1=-{\rm i}\left( \sinh
b\sin\theta\cos\varphi\frac{\partial}{\partial a}-\tanh a\cosh
b\sin\theta\cos\varphi\frac{\partial}{\partial b} \right.
\\   \left.
\phantom{P_1=}{} -\tanh a\frac{\cos\theta\cos\varphi}{\sinh
b}\frac{\partial}{\partial \theta}+\tanh a\frac{\sin\varphi}{\sinh
b\sin\theta}\frac{\partial}{\partial \varphi} \right),
\\
P_2=-{\rm i}\left( \sinh
b\sin\theta\sin\varphi\frac{\partial}{\partial a}-\tanh a\cosh
b\sin\theta\sin\varphi\frac{\partial}{\partial b} \right.
\\    \left.
\phantom{P_2=}{} -\tanh a\frac{\cos\theta\sin\varphi}{\sinh
b}\frac{\partial}{\partial \theta}-\tanh a\frac{\cos\varphi}{\sinh
b\sin\theta}\frac{\partial}{\partial \varphi} \right),
\\
P_3=-{\rm i}\left( \sinh b\cos\theta\frac{\partial}{\partial
a}-\tanh a\cosh b\cos\theta\frac{\partial}{\partial b}+\tanh
a\frac{\sin\theta}{\sinh b}\frac{\partial}{\partial \theta}
\right),
\\
N_1=-{\rm i}\left( \sin\theta\cos\varphi\frac{\partial}{\partial
b}+ \coth b\cos\theta\cos\varphi\frac{\partial}{\partial
\theta}-\coth
b\frac{\sin\varphi}{\sin\theta}\frac{\partial}{\partial \varphi}
\right),
\\
N_2=-{\rm i}\left( \sin\theta\sin\varphi\frac{\partial}{\partial
b}+\coth b\cos\theta\sin\varphi\frac{\partial}{\partial
\theta}+\coth
b\frac{\cos\varphi}{\sin\theta}\frac{\partial}{\partial \varphi}
\right),
\\
N_3=-{\rm i}\left( \cos\theta\frac{\partial}{\partial b}-\coth
b\sin\theta\frac{\partial}{\partial\theta} \right).
\end{gather*}

{\bf $O$-system:}
\begin{gather*}
M_1=-{\rm i}\left(-\sin\varphi\frac{\partial}{\partial
\theta}-\cot\theta\cos\varphi\frac{\partial}{\partial
\varphi}\right), \qquad M_2=-{\rm
i}\left(\cos\varphi\frac{\partial}{\partial
\theta}-\cot\theta\sin\varphi\frac{\partial}{\partial
\varphi}\right),\\ M_3=-{\rm i}\frac{\partial}{\partial \varphi},
\qquad P_0=-{\rm i}\left(-\frac{\partial}{\partial
a}+r\frac{\partial}{\partial r}\right),
\\
P_1=-{\rm i}\left\{-r\sin\theta\cos\varphi\frac{\partial}{\partial
 a}+\frac{e^{-a}}{2}[-e^{-a}+(r^2+1)e^a]
 \sin\theta\cos\varphi\frac{\partial}{\partial r} \right.
\\
\left.\phantom{P_1=}{}-\frac{e^{-a}}{2r}
 [e^{-a}+(r^2-1)e^a]\cos\theta\cos\varphi\frac{\partial}{\partial\theta}+
\frac{e^{-a}}{2r}[e^{-a}+(r^2-1)e^a]\frac{\sin\varphi}{\sin\theta}
\frac{\partial}{\partial\varphi}\right\},
\\
P_2=-{\rm i}\left\{-r\sin\theta\sin\varphi\frac{\partial}{\partial
 a}+\frac{e^{-a}}{2}[-e^{-a}+(r^2+1)e^a]
 \sin\theta\sin\varphi\frac{\partial}{\partial r} \right.
\\
\left.    \phantom{P_2=}{}  -\frac{e^{-a}}{2r}
 [e^{-a}+(r^2-1)e^a]\cos\theta\sin\varphi\frac{\partial}{\partial\theta}-
\frac{e^{-a}}{2r}[e^{-a}+(r^2-1)e^a]\frac{\cos\varphi}{\sin\theta}
\frac{\partial}{\partial\varphi}\right\},
\\
P_3=-{\rm i}\left\{-r\cos\theta\frac{\partial}{\partial
a}+\frac{e^{-a}}{2}[-e^{-a}+(r^2+1)e^a]\cos\theta\frac{\partial}{\partial
r}  \right.
\\
\left.
\phantom{P_3=}{}+\frac{e^{-a}}{2r}[e^{-a}+(r^2-1)e^a]\sin\theta\frac{\partial}{\partial
\theta}\right\},
\\
N_1=-{\rm i}\left\{r\sin\theta\cos\varphi\frac{\partial}{\partial
 a}-\frac{e^{-a}}{2}[-e^{-a}+(r^2-1)e^a]
 \sin\theta\cos\varphi\frac{\partial}{\partial r}  \right.
\\
 \left. \phantom{N_1=}{}+\frac{e^{-a}}{2r}
 [e^{-a}+(r^2+1)e^a]\cos\theta\cos\varphi\frac{\partial}{\partial
\theta}-\frac{e^{-a}}{2r}
 [e^{-a}+(r^2+1)e^a]\frac{\sin\varphi}{\sin\theta}
\frac{\partial}{\partial\varphi}\right\},
\\
N_2=-{\rm i}\left\{r\sin\theta\sin\varphi\frac{\partial}{\partial
 a}-\frac{e^{-a}}{2}[-e^{-a}+(r^2-1)e^a]
 \sin\theta\sin\varphi\frac{\partial}{\partial r}  \right.
\\
\left.\phantom{N_2=}{}+\frac{e^{-a}}{2r}
 [e^{-a}+(r^2+1)e^a]\cos\theta\sin\varphi\frac{\partial}{\partial\theta}+
\frac{e^{-a}}{2r}[e^{-a}+(r^2+1)e^a]\frac{\cos\varphi}{\sin\theta}
\frac{\partial}{\partial\varphi}\right\},
\\
N_3=-{\rm i}\left\{r\cos\theta\frac{\partial}{\partial
a}-\frac{e^{-a}}{2}[-e^{-a}+(r^2-1)e^a]\cos\theta\frac{\partial}{\partial
r} \right.
\\  \left. \phantom{N_3=}{}
-\frac{e^{-a}}{2r}
 [e^{-a}+(r^2+1)e^a]\sin\theta\frac{\partial}{\partial\theta}\right\}.
\end{gather*}

{\bf $OC$-system:}
\begin{gather*}
M_1=-{\rm
i}\left(-z\sin\varphi\frac{\partial}{\partial\xi}+\xi\sin\varphi
\frac{\partial}{\partial
z}-z\frac{\cos\varphi}{\xi}\frac{\partial}{\partial\varphi}\right),
\\
M_2=-{\rm
i}\left(z\cos\varphi\frac{\partial}{\partial\xi}-\xi\cos\varphi
\frac{\partial}{\partial
z}-z\frac{\sin\varphi}{\xi}\frac{\partial}{\partial\varphi}\right),\qquad
 M_3=-{\rm i}\frac{\partial}{\partial\varphi},
\\
P_0=-{\rm i}\left(-\frac{\partial}{\partial
a}+\xi\frac{\partial}{\partial\xi}+z\frac{\partial}{\partial
z}\right),
\\
P_1=-{\rm i}\left\{-\xi\cos\varphi\frac{\partial}{\partial
a}+\frac{e^{-a}}{2}[-e^{-a}+(\xi^2-z^2+1)e^a]\cos\varphi
\frac{\partial}{\partial\xi}  \right.
\\
 \left.\phantom{P_1=}{}+\xi
z\cos\varphi\frac{\partial}{\partial
z}+\frac{e^{-a}}{2\xi}[e^{-a}+(\xi^2+z^2-1)e^a]\sin\varphi
\frac{\partial}{\partial\varphi}\right\},
\\
P_2=-{\rm i}\left\{-\xi\sin\varphi\frac{\partial}{\partial
a}+\frac{e^{-a}}{2}[-e^{-a}+(\xi^2-z^2+1)e^a]\sin\varphi
\frac{\partial}{\partial\xi} \right.
\\
\left.\phantom{P_2=}{}+\xi z\sin\varphi\frac{\partial}{\partial
z}-\frac{e^{-a}}{2\xi}[e^{-a}+(\xi^2+z^2-1)e^a]\cos\varphi
\frac{\partial}{\partial\varphi}\right\},
\\
P_3=-{\rm i}\left\{-z\frac{\partial}{\partial
a}+z\xi\frac{\partial}{\partial\xi}+
\frac{e^{-a}}{2}[e^{-a}-(z^2-\xi^2+1)e^a] \frac{\partial}{\partial
z}\right\},
\\
N_1=-{\rm i}\left\{\xi\cos\varphi\frac{\partial}{\partial
a}+\frac{e^{-a}}{2}[e^{-a}+(z^2-\xi^2+1)e^a]\cos\varphi
\frac{\partial}{\partial\xi} \right.
\\
 \left.\phantom{N_1=}{}-z\xi
\cos\varphi\frac{\partial}{\partial
z}-\frac{e^{-a}}{2\xi}[e^{-a}+(\xi^2+z^2+1)e^a]\sin\varphi
\frac{\partial}{\partial\varphi}\right\},
\\
N_2=-{\rm i}\left\{\xi\sin\varphi\frac{\partial}{\partial
a}+\frac{e^{-a}}{2}[e^{-a}+(z^2-\xi^2+1)e^a]\sin\varphi
\frac{\partial}{\partial\xi} \right.
\\
\left.\phantom{N_2=}{}-z\xi \sin\varphi\frac{\partial}{\partial
z}+\frac{e^{-a}}{2\xi}[e^{-a}+(\xi^2+z^2+1)e^a]\cos\varphi
\frac{\partial}{\partial\varphi}\right\},
\\
N_3=-{\rm i}\left\{z\frac{\partial}{\partial
a}-z\xi\frac{\partial}{\partial\xi}+
\frac{e^{-a}}{2}[e^{-a}+(\xi^2-z^2+1)e^a] \frac{\partial}{\partial
z}\right\}.
\end{gather*}

{\bf $OT$-system:}
\begin{gather*}
M_1=-{\rm i}\left(y_2\frac{\partial}{\partial
y_3}-y_3\frac{\partial}{\partial y_2}\right),\qquad M_2=-{\rm
i}\left(y_3\frac{\partial}{\partial
y_1}-y_1\frac{\partial}{\partial y_3}\right),
\\
M_3=-{\rm i}\left(y_1\frac{\partial}{\partial
y_2}-y_2\frac{\partial}{\partial y_1}\right), \qquad P_0=-{\rm
i}\left(-\frac{\partial}{\partial a}+y_1\frac{\partial}{\partial
y_1}+y_2\frac{\partial}{\partial y_2}+y_3\frac{\partial}{\partial
y_3}\right),
\\
P_1=-{\rm i}\left\{-y_1\frac{\partial}{\partial
a}-\frac{e^{-a}}{2}[e^{-a}+(y^2-1)e^a]\frac{\partial}{\partial
y_1}+y_1(y_1\frac{\partial}{\partial
y_1}+y_2\frac{\partial}{\partial y_2}+y_3\frac{\partial}{\partial
y_3})\right\},
\\
 P_2=-{\rm i}\left\{-y_2\frac{\partial}{\partial
a}-\frac{e^{-a}}{2}[e^{-a}+(y^2-1)e^a]\frac{\partial}{\partial
y_2}+y_2(y_1\frac{\partial}{\partial
y_1}+y_2\frac{\partial}{\partial y_2}+y_3\frac{\partial}{\partial
y_3})\right\},
\\
 P_3=-{\rm i}\left\{-y_3\frac{\partial}{\partial
a}-\frac{e^{-a}}{2}[e^{-a}+(y^2-1)e^a]\frac{\partial}{\partial
y_3}+y_3(y_1\frac{\partial}{\partial
y_1}+y_2\frac{\partial}{\partial y_2}+y_3\frac{\partial}{\partial
y_3})\right\},
\\
 N_1=-{\rm i}\left\{y_1\frac{\partial}{\partial
a}-\frac{e^{-a}}{2}[e^{-a}+(y^2+1)e^a]\frac{\partial}{\partial
y_1}-y_1(y_1\frac{\partial}{\partial
y_1}+y_2\frac{\partial}{\partial y_2}+y_3\frac{\partial}{\partial
y_3})\right\},
\\
 N_2=-{\rm i}\left\{y_2\frac{\partial}{\partial
a}-\frac{e^{-a}}{2}[e^{-a}+(y^2+1)e^a]\frac{\partial}{\partial
y_2}-y_2(y_1\frac{\partial}{\partial
y_1}+y_2\frac{\partial}{\partial y_2}+y_3\frac{\partial}{\partial
y_3})\right\},
\\
 N_3=-{\rm i}\left\{y_3\frac{\partial}{\partial
a}-\frac{e^{-a}}{2}[e^{-a}+(y^2+1)e^a]\frac{\partial}{\partial
y_3}-y_3(y_1\frac{\partial}{\partial
y_1}+y_2\frac{\partial}{\partial y_2}+y_3\frac{\partial}{\partial
y_3})\right\},
\end{gather*}

{\bf $C$-system:}
\begin{gather*}
M_1=-{\rm i}\left(
{-}\sin\varphi\frac{\partial}{\partial\theta}-\cot\theta\cos\varphi
\frac{\partial}{\partial\varphi} \right),
\\
 M_2=-{\rm i}\left(
\cos\varphi\frac{\partial}{\partial\theta}- \cot
\theta\sin\varphi\frac{\partial}{\partial\varphi} \right),\qquad
M_3=-{\rm i}\frac{\partial}{\partial\varphi},\qquad P_0=-{\rm
i}\frac{\partial}{\partial b},
\\
P_1=-{\rm i}\left( -\sinh
b\sin\theta\cos\varphi\frac{\partial}{\partial a}{+}\tanh a\cosh
b\sin\theta\cos\varphi\frac{\partial}{\partial b} \right.
\\\left.
\phantom{P_1=}{} -\coth a\sinh
b\cos\theta\cos\varphi\frac{\partial}{\partial\theta}+\coth a\sinh
b\frac{\sin\varphi}{\sin\theta}\frac{\partial}{\partial\varphi}
\right),
\\
P_2= -{\rm i}\left( -\sinh
b\sin\theta\sin\varphi\frac{\partial}{\partial a}+\tanh a\cosh
b\sin\theta\sin\varphi\frac{\partial}{\partial b}  \right.
\\   \left.
\phantom{P_2=}{}-\coth a\sinh
b\cos\theta\sin\varphi\frac{\partial}{\partial\theta}-\coth a\sinh
b\frac{\cos\varphi}{\sin\theta}\frac{\partial}{\partial\varphi}
\right),
\\
P_3=-{\rm i}\left( -\sinh b\cos\theta\frac{\partial}{\partial
a}+\tanh a\cosh b\cos\theta\frac{\partial}{\partial b}+\coth
a\sinh b\sin\theta\frac{\partial}{\partial\theta} \right),
\\
N_1=-{\rm i}\left( \cosh
b\sin\theta\cos\varphi\frac{\partial}{\partial a}-\tanh a\sinh
b\sin\theta\cos\varphi\frac{\partial}{\partial b}  \right.
\\    \left.
\phantom{N_1=}{}+\coth a\cosh
b\cos\theta\cos\varphi\frac{\partial}{\partial\theta}-\coth a\cosh
b\frac{\sin\varphi}{\sin\theta}\frac{\partial}{\partial\varphi}
\right),
\\
N_2=-{\rm i}\left( \cosh
b\sin\theta\sin\varphi\frac{\partial}{\partial a}-\tanh a\sinh
b\sin\theta\sin\varphi\frac{\partial}{\partial b}  \right.
\\   \left.
\phantom{N_2=}{}+\coth a\cosh
b\cos\theta\sin\varphi\frac{\partial}{\partial\theta}+\coth a\cosh
b\frac{\cos\varphi}{\sin\theta}\frac{\partial}{\partial\varphi}
\right),
\\
 N_3=-{\rm i}
\left( \cosh b\cos\theta\frac{\partial}{\partial a}-\tanh a\sinh
b\cos\theta\frac{\partial}{\partial b}-\coth a\cosh
b\sin\theta\frac{\partial}{\partial\theta} \right).
\end{gather*}

{\bf $SH$-system:}
\begin{gather*}
M_1=-{\rm i}\left(\sinh
b\sin\varphi\cos\Phi\frac{\partial}{\partial a}-\tanh a\cosh
b\sin\varphi\cos\Phi\frac{\partial}{\partial b}   \right.
\\
\left.\phantom{M_1=}{}-\tanh a\frac{\cos\varphi\cos\Phi}{\sinh
b}\frac{\partial}{\partial\varphi}-\coth a\sinh
 b\sin\varphi\sin\Phi\frac{\partial}{\partial\Phi}\right),
\\
M_2=-{\rm i}\left(-\sinh
b\cos\varphi\cos\Phi\frac{\partial}{\partial a}+\tanh a\cosh
b\cos\varphi\cos\Phi\frac{\partial}{\partial b} \right.
\\
\left.\phantom{M_2=}{}-\tanh a\frac{\sin\varphi\cos\Phi}{\sinh
b}\frac{\partial}{\partial\varphi}+\coth a\sinh b
\cos\varphi\sin\Phi\frac{\partial}{\partial\Phi}\right),
\\
M_3=-{\rm i}\frac{\partial}{\partial\varphi},\qquad P_3=-{\rm
i}\frac{\partial}{\partial\Phi},
\\
 P_1=-{\rm i}\left(\sinh
b\cos\varphi\sin\Phi\frac{\partial}{\partial a}-\tanh a\cosh
b\cos\varphi\sin\Phi\frac{\partial}{\partial b}   \right.
\\
\left.\phantom{P_1=}{}+\tanh a\frac{\sin\varphi\sin\Phi}{\sinh
b}\frac{\partial}{\partial\varphi}+\coth a\sinh
 b\cos\varphi\cos\Phi\frac{\partial}{\partial\Phi}\right),
\\
P_2=-{\rm i}\left(\sinh
b\sin\varphi\sin\Phi\frac{\partial}{\partial a}-\tanh a\cosh
b\sin\varphi\sin\Phi\frac{\partial}{\partial b}  \right.
\\
\left.\phantom{P_2=}{}-\tanh a\frac{\cos\varphi\sin\Phi}{\sinh
b}\frac{\partial}{\partial\varphi}+\coth a\sinh b
\sin\varphi\cos\Phi\frac{\partial}{\partial\Phi}\right),
\\
N_1=-{\rm i}\left(\cos\varphi\frac{\partial}{\partial b}-\coth
b\sin\varphi\frac{\partial}{\partial\varphi}\right),\ \ \ \
N_2=-{\rm i}\left(\sin\varphi\frac{\partial}{\partial b}+\coth
b\cos\varphi\frac{\partial}{\partial\varphi}\right),
\\
N_3=-{\rm i}\left(\cosh b\cos\Phi\frac{\partial}{\partial a}-\tanh
a\sinh b\cos\Phi\frac{\partial}{\partial b} -\coth a\cosh
b\sin\Phi \frac{\partial}{\partial\Phi}\right),
\\
 P_0=-{\rm i}\left(\cosh
b\sin\Phi\frac{\partial}{\partial a}-\tanh a\sinh b\sin
\Phi\frac{\partial}{\partial b} +\coth a\cosh b\cos\Phi
\frac{\partial}{\partial\Phi}\right).
\end{gather*}

Below the corresponding expressions for the inf\/initesimal
generators $E_i$, $i=1,2,3$, in $O$-, $OC$- and $OT$-systems will
be used. They have the form

{\bf $O$-system:}
\begin{gather*}
E_1=-{\rm i}\left(\sin\theta\cos\varphi\frac{\partial}{\partial
r}+\frac{1}{r}\cos\theta\cos\varphi\frac{\partial}{\partial
\theta}-\frac{1}{r}\frac{\sin\varphi}{\sin\theta}
\frac{\partial}{\partial\varphi}\right),
\\
E_2=-{\rm i}\left(\sin\theta\sin\varphi\frac{\partial}{\partial
r}+\frac{1}{r}\cos\theta\sin\varphi\frac{\partial}{\partial
\theta}+\frac{1}{r}\frac{\cos\varphi}{\sin\theta}
\frac{\partial}{\partial\varphi}\right),
\\
E_3=-{\rm i}\left(\cos\theta\frac{\partial}{\partial
r}-\frac{1}{r}\sin\theta\frac{\partial}{\partial \theta}\right).
\end{gather*}

{\bf $OC$-system:}
\begin{gather*}
E_1=-{\rm i}\left(\cos\varphi\frac{\partial}{\partial
\xi}-\frac{\sin\varphi}{\xi}
\frac{\partial}{\partial\varphi}\right),\qquad E_2=-{\rm
i}\left(\sin\varphi\frac{\partial}{\partial\xi}+\frac{\cos\varphi}{\xi}
\frac{\partial}{\partial\varphi} \right),\qquad E_3=-{\rm
i}\frac{\partial}{\partial z}.
\end{gather*}

{\bf $OT$-system:}
\begin{gather*}
E_1=-{\rm i}\left(\frac{\partial}{\partial y_1}\right),\qquad
E_2=-{\rm i}\left(\frac{\partial}{\partial y_2}\right),\qquad
E_3=-{\rm i}\left(\frac{\partial}{\partial y_3}\right).
\end{gather*}

\section{Invariant operators on hyperboloid and their eigenfunctions}
\label{Eigen-hyperb}

\subsection{Introduction}
\label{Eigen-intr}

For each coordinate system on the hyperboloid $H^4_+$ we shall
f\/ind basis functions of the space $L^2(H^4_+)$. They are
constructed as common eigenfunctions of a full collection of
commuting self-adjoint operators. Casimir operators of the group
$SO_0(1,4)$ and of its subgroups are included into these
collections. We shall see that the coordinate systems, considered
above, are determined by the corresponding chains of subgroups of
the group $SO_0(1,4)$.

As we have seen, the Casimir operators
\begin{gather}
F=(P_0^2+{\bf N}^2)-({\bf P}^2+{\bf M}^2),
\\
W=({\bf M}\cdot{{\bf P}})^2-(P_0{\bf M}-{\bf P}\times{\bf N})^2-
({\bf M}\cdot{\bf N})^2
\end{gather}
are independent invariants of the Lie algebra ${\rm so}(1,4)$. The
second operator vanishes on the space~$L^2(H^4_+)$. We include the
operator $F$ into full collections of commuting self-adjoint
operators.

The quasi-regular representation \eqref{I-11} is reducible. Since
it is unitary, it decomposes into a direct integral of irreducible
unitary representations \cite{GG}.  Since $H^4_+\equiv
SO_0(1,4)/SO(4)$ and~$SO(4)$ is a compact subgroup, this
decomposition can be obtained from the Fourier transform and the
Plancherel formula for the regular representation of the group
$SO_0(1,4)$ \cite{Hir2}. The result of the decomposition of the
quasi-regular representation $\pi$ of $SO_0(1,4)$ on $L^2(H^4_+)$
is the following. {\it The representation $\pi$ on the space
$L^2(H^4_+)$ decomposes into the direct integral of the principal
unitary representations $\pi^{\sigma}$, $\sigma={\rm i}\rho-\frac
32$, $0\leqslant \rho<\infty$, and each of these irreducible
representations is contained in the decomposition only once.}

The spectrum of the Casimir operator $F$ on $L^2(H^4_+)$ is
determined by this decomposition, since on the representations
$\pi^{\sigma}$, $\sigma={\rm i}\rho-\frac 32$, $0\leqslant
\rho<\infty$, this operator is multiple to the unit operator.

Since on the representation $\pi^{\sigma}$ the operator $F$ takes
the value $-\sigma(\sigma+3)$, its spectrum on $L^2(H^4_+)$
consists of the points
\[
-({\rm i}\rho-3/2)({\rm i}\rho+3/2)=\rho^2+9/4,\qquad
0\leqslant\rho<\infty .
\]
The operator $-F$ on $L^2(H^4_+)$ coincides with the Laplace
operator $\Delta(H^4_+)$ in the corresponding coordinate system
(see~\cite{LNR}).

If we have a collection of commuting self-adjoint operators for
each of coordinate systems \eqref{I-26}--\eqref{I-32}, we can
f\/ind their common eigenfunctions, which constitute a basis of
the space $L^2(H^4_+)$. It is not a basis in the usual sense. It
is rather a ``continuous'' basis (corresponding to continuous
spectra of commuting operators). A strict mathematical meaning to
such bases can done by using the results of~\cite{Maur}.
Eigenfunctions of collections of self-adjoint operators will be
found by means of the corresponding dif\/ferential
equations~\cite{VSm}.

A dif\/ferential form of commuting operators, which are contained
in a collection, is found by means of dif\/ferential form of the
inf\/initesimal operators calculated above.

\subsection[$H$-system]{$\boldsymbol{H}$-system}
\label{eig-h}

The operator $\Delta(H_+^4)=-F$ in the coordinates $a$, $b$,
$\theta$, $\varphi$ has the form
\begin{gather}
\Delta(H_+^4)=\frac{1}{\cosh^3a}
\frac{\partial}{\partial{a}}\cosh^3a\frac{\partial}{\partial{a}}+
\frac{1}{\cosh^2a\sinh^2b}\left[
\frac{\partial}{\partial{b}}\sinh^2b\frac{\partial}{\partial{b}}\right.\nonumber\\
\label{I-35}      \left.\phantom{\Delta(H_+^4)=}{}
+\frac{1}{\sin{\theta}}\left(
\frac{\partial}{\partial{\theta}}\sin{\theta}
\frac{\partial}{\partial{\theta}}+
\frac{1}{\sin{\theta}}\frac{\partial^2}{\partial\varphi^2}\right)
\right].
\end{gather}
The operators
\begin{gather}
{\bf N}^2-{\bf M}^2\equiv{N_1^2+N_2^2+N_3^2-M_1^2-M_2^2-M_3^2}
\nonumber\\
\label{I-36}      \phantom{{\bf N}^2-{\bf M}^2}{}
=-\frac{1}{\sinh^2b}\left(\frac{\partial}{\partial{b}}\sinh^2b
\frac{\partial}{\partial{b}}+
\frac{1}{\sin\theta}\frac{\partial}{\partial{\theta}}\sin{\theta}
\frac{\partial}{\partial{\theta}}+
\frac{1}{\sin^2\theta}\frac{\partial^2}{\partial{\varphi^2}}\right),
\\
\label{I-37} {\bf
M}^2=-\left(\frac{1}{\sin\theta}\frac{\partial}{\partial{\theta}}
\sin{\theta}\frac{\partial}{\partial{\theta}}+
\frac{1}{\sin^2\theta}\frac{\partial^2}{\partial{\varphi^2}}\right),\qquad
M_3=-{\rm i}\frac{\partial}{\partial{\varphi}}
\end{gather}
commute with the operator $\Delta(H_+^4)$.

The operators $\Delta(H_+^4)$, ${\bf N}^2-{\bf M}^2$, ${\bf M}^2$,
$M_3^2$ are quadratic Casimir operators for the groups
$SO_0(1,4)$, $SO_0(1,3)$, $SO(3)$ and $SO(2)$, respectively.

Let us f\/ind spectra and common eigenfunctions of self-adjoint
operators $\Delta(H_+^4)$, ${\bf N}^2-{\bf M}^2$, ${\bf M}^2$ and
$M_3^2$ on the space $L^2(H^4_+)$. (If the operators~\eqref{I-35}
and~\eqref{I-36} are considered as a product of the
inf\/initesimal operators of the quasi-regular representation
$\pi$, then they are symmetric operators and have self-adjoint
extensions. If we speak about self-adjoint operators~\eqref{I-35}
and~\eqref{I-36}, then we understand that they are these
self-adjoint extensions. This remark concerns also to other
collections of operators considered below.)

The spectrum of the operator $\Delta(H_+^4)$ is described above.
Let us discuss the spectrum of the operator $-({\bf N}^2-{\bf
M}^2)$. Its form coincides with the Laplace operator
$\Delta(H_+^3)$ on the space $L^2(H^3_+)$ in the spherical
coordinate system on $H^3_+$, where $H^3_+$ is the upper sheet of
the two-sheeted hyperboloid in 4-dimensional Minkowski space, that
is, $H^3_+ \equiv SO_0(1,3)/SO(3)$. However, we have to consider
$H^3_+$ as a subset of~$H^4_+$. Let us take the hyperbolic
coordinate system~$H$ on~$H^4_+$. The operator ${\bf N}^2-{\bf
M}^2$ does not depend on the coordinate $a$. At each f\/ixed~$a$
a~point $(x_0,x_1, x_2,x_3,x_4=\sinh a)\in H^4_+$ runs over the
upper sheet of two-sheeted hyperboloid
$x_0^2-x_1^2-x_2^2-x_3^2=\cosh^2 a$ in the 4-dimensional space of
points $(x_0,x_1,x_2,x_3,x_4)$. The coordinate~$a$ runs over the
real line $\mathbb{R}$. It is evident that there are identical
3-dimensional hyperboloids $x_0^2-x_1^2-x_2^2-x_3^2=\cosh^2 a$
with dif\/ferent values of the coordinate~$x_4$: $x_4=\sinh a$ and
$x_4=-\sinh a$ corresponding to the points~$a$ and $-a$
respectively.

The above reasoning shows that the operator ${\bf M}^2-{\bf N}^2$
leads to two Laplace operators $\Delta(H_+^3)$ on $L^2(H^4_+)$.
One of them corresponds to the value $a\in (-\infty,0)$ and the
other to the value $a\in (0,\infty)$.

Thus, in order to f\/ind a whole spectrum of eigenvalues and the
corresponding eigenfunctions of the operators
\eqref{I-35}--\eqref{I-37} it is necessary to solve two systems of
equations
\begin{gather}
\Delta(H_+^4)\Phi_{\rho\nu lm}^\varepsilon(a,b,\theta,\varphi)=
-(\rho^2+^9/_4)\Phi_{\rho\nu
lm}^\varepsilon(a,b,\theta,\varphi),\nonumber\\
\Delta(H_+^3)\Phi_{\rho\nu lm}^\varepsilon(a,b,\theta,\varphi)=
-(\nu^2+1)\Phi_{\rho\nu lm}^\varepsilon(a,b,\theta,\varphi),
\nonumber\\
\label{I-38} {\bf M}^2\Phi_{\rho\nu
lm}^\varepsilon(a,b,\theta,\varphi)= l(l+1)\Phi_{\rho\nu
lm}^\varepsilon(a,b,\theta,\varphi),
\\
M_3\Phi_{\rho\nu lm}^\varepsilon(a,b,\theta,\varphi)=
m\Phi_{\rho\nu lm}^\varepsilon(a,b,\theta,\varphi).\nonumber
\end{gather}
One system corresponds to $\varepsilon=+$ and the other to
$\varepsilon=-$.

We try to f\/ind solutions of the system \eqref{I-38} in the form
of products of functions of each variables $a$, $b$, $\theta$,
$\varphi$:
\begin{equation}\label{I-39}
\Phi_{\rho\nu lm}^\varepsilon(a,b,\theta,\varphi)\equiv \langle
a,b,\theta,\varphi\mid\rho,\nu,l,m\rangle^\varepsilon=\langle
a\mid\rho,\nu\rangle^\varepsilon\langle
b\mid\nu,l\rangle\langle\theta\mid l,m\rangle\langle\varphi\mid
m\rangle.
\end{equation}
According to \eqref{I-35}--\eqref{I-37} the functions $\langle
a\mid\rho,\nu\rangle^\varepsilon$, $\langle b\mid\nu,l\rangle$,
$\langle\theta\mid l,m\rangle$ and $\langle\varphi\mid m\rangle$
satisfy the equa\-tions
\begin{gather}\label{I-40}
\left(\frac{\partial^2}{\partial a^2}+\frac{3}{\coth
a}\frac{\partial}{\partial
a}-\frac{\nu^2+1}{\cosh^2a}+\rho^2+\frac{9}{4}\right)\langle
a\mid\rho,\nu\rangle^\varepsilon=0,
\\
\label{I-41} \left(\frac{1}{\sinh^2b}\frac{\partial}{\partial
b}\sinh^2b\frac{\partial}{\partial
b}-\frac{l(l+1)}{\sinh^2b}+\nu^2+1\right)\langle
b\mid\nu,l\rangle=0,
\\
\label{I-42} \left(\frac{1}{\sin\theta}\frac{\partial}{\partial
\theta}\sin\theta\frac{\partial}{\partial
\theta}-\frac{m^2}{\sin^2\theta}+l(l+1)\right)\langle\theta\mid
l,m\rangle=0,
\\
\label{I-43}
\left(i\frac{\partial}{\partial\varphi}+m\right)\langle\varphi\mid
m\rangle=0.
\end{gather}
The equations \eqref{I-41}--\eqref{I-43} give a system of
equations for eigenvalues and eigenfunctions for the collection of
self-adjoint operators on the space $L^2(H^3_+)$. A solution of
this problem is given in~\cite{VSm}. Therefore, solutions of the
system of equations \eqref{I-40}--\eqref{I-43} are the functions
\begin{gather*}
\langle a\mid\rho,\nu\rangle^\varepsilon=(\cosh
a)^{-3/2}P_{-1/2+i\nu}^{-i\rho} (\varepsilon\tanh a),\qquad
\varepsilon=\pm,\qquad 0\leqslant\rho<\infty,
\\
 \langle
b\mid\nu,l\rangle=(\sinh b)^{-1/2}P_{-1/2+i\nu}^{-1/2-l}(\cosh
b),\qquad 0\leqslant\nu<\infty,
\\
\langle\theta \mid l,m\rangle=P_l^m(\cos\theta),\qquad
\langle\varphi\mid m\rangle=e^{im\varphi},\qquad l=0, 1,
2,\dots,\quad -l\leqslant m\leqslant l,
\end{gather*}
where $P^\mu_\nu (z)$ are associated Legendre functions and
$P^m_l(\cos \theta)$ are associated Legendre functions on the cut.
Thus, the functions
\begin{gather}
\langle a,b,\theta,\varphi\mid\rho,\nu,l,m\rangle^\varepsilon
=(\cosh a)^{-3/2}(\sinh b)^{-1/2} P_{-1/2+i\nu}^{-i\rho}
(\varepsilon\tanh a)\nonumber
\\
\label{I-44}  \phantom{\langle
a,b,\theta,\varphi\mid\rho,\nu,l,m\rangle^\varepsilon=}{} \times
P_{-1/2+i\nu}^{-1/2-l} (\cosh b)P_l^m(\cos\theta)e^{im\varphi},
\\
\varepsilon=\pm,\qquad 0\leqslant \rho, \nu<\infty,\qquad l=0, 1,
2,\dots,\qquad -l\leqslant m\leqslant l,\nonumber
\end{gather}
constitute a basis (not normed) of the space $L^2(H^4_+)$ in the
$H$-coordinate system. This basis corresponds to the chain of
subgroups
\[
SO_0(1,4)\supset SO_0(1,3)\supset SO(3)\supset SO(2).
\]

\noindent {\bf Remark.} The operators \eqref{I-35}--\eqref{I-37}
do not constitute a full collection of commuting self-adjoint
operators in the space $L^2(H^4_+)$. For obtaining a full
collection of such operators we have to add to the collection
\eqref{I-35}--\eqref{I-37} an operator, which separates
eigenfunctions with dif\/ferent values of~$\varepsilon$. However,
we have found the whole system of eigenfunctions without use of
this operator. Collections of self-adjoint operators, considered
below for other coordinate systems, will constitute a full
collections of self-adjoint operators.

\subsection[$O$-system]{$\boldsymbol{O}$-system}
\label{eig-o}

In this coordinate system, the Laplace operator is of the form
\[
\Delta(H^4_+)=\frac{\partial^2}{\partial
a^2}+3\frac{\partial}{\partial
a}+e^{-2a}\left[\frac{\partial^2}{\partial
r^2}+\frac{2}{r}\frac{\partial}{\partial r}
+\frac{1}{r^2}\left(\frac{\partial^2}{\partial
\theta^2}+\cot\theta
\frac{\partial}{\partial\theta}+\frac{1}{\sin^2\theta}
\frac{\partial^2}{\partial \varphi^2}\right)\right].
\]
If we introduce the variable $b=e^{-a}$, then  $\Delta(H^4_+)$ can
be written in the form
\begin{equation}\label{I-48}
\Delta(H^4_+)=b^2\frac{\partial^2}{\partial
b^2}-2b\frac{\partial}{\partial
b}+b^2\left[\frac{\partial^2}{\partial
r^2}+\frac{2}{r}\frac{\partial}{\partial r} \right.
 \left.
+\frac{1}{r^2}\left(\frac{\partial^2}{\partial
\theta^2}+\cot\theta
\frac{\partial}{\partial\theta}+\frac{1}{\sin^2\theta}\frac{\partial^2}{\partial
\varphi^2}\right)\right].
\end{equation}
The operators
\begin{gather}\label{I-49}
-{\bf E}^2 \equiv\Delta(\mathbb{R}^3)=\frac{\partial^2}{\partial
r^2}+\frac{2}{r}\frac{\partial}{\partial
r}+\frac{1}{r^2}\left(\frac{\partial^2}{\partial
\theta^2}+\cot\theta
\frac{\partial}{\partial\theta}+\frac{1}{\sin^2\theta}\frac{\partial^2}{\partial
\varphi^2}\right),
\\
\label{I-50} {\bf
M}^2=-\left(\frac{1}{\sin\theta}\frac{\partial}{\partial\theta}
\sin\theta\frac{\partial}{\partial\theta}+
\frac{1}{\sin^2\theta}\frac{\partial^2}{\partial\varphi^2}\right),\qquad
M_3=-i\frac{\partial}{\partial\varphi}
\end{gather}
commute with the operator \eqref{I-48}. The operators ${\bf E}^2$,
${\bf M}^2$ and $M^2_3$ are Casimir operators for the subgroups
$ISO(3)$, $SO(3)$ and $SO(2)$, respectively. Therefore, a basis of
the space $L^2(H^4_+)$ in the $O$-coordinate system, which will be
found below, corresponds to the chain of subgroups
\begin{equation}\label{I-51}
SO_0(1,4)\supset ISO(3)\supset SO(3)\supset SO(2).
\end{equation}

The operator $\Delta(\mathbb{R}^3)$ coincides with the
3-dimensional Laplace operator on the Euclidean space
$\mathbb{R}^3$ in the spherical coordinate system. The space
$\mathbb{R}^3$ is obtained  from $H^4_+$ by cutting this
hyperboloid by the hyperplane $e^{-a}={\rm const}$. At each
f\/ixed $a$, functions $f(a,r,\theta,\varphi)$ of  $L^2(H^4_+)$
lead to functions of $L^2(\mathbb{R}^3)$. This means that the
space $L^2(H^4_+)$ is a direct integral over values of $a$ of the
spaces $L_a^2(\mathbb{R}^3)\equiv L^2(\mathbb{R}^3)$.
Eigenfunctions and eigenvalues of the operator
$\Delta(\mathbb{R}^3)$ on the space $L^2(\mathbb{R}^3)$ are known.
Since $\mathbb{R}^3 \equiv ISO(3)/SO(3)$, they can be obtained,
for example, from the Fourier transform and Plancherel formula for
the regular representation of the group $ISO(3)$.

The system of dif\/ferential equations for eigenfunctions and
eigenvalues of the operators \eqref{I-48}--\eqref{I-50} in the
$O$-coordinate system is of the form
\begin{gather}
\Delta(H^4_+)\Phi^{\rho\kappa}_{lm}(b,r,\theta,\varphi)=
-(\rho^2+9/4)\Phi^{\rho\kappa}_{lm}(b,r,\theta,\varphi),
\nonumber\\
\Delta(\mathbb{R}^3)\,\Phi^{\rho\kappa}_{lm}(b,r,\theta,\varphi)=
-\kappa^2\,\Phi^{\rho\kappa}_{lm}(b,r,\theta,\varphi),
\nonumber\\
\label{I-52} {\bf M}^2\Phi^{\rho\kappa}_{lm}(b,r,\theta,\varphi)=
l(l+1)\Phi^{\rho\kappa}_{lm}(b,r,\theta,\varphi),
\\
M_3\,\Phi^{\rho\kappa}_{lm}(b,r,\theta,\varphi)=
m\,\Phi^{\rho\kappa}_{lm}(b,r,\theta,\varphi).\nonumber
\end{gather}
We try to f\/ind solutions in the form of separated variables:
\[
\Phi^{\rho\kappa}_{lm}(b,r,\theta,\varphi)\equiv\langle
b,r,\theta,\varphi\mid\rho,\kappa,l,m\rangle=\langle
b\mid\rho,\kappa\rangle\langle r\mid
\kappa,l\rangle\langle\theta\mid l,m\rangle\langle\varphi\mid
m\rangle.
\]
For these solutions the system \eqref{I-52} take the form
\begin{gather}\label{I-53}
\left(\frac{\partial^2}{\partial
b^2}-\frac{2}{b}\frac{\partial}{\partial
b}+\left(-\kappa^2+\frac{\rho^2+9/4}{b^2}\right)\right)\langle
b\mid\rho,\kappa\rangle=0,\qquad 0\leqslant\rho<\infty,
\\
\label{I-54} \left(\frac{\partial^2}{\partial
r^2}+\frac{2}{r}\frac{\partial}{\partial
r}+\kappa^2-\frac{l(l+1)}{r^2}\right)\langle r\mid
\kappa,l\rangle=0,\qquad 0\leqslant\kappa<\infty,
\\
\label{I-55}
\left(\frac{1}{\sin\theta}\frac{\partial}{\partial\theta}
\sin\theta\frac{\partial}{\partial\theta}-
\frac{m^2}{\sin^2\theta}-l(l+1)\right)\langle\theta\mid
l,m\rangle=0,\qquad l=0,1,2,\dots,
\\
\label{I-56} \left({\rm
i}\frac{\partial}{\partial\varphi}+m\right)\langle\varphi\mid
m\rangle=0,\qquad -l\leqslant m\leqslant l.
\end{gather}
The functions
\begin{gather*}
\langle b\mid\rho,\kappa\rangle=(\kappa b)^{3/2}K_{{\rm
i}\rho}(\kappa b),\ \ \langle r\mid \kappa,l\rangle=(\kappa
r)^{-1/2}J_{l+1/2}(\kappa r),
\\
 \langle\theta\mid
l,m\rangle=P_l^m(cos\theta),\qquad \langle\varphi\mid
m\rangle=e^{{\rm i}m\varphi},
\end{gather*}
are solutions of the equations \eqref{I-53}--\eqref{I-56}, where
$K_{{\rm i}\rho}(\kappa b)$ is the Macdonald function and
$J_{l+1/2}(\kappa r)$ is the Bessel function.

Thus, the functions
\begin{gather}\label{I-57}
\langle b,r,\theta,\varphi\mid\rho,\kappa,l,m\rangle=(\kappa
b)^{3/2}(\kappa r)^{-1/2}K_{{\rm i}\rho}(\kappa
b)\,J_{l+1/2}(\kappa r)P_l^m(\cos\theta)\,e^{{\rm i}m\varphi},
\\
0\leqslant\rho,\kappa<\infty,\qquad l=0,1,2,\dots,\qquad
-l\leqslant m\leqslant l,\qquad b=e^{-a},\nonumber
\end{gather}
constitute a basis of the space $L^2(H_+^4)$ in the $O$-system of
coordinates.

\subsection[$OC$-system]{$\boldsymbol{OC}$-system}
\label{eig-oc}

In the $OC$-coordinate system, the Laplace operator is of the form
\begin{equation}\label{I-61}
\Delta(H^4_+)=b^2\frac{\partial^2}{\partial
b^2}-2b\frac{\partial}{\partial b}+
b^2\left(\frac{\partial^2}{\partial\xi^2}+\frac{1}{\xi}\frac{\partial}{\partial\xi}+
\frac{1}{\xi^2}\frac{\partial^2}{\partial\varphi^2}+\frac{\partial^2}{\partial
z^2}\right).
\end{equation}
The operators
\begin{gather}\label{I-62}
{\bf E}^2=-\Delta(\mathbb{R}^3) =
-\left(\frac{\partial^2}{\partial\xi^2}+\frac{1}{\xi}\frac{\partial}{\partial\xi}+
\frac{1}{\xi^2}\frac{\partial^2}{\partial\varphi^2}+\frac{\partial^2}{\partial
z^2}\right),
\\
\label{I-63}
 E_3=-i\frac{\partial}{\partial
z},\qquad M_3=-i\frac{\partial}{\partial\varphi},
\\
\label{I-64} \tilde{\bf E}^2=E_1^2+E_2^2=-\Delta(\mathbb{R}^2)=
-\left(\frac{\partial^2}{\partial\xi^2}+\frac{1}{\xi}\frac{\partial}{\partial\xi}+
\frac{1}{\xi^2}\frac{\partial^2}{\partial\varphi^2}\right)
\end{gather}
commute with the operator \eqref{I-61}. The operator
$\Delta(\mathbb{R}^3)$ is the Laplace operator on the
space~$\mathbb{R}^3$ in the cylindric coordinates. The operators
$\tilde{\bf E}^2$ and $E^2_3$ are Casimir operators of the group
$ISO(2)\otimes T_\perp$, and the operators $E_3^2$ and $M_3^2$ are
Casimir operators of $SO(2)\otimes T_\perp$. Eigenvalues of the
operators \eqref{I-62}--\eqref{I-64} in $L^2(H^4_+)$ are
\begin{gather*}
\nu({\bf E}^2)=\kappa^2,\qquad \nu(E_3)=q,\qquad \nu(M_3)=m,\qquad
\nu(\tilde{{\bf E}}^2)=\eta^2,
\\
 \kappa^2=q^2+\eta^2,\qquad
-\infty<q<\infty,\qquad 0\leqslant\eta<\infty,\qquad
m=0,\pm1,\pm2,\dots .
\end{gather*}
We have the following system of dif\/ferential equations for
eigenfunctions:
\begin{gather}
\Delta(H^4_+)\Phi^{m}_{\rho\eta q}(b,\xi,z,\varphi)=
-(\rho^2+9/4)\Phi^{m}_{\rho\eta q}(b,\xi,z,\varphi),\qquad
0\leqslant\rho<\infty,\nonumber\\
\label{I-65}
 \Delta(\mathbb{R}^2)\Phi^{m}_{\rho\eta q
}(b,\xi,z,\varphi)= -\eta^2\,\Phi^{m}_{\rho\eta q
}(b,\xi,z,\varphi),\qquad 0\leqslant\eta<\infty,
\\
E_3\Phi^{m}_{\rho\eta q}(b,\xi,z,\varphi)=q\Phi^{m}_{\rho\eta
q}(b,\xi,z,\varphi),\qquad -\infty<q<\infty,\nonumber
\\
M_3\Phi^{m}_{\rho\eta q}(b,\xi,z,\varphi)=m\,\Phi^{m}_{\rho\eta
q}(b,\xi,z,\varphi),\qquad m=0,\pm1,\pm2,\dots.\nonumber
\end{gather}
Solutions of this system can be represented in the form of
separated variables:
\[
\Phi^{m}_{\rho\eta q}(b,\xi,z,\varphi)\equiv\langle
b,\xi,z,\varphi\mid\rho,\eta,q,m\rangle=\langle
b\mid\rho,\kappa\rangle\langle \xi\mid \eta,m\rangle\langle z\mid
q\rangle\langle\varphi\mid m\rangle,
\]
where $\kappa^2=\eta^2+q^2$. After separation of variables the
system \eqref{I-65} takes the form
\begin{gather}\label{I-66}
\left(\frac{\partial^2}{\partial
b^2}-\frac{2}{b}\frac{\partial}{\partial
b}+\frac{\rho^2+9/4}{b^2}-\kappa^2\right)\langle
b\mid\rho,\kappa\rangle=0,
\\
\label{I-67}
\left(\frac{\partial^2}{\partial\xi^2}+\frac{1}{\xi}\frac{\partial}{\partial\xi}-
\frac{m^2}{\xi^2}+\eta^2\right)\langle \xi\mid\eta,m\rangle=0,
\\
\label{I-68} \left(i\frac{\partial}{\partial z}+q\right)\langle
q\mid z\rangle=0,\ \ \ \left( {\rm i}\frac{\partial}{\partial
\varphi}+m\right)\langle \varphi\mid m\rangle=0.
\end{gather}
Solutions of these equations are of the form
\begin{gather*}
\langle b\mid\rho,\kappa\rangle=(\kappa b)^{3/2}K_{{\rm
i}\rho}(\kappa b),\qquad \langle
\xi\mid\eta,m\rangle=J_m(\eta\xi),\qquad \eta^2=\kappa^2-q^2,
\\
\langle z\mid q\rangle=e^{{\rm i}qz},\quad\langle \varphi\mid
m\rangle=e^{{\rm i}m\varphi}.
\end{gather*}
Therefore, the functions
\begin{gather}\label{I-69}
\langle b,\xi,z,\varphi\mid\rho,\eta,q,m\rangle=(\kappa b)^{
3/2}K_{{\rm i}\rho}(\kappa b)J_m(\eta\xi) e^{{\rm i}qz}e^{{\rm
i}m\varphi},
\\
\eta^2=\kappa^2-q^2,\qquad 0\leqslant\rho,\eta<\infty,\qquad
-\infty<q<\infty,\qquad m=0,\pm 1,\pm 2,\dots,\nonumber
\end{gather}
constitute a basis of the space $L^2(H_+^4)$ in the
$OC$-coordinate system. This basis corresponds to the chain of
subgroups
\[
SO_0(1,4)\supset ISO(3)\supset ISO(2)\otimes T_\bot\supset
SO(2)\otimes T_\bot.
\]

\subsection[$OT$-system]{$\boldsymbol{OT}$-system}
\label{eig-ot}

In this coordinate system
\begin{equation}\label{I-72}
\Delta(H^4_+)=b^2\frac{\partial^2}{\partial
b^2}-2b\frac{\partial}{\partial b}+
b^2\left(\frac{\partial^2}{\partial y_1^2}+
\frac{\partial^2}{\partial y_2^2}+\frac{\partial^2}{\partial
y_3^2}\right),\qquad b=e^{-a}.
\end{equation}
The operators
\begin{equation}\label{I-73}
{\bf E}^2=\Delta(\mathbb{R}^3)=-\left(\frac{\partial^2}{\partial
y_1^2}+ \frac{\partial^2}{\partial
y_2^2}+\frac{\partial^2}{\partial y_3^2}\right),\qquad
E_j=-i\frac{\partial}{\partial y_j},\qquad j=1,2,3,
\end{equation}
commute with $\Delta(H^4_+)$. The operator $\Delta(\mathbb{R}^3)$
is an invariant for the group $ISO(3)$. It is the Laplace operator
on the 3-dimensional Euclidean space in the homogeneous
coordina\-tes~$y_1$,~$y_2$,~$y_3$. The $ISO(3)$ is a motion group
of this Euclidean space. Eigenvalues of the operators \eqref{I-73}
are
\[
\nu({\bf E}^2)=\kappa^2,\qquad \nu(E_j)=\kappa_j,\qquad
-\infty<\kappa_j<\infty,\qquad 0\leqslant\kappa<\infty,\qquad
\kappa_1^2+\kappa_2^2+\kappa_3^2=\kappa^2.
\]
The spectrum of the collection of operators \eqref{I-72} and
\eqref{I-73} is simple. We represent eigenfunctions $\langle
b,{\bf y}\mid\rho,\boldsymbol{\kappa}\rangle$,
$\boldsymbol{\kappa}=(\kappa_1,\kappa_2,\kappa_3)$, ${\bf
y}=(y_1,y_2,y_3,)$, of this collection of operators in the form
\[
\langle b,{\bf y}\mid\rho,\boldsymbol{\kappa}\rangle =\langle
b\mid\rho,\kappa\rangle \langle {\bf y}\mid
\boldsymbol{\kappa}\rangle,\qquad \kappa=| {\boldsymbol{\kappa}}|.
\]
For $\langle b\mid\rho,\kappa\rangle$ and $\langle {\bf y}\mid
\boldsymbol{\kappa}\rangle$ we have the equations
\begin{equation}\label{I-74}
\left(\frac{\partial^2}{\partial
b^2}-\frac{2}{b}\frac{\partial}{\partial
b}+\frac{\rho^2+9/4}{b^2}-\kappa^2\right)\langle
b\mid\rho,\kappa\rangle=0,\qquad \left({\rm
i}\frac{\partial}{\partial {\bf
y}}+\boldsymbol{\kappa}\right)\langle {\bf y}\mid
\boldsymbol{\kappa}\rangle=0.
\end{equation}
Solutions of these equations are
\[
\langle b\mid\rho,\kappa\rangle=(\kappa b)^{3/2}K_{{\rm
i}\rho}(\kappa b),\qquad \langle {\bf
y}\mid\boldsymbol{\kappa}\rangle =e^{{\rm i}\boldsymbol{\kappa}
{\bf y}},\qquad b=e^{-a},
\]
where $\boldsymbol{\kappa} {\bf y}=\kappa_1y_1+\kappa_2y_2+
\kappa_3y_3$. Therefore, the functions
\begin{gather}\label{I-75}
\langle b,{\bf y}\mid\rho,\boldsymbol{\kappa}\rangle=(\kappa
b)^{3/2}K_{{\rm i}\rho}(\kappa b)e^{{\rm i}\boldsymbol{\kappa}
{\bf y}}, \qquad b=e^{-a},
\\
\kappa^2=\kappa_1^2+\kappa_2^2+\kappa_3^2,\qquad 0\leqslant\rho,
\kappa<\infty,\quad-\infty<\kappa_j<\infty,\nonumber
\end{gather}
constitute a basis of the space $L^2(H_+^4)$ in the
$OT$-coordinate system.

\subsection[$C$-system]{$\boldsymbol{C}$-system}
\label{eig-c}

In this coordinate system the Laplace operator on $H_+^4$ is of
the form
\begin{gather}
\Delta(H_+^4)=\frac{1}{\cosh a \sinh^2a}\frac{\partial}{\partial
a}\cosh a \sinh^2a\frac{\partial}{\partial
a}+\frac{1}{\cosh^2a}\frac{\partial^2}{\partial b^2}
\nonumber\\
\label{I-76}  \phantom{\Delta(H_+^4)=}{}
+\frac{1}{\sinh^2a}\left(\frac{1}{\sin\theta}
\frac{\partial}{\partial\theta}\sin\theta\frac{\partial}{\partial\theta}
+\frac{1}{\sin^2\theta}\frac{\partial^2}{\partial\varphi^2}\right).
\end{gather}
The operators
\begin{gather}
\mathbf{M}^2=-\frac{1}{\sin\theta}\frac{\partial}{\partial\theta}
\sin\theta\frac{\partial}{\partial\theta}
-\frac{1}{\sin^2\theta}\frac{\partial^2}{\partial\varphi^2},
\nonumber\\
\label{I-77} M_3=-{\rm i}\frac{\partial}{\partial\varphi},\qquad
P_0=-{\rm i}\frac{\partial}{\partial b}
\end{gather}
commute with $\Delta(H_+^4)$. They are invariants of the subgroups
$SO(3)$, $SO(2)$ and $SO_0(1,1)$ of the group $SO_0(1,1)\otimes
SO(3)$. Their eigenvalues are
\begin{gather}\label{I-78}
\nu(\mathbf{M}^2)=l(l+1),\qquad \nu(M_3)=m,\qquad \nu(P_0)=\tau,
\\
-\infty<\tau<\infty, \qquad l=0,1,2,\dots, \qquad -l\leqslant m
\leqslant l.\nonumber
\end{gather}
We try to f\/ind eigenfunctions of the operators \eqref{I-76} and
\eqref{I-77} in the form
\begin{equation}\label{I-79}
\langle a,b,\theta,\varphi\mid\rho,\tau,l,m\rangle=\langle a
\mid\rho,\tau,l\rangle\langle b\mid\tau\rangle\langle\theta\mid
l,m\rangle\langle\varphi\mid m\rangle.
\end{equation}
The functions on the right hand side satisfy the equations
\begin{gather}\label{I-80}
\left\{\frac{\partial^2}{\partial a^2}+(\tanh a+2\coth
a)\frac{\partial}{\partial
a}+\rho^2+\frac{9}{4}-\frac{\tau^2}{\cosh^2a}-\frac{l(l+1)}{\sinh^2a}\right\}
\langle a \mid\rho,\tau,l\rangle=0,
\\
\label{I-81}
\left\{\frac{1}{\sin\theta}\frac{\partial}{\partial\theta}
\sin\theta\frac{\partial}{\partial\theta}-\frac{m^2}{\sin^2\theta}
+l(l+1)\right\}\langle\theta\mid l,m\rangle=0,
\\
\label{I-82} \left(i\frac{\partial}{\partial b}+\tau\right)\langle
b\mid\tau\rangle=0,\quad
\left(i\frac{\partial}{\partial\varphi}+m\right)
\langle\varphi\mid m\rangle=0.
\end{gather}
These equations have solutions
\begin{gather*}
\langle a\mid\rho,\tau,l\rangle= \tanh^la(\cosh a)^{{\rm
i}\rho-3/2} {}_2F_1\!\left(\frac{l-{\rm i}\rho+{\rm
i}\tau+3/2}{2}, \frac{l-{\rm i}\rho-{\rm i}\tau+3/2}{2}
;l+\frac{3}{2};\tanh^2a\right)\!,\!
\\
\langle\theta\mid l,m\rangle=P^m_l(\cos\theta),\qquad  \langle
b\mid\tau\rangle=e^{{\rm i}\tau b},\qquad \langle\varphi\mid
m\rangle=e^{{\rm i}m\varphi},
\end{gather*}
where ${}_2F_1$ is a Gauss hypergeometric function. Thus, the
functions
\begin{gather}
\langle a,b,\theta,\varphi\mid\rho,\tau,l,m\rangle =\tanh^l
a(\cosh a)^{{\rm i}\rho-3/2}\nonumber
\\
\label{I-83} \qquad{} \times {}_2F_1 \left(\frac{l-{\rm i}\rho
+{\rm i}\tau+3/2}{2}, \frac{l-{\rm i}\rho-{\rm
i}\tau+3/2}{2};l{+}\frac 32;\, \tanh^2a\right) P^m_l(\cos\theta)
e^{{\rm i}\tau b}e^{{\rm i}m\varphi},
\\
0\leqslant\rho<\infty, \qquad  -\infty<\tau<\infty, \qquad
l=0,1,2,\dots, \qquad -l\leqslant m\leqslant l,\nonumber
\end{gather}
constitute a basis of the space $L^2(H^4_+)$ in the $C$-coordinate
system. This basis corresponds to the chain of subgroups
\[
SO_0(1,4)\supset SO_0(1,1)\otimes SO(3)\supset SO(2).
\]

\subsection[$SH$-system]{$\boldsymbol{SH}$-system}
\label{eig-sh}

In this coordinate system the Laplace operator on $H_+^4$ is of
the form
\begin{gather}
\Delta(H^4_+)=\frac{1}{\cosh^2a \sinh a}\frac{\partial}{\partial
a} \cosh^2a \sinh a\frac{\partial}{\partial
a}+\frac{1}{\sinh^2a}\frac{\partial^2}{\partial\Phi^2}
\nonumber\\
\label{I-85}  \phantom{\Delta(H^4_+)=}{} +\frac{1}{\cosh^2a}
\left(\frac{\partial^2}{\partial b^2}+\coth b
\frac{\partial}{\partial
b}+\frac{1}{\sinh^2b}\frac{\partial^2}{\partial \varphi^2}\right).
\end{gather}
The operators
\begin{gather}
M^2_3-N^2_1-N^2_2=\Delta(H^2_+)=\frac{\partial^2}{\partial
b^2}+\coth b \frac{\partial}{\partial
b}+\frac{1}{\sinh^2b}\frac{\partial^2}{\partial \varphi^2},
\nonumber\\
\label{I-86}
 M_3=-{\rm i}\frac{\partial}{\partial
\varphi},\qquad P_3=-{\rm i}\frac{\partial}{\partial \Phi}
\end{gather}
commute with the operator $\Delta(H^4_+)$. They are invariants of
the subgroups $SO_0(1,2)$, $SO(2)$ and $SO' (2)$ of the group
$SO_0(1,2)\otimes SO' (2)$, where $SO' (2)\backsim SO(2)$. The
subgroup $SO' (2)$ is generated by the generator $P_3$. The
operator $\Delta(H^2_+)$ is the Laplace operator on $H^2_+ \equiv
SO_0(1,2)/SO(2)$ in the spherical coordinates. Eigenvalues of the
above operators are
\begin{gather*}
\nu(\Delta(H^4_+))=-(\rho^2+9/4),\quad
\nu(\Delta(H^2_+))=-(\omega^2 +1/4),\quad \nu(M_3)=m,\quad
\nu(P_3)=m',
\\
0\leqslant\rho<\infty,\quad 0\leqslant\omega<\infty,\quad
m,m'=0,\pm1,\pm2,\dots.
\end{gather*}
We represent solutions of the system of equations \eqref{I-85} and
\eqref{I-86} in the form
\[
\langle a,b,\Phi,\varphi\mid\rho,\omega,m',m\rangle =\langle
a\mid\rho,\omega\rangle\langle b\mid\omega,m\rangle\langle\Phi\mid
m'\rangle\langle\varphi\mid m\rangle.
\]
Then this separation of variables leads to the equations
\begin{gather}\label{I-87}
\left\{\frac{\partial^2}{\partial a^2}+(2\tanh a+\coth
a)\frac{\partial}{\partial
a}-\frac{\omega^2+1/4}{\cosh^2a}-\frac{m^{\shortmid
2}}{\sinh^2a}+\rho^2+\frac94 \right\}\langle
a\mid\rho,\omega,m'\rangle=0,
\\
\label{I-88}
 \left(\frac{\partial^2}{\partial
b^2}+\coth b\frac{\partial}{\partial
b}-\frac{m^2}{\sinh^2b}+\omega^2+1/4\right)\langle
b\mid\omega,m\rangle=0,
\\
\label{I-89}
 \left({\rm i}\frac{\partial}{\partial
\Phi}+m'\right)\langle\Phi\mid m'\rangle=0,\qquad \left({\rm
i}\frac{\partial}{\partial \varphi}+m\right)\langle\varphi\mid
m\rangle=0.
\end{gather}
Solutions of these equations are the functions
\begin{gather*}
\langle a\mid\rho,\omega,m'\rangle=(\tanh a)^{m'} (\cosh
a)^{{\rm i}\rho-3/2}\\
\phantom{\langle a\mid\rho,\omega,m'\rangle=}{}\times
{}_2F_1\left(\frac{m'-{\rm i}\rho+{\rm i}\omega+1}{2},
\frac{m'-{\rm i}\rho-{\rm i}\omega+1}{2}; m'+1; \tanh^2a\right),
\\
\langle b\mid\omega,m\rangle=P^m_{{\rm i}\omega-1/2}(\cosh
b),\qquad \langle\Phi\mid m'\rangle=e^{{\rm i}m'\Phi},\qquad
\langle\varphi\mid m\rangle=e^{{\rm i}m\varphi}.
\end{gather*}
Thus, the functions
\begin{gather}
\langle a,b,\Phi,\varphi\mid \rho,\omega,m',m\rangle=(\tanh
a)^{m'} (\cosh a)^{i\rho-3/2}P^m_{{\rm i}\omega-1/2}(\cosh b)
e^{{\rm i}m'\Phi} e^{{\rm i}m\varphi}\label{I-90}
\\
  \phantom{\langle a,b,\Phi,\varphi\mid \rho,\omega,m',m\rangle=}{}
\times{}_2F_1\left( \frac{m'-{\rm i}\rho+{\rm
i}\omega+1}{2},\,\frac{m'-{\rm i}\rho
-{\rm i}\omega+1}{2};\,m'+1;\,\tanh^2 a \right),  \nonumber\\
 0\leqslant\rho,\omega<\infty,\qquad m,m'=0,\pm
1,\pm 2, \dots , \nonumber
\end{gather}
constitute a basis of the space $L^2(H^4_+)$ in the
$SH$-coordinate system. This basis corresponds to the chain of
subgroups
\[
SO_0(1,4)\supset SO_0(1,2)\otimes SO'(2)\supset SO(2).
\]

\subsection[$S$-system]{$\boldsymbol{S}$-system}
\label{eig-s}

The problem of f\/inding basis functions of the Hilbert space
$L^2(H^4_+)$ in the spherical coordinates is solved in \cite{Vil1}
(see also \cite{VKII}). The Laplace operator in this coordinate
system has the form
\begin{gather}
\Delta(H_+^4)=\frac{1}{\sinh^3 a}\frac{\partial}{\partial a}
\sinh^3a\frac{\partial}{\partial a}  \nonumber\\
 \phantom{\Delta(H_+^4)=}{}+\frac{1}{\sin^ 2 \beta}
\left(\frac{\partial}{\partial\beta}\sin^2\beta
\frac{\partial}{\partial\beta} +\frac{1}{\sin\theta}
\frac{\partial}{\partial\theta}\sin\theta\,
\frac{\partial}{\partial\theta}+\frac{1}{\sin^2\theta}\,
\frac{\partial^2}{\partial\varphi^2}\right).\label{I-91}
\end{gather}
The following operators commute with $\Delta_L(H_+^4)$:
\begin{gather}
{\bf J}^2\equiv {\bf M}^2+ {\bf P}^2=
-\Delta(S^3)\nonumber\\
\phantom{{\bf J}^2}{}=-\frac{1}{\sin^2\beta}\left(
\frac{\partial}{\partial\beta}\sin^2\beta\frac{\partial}{\partial\beta}
  +\frac{1}{\sin\theta}\frac{\partial}{\partial\theta}\sin\theta\frac{\partial}
{\partial\theta}+\frac{1}{\sin^2\theta}\frac{\partial^2}{\partial\varphi^2}
\right),\label{I-92}
\\
\label{I-93}
 {\bf M}^2=-\left(
\frac{1}{\sin\theta}\frac{\partial}{\partial\theta}\sin\theta\frac{\partial}
{\partial\theta}{+}\frac{1}{\sin^2\theta}\frac{\partial^2}{\partial\varphi^2}
\right),\qquad M_3=-{\rm i}\frac{\partial}{\partial\varphi}.
\end{gather}
They are invariants of the groups $SO(4)$, $SO(3)$ and $SO(2)$,
respectively. The operator $\Delta(S^3)$ in~\eqref{I-92} is the
Laplace operator on the sphere $S^3$ in the 3-dimensional
Euclidean space. Eigenvalues $\nu(I)$ of the operators
\eqref{I-91}--\eqref{I-93} are
\begin{gather*}
\nu(\Delta(H_+^4))=-(\rho^2+9/4),\qquad \nu({\bf
J}^2)=j(j+2),\qquad \nu({\bf M}^2)=l(l+1),\qquad \nu(M_3)=m,
\\
 0\leqslant\rho<\infty,\qquad
j=0,1,2,\dots,\qquad 0\leqslant l\leqslant j,\qquad -l\leqslant
m\leqslant l,
\end{gather*}
and the spectrum of this collection of operators is simple.
Eigenfunctions of the collection \eqref{I-91}--\eqref{I-93} are
\cite{Vil1}
\begin{gather}\label{I-94}
\langle a,\beta,\theta,\varphi|\rho,j,l,m \rangle =\sinh^{{-}1}\!
a\sin^{-1/2} \beta P_{{\rm i}\rho-1/2}^{{-}j{-}1}(\cosh a)
P_{j{+}1/2}^{{-}l{-}1/2}(\cos\beta) P_l^m(\cos\theta)e^{{\rm
i}m\varphi}.\!\!\!
\end{gather}
They form a basis of the space $L^2(H_+^4)$ in the $S$-coordinate
system. This basis corresponds to the chain of subgroups
\[
SO_0(1,4)\supset SO(4)\supset SO(3)\supset SO(2).
\]

\section{The method of orispherical transforms}
\label{orispher}
Now we consider another problem -- expansion of functions
$\psi(x)\in L^2(H^4_+)$ in eigenfunctions of a full collection of
self-adjoint operators. Derivation of the inverse formulas will
allow us to represent these eigenfunctions in a normed form. Since
functions $\psi(x)$ are determined on the hyperboloid $H^4_+$,
then for construction of expansion we can use the method of
orispheres, which was worked out by Gel'fand and Graev on the base
of integral geometry \cite{GG}. This method allows us to go from
studying functions $\psi(x)$ on $H^4_+$ to studying corresponding
functions $h(k)$ on the upper sheet $C^4_+$ of the cone $C^4$.
Consideration of functions on the cone is convenient, since we can
expand them in functions $\Psi(k,\sigma)$ homogeneous in the set
of the homogeneous coordinates $k=(k_0,k_1,k_2,k_3,k_4)$. Under
shifts $\psi(x)\to \psi(g^{-1}x)$, $g\in SO_0(1,4)$, the functions
$\Psi(k,\sigma)$ transform under irreducible representations
$\pi^\sigma$ of the group $SO_0(1,4)$. Let us give an information
on the method of orispheres \cite{GG, Vil1}.

An orisphere of the hyperboloid $H^4_+$ is a cut of $H^4_+$ by the
plane
\[
[x,k]\equiv x_0k_0-x_1k_1-x_2k_2-x_3k_3-x_4k_4=1,
\]
where $k$ is a f\/ixed point of the cone $C^4_+$ given by the
equation
\[
[k,k]\equiv k_0k_0-k_1k_1-k_2k_2-k_3k_3-k_4k_4=0, \qquad k_0>0.
\]
Thus, an orisphere on $H^4_+$ is given by a point $k\in C^4_+$.
Then the set of orispheres can be considered as the set of all
points of $C^4_+$.

According to the method of orispheres, to each f\/inite function $\psi(x)$ on
$H^4_+$ there corresponds a function
$h(\omega)$, determined on the set of orispheres $\omega$ of
$H^4_+$. This correspondence is given by the Gel'fand--Graev
formula
\begin{equation}\label{I-95}
h(\omega)=\int_\omega d\omega\, \psi(x),
\end{equation}
where the integral of the function $\psi(x)$ on the orisphere
$\omega:[x,k]=1$ is def\/ined by the formula
\begin{equation}\label{I-96}
\int_\omega
d\omega\,\psi(x)=\int_{H^4_+}\frac{d^4x}{x_0}\psi(x)\delta([x,k]-1).
\end{equation}
Here $d^4x/x_0=dx_1dx_2dx_3dx_4/x_0$ is an invariant (with respect
to $SO_0(1,4)$) measure on $H^4_+$. Under shifts of points $x$ and
$k$ by an element  $g\in SO_0(1,4)$ the bilinear form $[x,k]$ and
the measure $d^4x/x_0$ are conserved. For this reason, the measure
$d\omega$, determined by formula \eqref{I-96}, is conserved under
shifts $\omega\to g\omega$. This means that
\[
\int_\omega d\omega\, \psi(gx)=\int_{g\omega}
d\omega_g\psi(x),\qquad g\in SO_0(1,4),
\] where $d\omega_g$ is a measure on the orisphere $g\omega$.
In particular, if a shift $g$ leaves an orisphere $\omega$
invariant, then
\[
\int_\omega d\omega\, \psi(gx)=\int_\omega\, d\omega\, \psi(x).
\]

Since there exists a one-to-one correspondence between points of
the set of orispheres $\omega$ of~$H^4_+$ and points $k\in C^4_+$,
the functions $h(\omega)$ can be considered as functions
on~$C^4_+$. Instead of $h(\omega)$ we shall write~$h(k)$. Then
according to \eqref{I-95} and \eqref{I-96} we have
\begin{equation}\label{I-97}
h(k)=\int_ {H^4_+}\frac{d^4x}{x_0}\psi(x)\delta([x,k]-1).
\end{equation}
This transform, turning a function $\psi(x)$ into a function
$h(k)$, is called {\it Gel'fand--Graev integral transform}.

It is proved in \cite{Gelf} that if $\psi(x)$ is inf\/initely
dif\/ferentiable f\/inite function on $H^4_+$, then the function
$h(k)$ on $C^4_+$ is inf\/initely dif\/ferentiable, f\/inite, and
vanishes inside of some neighborhood of the point $k=0$ of the
cone $C^4_+$.

The inverse formula for the integral transform \eqref{I-97} is of
the form \cite{Gelf}
\begin{equation}\label{I-98}
\psi(x)=\frac{\Gamma(4)}{(2\pi)^4}\int_
{C^4_+}\frac{d^4k}{k_0}h(k)([x,k]-1)^{-4},
\end{equation}
where $d^4k/k_0=dk_1dk_2dk_3dk_4/k_0$ is an invariant (with
respect to $SO_0(1,4)$) measure on the cone~$C^4_+$. Here the
integral is understood in the sense of a value, regularized by
analytic conti\-nua\-tion in a power:
\[
\int_{C^4_+}\frac{d^4k}{k_0}h(k)([x,k]-1)^{-4}=\int_
{C^4_+}\frac{d^4k}{k_0}h(k)([x,k]-1)^{\tau}\vert_{\tau\rightarrow{-4}}.
\]

We wish to know a form of the operator $\pi(g)$ of the
quasi-regular representation \eqref{I-11} of $SO_0(1,4)$
(restricted onto the space of inf\/initely dif\/ferentiable
f\/inite functions on $H^4_+$) under the transform \eqref{I-97}.
It is easy to see that this transform turns the function
$\pi(g)\psi(x)=\psi(g^{-1}x)$ on $H^4_+$ into the function
$h_g(k)\equiv h(g^{-1}k)$ on $C^4_+$. Therefore, to the shift operators $\pi(g)$ on $H^4_+$
there corresponds shift
operators $\hat \pi(g)$ on the space of functions $h(k)$ on
$C^4_+$:
\[
\hat\pi(g)h(k)=h(g^{-1}k),\qquad g\in SO_0(1,4).
\]
This equality determines the quasi-regular representation of
$SO_0(1,4)$ on the space of inf\/initely dif\/ferentiable
functions $h(k)$ on the cone $C_+^4$, satisfying an additional
symmetry condition \cite{Gelf}.

A function $h(k)$ on $C_+^4$ can be expanded in functions on the
cone $C^4_+$ homogeneous in coordinates $k=(k_0,k_1,k_2,k_3,k_4)$:
\begin{equation}\label{I-99}
\Psi(k,\sigma)=\int_0^\infty dt\, h(tk)t^{-\sigma-1},\qquad
\sigma\in \mathbb{C}.
\end{equation}
These functions have homogeneity degrees $\sigma$:
\[
\Psi(ak,\sigma)=a^\sigma\Psi(k,\sigma),\qquad a>0.
\]
This expansion has the form (see, for example, \cite{Vil1})
\begin{equation}\label{I-100}
h(k)=\frac{1}{2\pi {\rm i}}\int_{\delta-{\rm i}\infty}^{\delta
+{\rm i}\infty}d\sigma\, t^\sigma\Psi(k,\sigma),
\end{equation}
where a value $\delta$ is taken in such a way that the function
$\Psi(k,\sigma)$ does not have poles on the strip $0\leqslant {\rm
Re}\; \sigma\leqslant \delta$.

It is easy to check that to the function $h_g(k)=\hat \pi(g)
h(k)=h(g^{-1}k)$ on the cone $C^4_+$ there corresponds a homogeneous function
$\Psi_g(k,\sigma)=\Psi(g^{-1}k,\sigma)$, $\sigma\in \mathbb{C}$,
on $C^4_+$. Thus,  to the quasi-regular representation $\hat \pi$ on
the space of functions $h(k)$ on $C^4_+$ there corresponds a collection
of representations $T^\sigma$, $\sigma\in \mathbb{C}$, and
$T^\sigma$ acts on the space of homogeneous functions
$\Psi(k,\sigma)$ on $C^4_+$ by the formula
\[
T^\sigma(g)\Psi(k,\sigma)=\Psi(g^{-1}k,\sigma),\qquad g\in
SO_0(1,4).
\]
The representation $T^\sigma$ is irreducible for all values of
$\sigma$ except for the case when $\sigma$ or $-\sigma-3$ is a
non-negative integer. Therefore, the representations $T^\sigma$
are irreducible components of the representation $\hat \pi$ or of
the representation $\pi$, equivalent to $\hat \pi$. Note that for
$\sigma={\rm i}\rho-3/2$, $\rho\in \mathbb{R}$, the representation
$T^\sigma$ is unitary.

The representation $T^\sigma$, $\sigma\in \mathbb{C}$, in fact,
coincides with the representation $\pi^\sigma$ considered in
Section~2. To show this, we note that the representation
$\pi^\sigma$ acts in the space of functions $f(g)$ on $SO_0(1,4)$
such that
\[
f(gmnh)=\exp (-\lambda(\log h))f(g),\qquad  m\in M=SO(3),\qquad
n\in N,\qquad h\in A=\exp \mathfrak{a},
\]
where $\lambda$ is a linear form on the commutative subalgebra
$\mathfrak{a}$ of the Lie algebra ${\rm so}(1,4)$, determined by
the number $\sigma$ (see Section~2). This representation is given
by the formula
\[
\pi^\sigma(g_0)f(g)=f(g_0^{-1}g).
\]
The product $mnh$, $m\in M$, $n\in N$, $h\in A$, has the property
$mnh=hm'n'$, $m'\in M$, $n'\in N$. Thus, the functions $f(g)$, on
which the representation $\pi^\sigma$ is def\/ined, satisfy the
condition
\[
f(ghmn)=\exp(-\lambda(\log h))f(g),\qquad m\in M,\qquad n\in
N,\qquad h\in A.
\]
This means that the representation $\pi^\sigma$ can be realized on
functions $f(x)$ on the space $X=SO_0(1,4)/MN$, satisfying the
condition $f(xh)=\exp(-\lambda(\log h))f(x)$. Since
$MN=SO(3)\times T(3)$, then $X=C_+^4$. Besides, the latter
condition is the condition of homogeneity (of degree $\sigma$) of
the function $f(x)$ with respect to the collection of variables
$k_0$, $k_1$, $k_2$, $k_3$, $k_4$. This shows that the
representations $T^\sigma$ and $\pi^\sigma$ are equivalent.

Using the formulas \eqref{I-98}--\eqref{I-100} we can f\/ind a
connection between the functions  $\psi(x)$ and $\Psi(k,\sigma)$
\cite{VSm}:
\begin{gather}\label{I-101}
\psi(x)=\frac{\rm
i}{2(2\pi)^4}\int_{\delta-i\infty}^{\delta+i\infty}
\frac{\Gamma(\sigma+3)}{\Gamma(\sigma)} \cot\pi\sigma\,
\int_\Gamma d\mu(k')\Psi(k',\sigma)[x,k']^{-\sigma-3},
\\
\label{I-102}
\Psi(k',\sigma)=\int_{H_+^4}\frac{d^4x}{x_0}\psi(x)[x,k']^\sigma,
\end{gather}
where $-3{<}\delta{<}1$ (this condition for $\delta$ is connected
with poles of the function
$\Gamma(\sigma+3)\cot\pi\sigma/\Gamma(\sigma))$. Integration
contour $\Gamma$ in the formula \eqref{I-101} is any surface on
the cone $C_+^4$ intersecting once each generator of the cone, and
$d\mu(k')$ is an invariant measure on $\Gamma$ determined by means
of the invariant measure $d^4k/k_0$ on the cone $C_+^4$ using the
equality
\[
d^4k/k_0=t^2dtd\mu(k'),
\]
where $k\in C_+^4$ and $k'\in\Gamma$ are connected by $k=tk'$.

Below we shall consider cuts of the cone by the surfaces $k_4=\pm
1$, $k_0-k_4=1$, $k_0^2-k_4^2=1$, $k_0^2-k_1^2-k_2^2=1$, $k_0=1$.
To these choices of $\Gamma$ there will correspond expansions of the function~$\psi(x)$ in basis
functions in the respective coordinate systems on $H_+^4$. In order to obtain an
expansion of $\psi(x)$ in basis functions on $H_+^4$, we use
formulas \eqref{I-101} and \eqref{I-102}. First we expand the
functions $\Psi(k',\sigma)$ in basis functions on $\Gamma$ and
then take the corresponding integrals over $\Gamma$. Let us
consider expansions of $\psi(x)$ for each of the cuts, given
above.

\section{Expansion of functions on the hyperboloid}
\label{expan-hyp}

\subsection[Expansion in $H$-coordinate system]{Expansion in $\boldsymbol{H}$-coordinate system}
\label{expan-hyp-h}
The cone $C_+^4$ is determined by the relations
\[
C_+^4: \ [k,k]=k_0^2-k_1^2-k_2^2-k_3^2-k_4^2=0,\qquad k_0>0.
\]
We consider the cut of $C^4_+$ by the surfaces $k_4=\pm 1$. We
obtain the contour $\Gamma$ coinciding with upper sheets of two
hyperboloids
\[
H_\varepsilon^3: \ k_0^2-k_1^2-k_2^2-k_3^2=1,\qquad
k_4=\varepsilon=\pm 1.
\]
The motion group of the hyperboloid $k_0^2-k_1^2-k_2^2-k_3^2=1$ is
isomorphic to the Lorentz group $SO_0(1,3)$ (see \cite{Vil1}). We
denote points of this hyperboloid by $k':
k_0^{'2}-k_1^{'2}-k_2^{'2}-k_3^{'2}=1$. If $k_4=+1$ we have the
hyperboloid $H_+^3$, and if $k_4=-1$ then we have the hyperboloid
$H_-^3$. Thus, the contour $\Gamma$ consists of two parts:
$\Gamma=\Gamma_+\cup\Gamma_-$.

Expansion of functions given on the 3-dimensional hyperboloid in
dif\/ferent coordinate systems is derived in \cite{VSm}. In the
spherical coordinate system, corresponding to the reduction
$SO_0(1,3)\supset SO(3)\supset SO(2)$, the expansion is of the
form
\begin{gather}
\phi(k'){=}\frac{\rm i}{2}\sum_{l=0}^\infty \sum_{m=-l}^l\,
\int_{\gamma-{\rm i}\infty}^{\gamma+i\infty}d\tau
C_{lm}(\tau)\frac{\Gamma({-}\tau{+}1)}{\Gamma({-}\tau{-}l{-}1)}(\sinh
c)^{-1/2} P_{\tau+1/2}^{-l-1/2}(\cosh c)Y_{lm}(\Theta,\Phi),
\nonumber\\
C_{lm}(\tau)=\frac{\Gamma(\tau+1)}{\Gamma(\tau-l+1)}\int_{H_+^3}dk'
\phi(k')(\sinh c)^{-1/2}P_{-\tau-3/2}^{-l-1/2}(\cosh
c)\overline{Y_{lm}(\Theta,\Phi)},\label{I-103}
\end{gather}
where $Y_{lm}(\theta,\Phi)$ is the $SO(3)$-spherical function
\[
Y_{lm}(\Theta,\Phi)=\left[\frac{2l+1}{4\pi}\frac{(l-m)!}{(l+m)!}\right]^{1/2}
P_l^m(\cos\Theta)e^{{\rm i}m\Phi}
\]
and
\begin{gather*}
k_0'=\cosh c,\qquad k_1'=\sinh c\sin\Theta\cos\Phi,\qquad
k_2'=\sinh c\sin\Theta\sin\Phi,\qquad k_3'=\sinh c\cos\Theta,
\\
0\leqslant c<\infty,\qquad 0\leqslant \Theta<\pi,\qquad
0\le\Phi<2\pi,\qquad dk'\equiv\sinh^2c\sin\Theta\,
dc\,d\Theta\,d\Phi.
\end{gather*}
Applying this decomposition to the functions $\Psi_+(k',\sigma)$
(with $k'\in\Gamma_+\equiv H_+^3$) and $\Psi_-(k',\sigma)$ (with
$k'\in\Gamma_-\equiv H_-^3$), substituting this expressions into
\eqref{I-101} and permuting an order of integrals, we obtain
\begin{gather}
\psi(x)=-\frac{1}{4(2\pi)^4}\sum_{l,m,\varepsilon}\,
\int_{\delta-{\rm i}\infty}^{\delta+{\rm i}\infty} d\sigma
(\sigma{+}2)(\sigma{+}1)\sigma\,\cot\pi\sigma \int_{\gamma-{\rm
i}\infty}^{\gamma+{\rm i}\infty} d\tau\frac{\Gamma(-\tau+2)}
{\Gamma({-}\tau{-}l{-}1)}C_{lm}^\varepsilon(\tau,\sigma)
\nonumber\\
\label{I-104}  \phantom{\psi(x)=}{} \times
\int_{\Gamma_\varepsilon}\,dk'[x,k']^{-\sigma-3}(\sinh c)^{-1/2}
 P_{\tau+1/2}^{-l-1/2}(\cosh c)Y_{lm}(\Theta,\Phi).
\end{gather}
Note that the point $x$ belongs to  $H_+^4,$ and $c$, $\Theta$,
$\Phi$ parametrize points of the hyperboloids $H_\varepsilon^3$.

Let us calculate the integrals over the contours
$\Gamma_\varepsilon$ in \eqref{I-104}:
\begin{equation}\label{I-105}
J_{\sigma\tau lm}^\varepsilon(x)=
\int_{\Gamma_\varepsilon}\,dk'[x,k']^{-\sigma-3}(\sinh c)^{-1/2}
P_{\tau+1/2}^{-l-1/2}(\cosh c)P_l^m(\cos\Theta)e^{{\rm i}m\Phi}.
\end{equation}
First we calculate the integrals for the point $x^0=(\cosh
a,0,0,0,\sinh a)$ corresponding to the hyperbolic rotation
$g_{04}(a)\in SO_0(1,4)$, and then, by means of transformations of
the subgroup $SO_0(1,3)$, we go from $x^0$ to $x$. For $x^0$ we
have
\[
J_{\sigma\tau}^\varepsilon(a)=4\pi\int_1^\infty d(\cosh c)(\cosh
a\cosh c-\varepsilon\sinh a)^{-\sigma-3} \sinh^{1/2}c \times
P_{\tau+1/2}^{-1/2}(\cosh c),
\]
where $J_{\sigma\tau}^\varepsilon(a)\equiv J_{\sigma\tau
00}^\varepsilon(x^0)$. We calculate the integral
\begin{equation}\label{I-106}
J_{\sigma\tau}(u)=\int_1^\infty dy\,
(y^2-1)^{1/4}P_{\tau+1/2}^{-1/2}(y) (y+u)^{-\sigma-3},\qquad
u\equiv-\varepsilon\tanh a.
\end{equation}
For this we substitute into \eqref{I-106} the expression
\[
P_{\tau+1/2}^{-1/2}(y)=\frac{(y^2-1)^{-1/4}}{\sqrt{\pi}}\int_1^y\,
dt(y-t)^{-1/2} P_{\tau+1/2}(t)
\]
(see formula 3.7(29) in \cite{RG}) for $P_{\tau+1/2}^{-1/2}(y)$:
\[
J_{\sigma\tau}(u)=\frac{1}{\sqrt{\pi}}\,\int_1^\infty\,dy\int_1^y\,dt(y+u)^{-\sigma-3}
(y-t)^{-1/2}P_{\tau+1/2}(t).
\]
The following integral representation holds for the function
$P_{\tau+1/2}(t)$:
\[
P_{\tau+1/2}(t)=\frac{2^{-\tau-1/2}}{2\pi {\rm i}}
\int_C\,dt'\,\frac{(t^{'2}-1)^{\tau+1/2}}{(t'-t)^{\tau+3/2}},
\]
where the contour $C$ is a closed curve in the plane of complex
variable $t'$, containing inside the points 1 and $t$; integration
is in the counter-clockwise direction. Therefore,
\begin{gather*}
J_{\sigma\tau}(u)=\frac{2^{-\tau-3/2}}{i\pi^{3/2}}\,\int_1^\infty
dy(y+u)^{-\sigma-3} \int_1^y\,dt(y-t)^{-1/2}
\int_C\,dt'(t^{'2}-1)^{\tau+1/2}(t'-t)^{-\tau-3/2}.
\end{gather*}
Making some transformations (see \cite{KK86}) we reduce this
expression for $J_{\sigma\tau}(u)$ to the form
\begin{gather*}
J_{\sigma\tau}(u)=\lambda \int_1^\infty
dy\,\int_1^y\,dt'\,\frac{(t^{'2}-1)^{\tau+1/2}}
{(y-t')^{\tau+1}(y+u)^{\sigma+3}}
\\
\phantom{J_{\sigma\tau}(u)}{}
=\lambda\int_1^\infty\,dt'\,(t^{'2}-1)^{\tau+1/2}
\int_{t'}^\infty\,dy\,(y-t')^{-\tau-1}(y+u)^{-\sigma-3},
\end{gather*}
where $\lambda=2^{-\tau-1/2}\cos\pi\tau\, \Gamma(-\tau-1/2)/
\pi\Gamma(-\tau)$. Using the formula 3.196(2) from \cite{RG}, we
get
\[
J_{\sigma\tau}(u)=\lambda \frac{\Gamma(-\tau)
\Gamma(\sigma+\tau+3)}{\Gamma(\sigma+3)}\int_1^\infty dt'\,
\frac{(t^{'2}-1)^{\tau+1/2}}{(t'+u)^{\sigma+\tau+3}}.
\]
Since
\[
2^{-\tau-1/2}\int_1^\infty\,dt'\,
\frac{(t^{'2}{-}1)^{\tau+1/2}}{(t'{+}u)^{\sigma+\tau+3}}=
\frac{\Gamma(\sigma-\tau+1)}{[\Gamma(\tau{+}3/2)]^{-1}}
(1{-}u^2)^{-(\sigma+3/2)/2}P_{\tau+1/2}^{-\sigma-3/2}(u),
\]
then
\begin{equation}\label{I-112}
J_{\sigma\tau}(u)=-\frac{\Gamma(\sigma+\tau+3)\Gamma(\sigma-\tau+1)}
{\Gamma(\sigma+3)}(1-u^2)^{-(\sigma+3/2)/2}P_{\tau+1/2}^{-\sigma-3/2}(u),
\end{equation}
where the equality
$\Gamma(-\tau+1/2)\Gamma(\tau+1/2)=\pi\cos^{-1}(\pi\tau)$ was
used. Thus, we have
\[
J_{\sigma\tau}^\varepsilon(a)=-4\pi\frac{\Gamma(\sigma+\tau+3)\Gamma(\sigma-\tau+1)}
{\Gamma(\sigma+3)}(\cosh a)^{-3/2}
P_{\tau+1/2}^{-\sigma-3/2}(-\varepsilon\tanh a).
\]

For any $x=x(a,b,\theta,\varphi)$, where $a,b,\theta,\varphi$ are
the parameters from \eqref{I-27}, the integral $J_{\sigma\tau lm
}^\varepsilon(x)$ is easily calculated as in \cite{VSm} and we
have
\begin{gather*}
J_{\sigma\tau
lm}^\varepsilon(x)=-(2\pi)^{3/2}\frac{\Gamma(\sigma+\tau+3)
\Gamma(\sigma-\tau+1)}
{\Gamma(\sigma+3)}(\cosh a)^{-3/2}(\sinh b)^{-1/2}\\
\phantom{J_{\sigma\tau lm}^\varepsilon(x)=}{} \times
P_{\tau+1/2}^{-\sigma-3/2}(-\varepsilon\tanh a)
P_{\tau+1/2}^{-l-1/2}(\cosh b) P_l^m(\cos\theta)e^{{\rm
i}m\varphi}.
\end{gather*}
Thus, according to the formula \eqref{I-104} we get
\begin{gather}
\psi(x)=\frac{1}{4(2\pi)^{5/2}}\sum_{l=0}^\infty \sum_{m=-l}^l
\int_{\delta-{\rm i}\infty}^{\delta+{\rm i}\infty} d\sigma\frac{\cot\pi\sigma}{\Gamma(\sigma)}\nonumber\\
\phantom{\psi(x)=}{} \times \int_{\gamma-{\rm
i}\infty}^{\gamma+{\rm i}\infty} d\tau\, (\tau{+}1)\tau
\Gamma(\sigma{-}\tau{+}1) \Gamma(\sigma{+}\tau{+}3) V_{\tau
lm}(b,\theta,\varphi)
\nonumber\\
\label{I-114} \phantom{\psi(x)=}{} \times \left[
C_{lm}^+(\tau,\sigma) P_{\tau+1/2}^{-\sigma-3/2}(-\tanh a)
+C_{lm}^-(\tau,\sigma)P_{\tau+1/2}^{-\sigma-3/2}(\tanh a)\right]
(\cosh a)^{-3/2},
\end{gather}
where{\samepage
\begin{equation}\label{I-115}
V_{\tau lm}(b,\theta,\varphi)=
\frac{\Gamma(-\tau-1)}{\Gamma(-\tau-l-1)} (\sinh b)^{-1/2}
P_{\tau+1/2}^{-l-1/2}(\cosh b)Y_{lm}(\theta,\varphi).
\end{equation}
The functions \eqref{I-115} coincide in a form with the functions
from \eqref{I-103}.}

The formula \eqref{I-114} expresses the functions $\psi(x)$ in
terms of the functions $C_{lm}^\pm (\tau,\sigma)$. Let us f\/ind
an expression for $C_{lm}^\pm (\tau,\sigma)$ in terms of
$\psi(x)$. Using the expansions \eqref{I-102} and \eqref{I-103} we
have
\begin{gather*}
C_{lm}^\varepsilon(\tau,\sigma)=\frac{\Gamma(\tau+1)}{\Gamma(\tau-l+1)}
\,\int_{H_+^4}\,\frac{d^4x}{x_0}\psi(x)\\
\phantom{C_{lm}^\varepsilon(\tau,\sigma)=}{}
\times\int_{\Gamma_\varepsilon}\, dk'[x,k']^\sigma(\sinh c)^{-1/2}
 P_{-\tau-3/2}^{-l-1/2}(\cosh c)
\overline{Y_{lm}(\Theta,\Phi)}.
\end{gather*}
Integration over $k'$ is fulf\/illed in the same way as in the
case of the integral \eqref{I-105}. We obtain
\begin{gather}
C_{lm}^\varepsilon(\tau,\sigma)=(2\pi)^{3/2}
\frac{\Gamma(\tau{+}1)\Gamma({-}\sigma{-}\tau{-}2)\Gamma(\tau{-}\sigma)}
{\Gamma(\tau{-}l{+}1)\Gamma({-}\sigma)}
\int_{H_+^4}\frac{d^4x}{x_0} \psi(x)(\cosh a)^{-3/2} (\sinh
b)^{-1/2}
\nonumber\\
\label{I-116} \phantom{C_{lm}^\varepsilon(\tau,\sigma)=}{} \times
P_{-\tau-3/2}^{\sigma+3/2}(-\varepsilon\tanh a)
P_{-\tau-3/2}^{-l-1/2}(\cosh b)\overline{Y_{lm}(\theta,\phi)},
\end{gather}
where the measure $d^4x/x_0$ in the coordinates $a$, $b$,
$\theta$, $\varphi$ has the form
\[
d^4x/x_0=\cosh^3a\sinh^2b\sin\theta\, da\, db\, d\theta\,
d\varphi.
\]

For the unitary case when $\sigma={\rm i}\rho-3/2$ (that is,
$\delta=-3/2$) and $\tau={\rm i}\nu-1$ (that is $\gamma=-1$),
where $\rho$ and $\nu$ are real numbers, we receive
\begin{gather}
\psi(x)=\frac{1}{(2\pi)^{5/2}}\sum_{l=0}^\infty\,\sum_{m=-l}^l\,\int_{0}^{\infty}
\,d\rho(\rho^2+1/4)\rho\tanh\pi\rho\,\int_{0}^{\infty}\,d\nu\,
\nu^2
\nonumber\\
\phantom{\psi(x)=}{}\times \frac{\Gamma({\rm i}\rho+{\rm
i}\nu+1/2)\Gamma({\rm i}\rho-{\rm i}\nu+1/2)} {\Gamma({\rm
i}\rho+3/2)}V_{\nu lm}(b,\theta,\varphi)(\cosh a)^{-3/2}
\nonumber\\
\label{I-117} \phantom{\psi(x)=}{}\times \left[
C_{lm}^+(\nu,\rho)P_{{\rm i}\nu-^1/_2}^{-{\rm i}\rho}(-\tanh a)+
C_{lm}^-(\nu,\rho)P_{{\rm i}\nu-^1/_2}^{-{\rm i}\rho}(\tanh
a)\right] ,
\\
C_{lm}^\varepsilon(\nu,\rho)=(2\pi)^{3/2} \frac{\Gamma(-{\rm
i}\rho-{\rm i}\nu+1/2)\Gamma(-{\rm i}\rho+{\rm i}\nu+1/2)}
{\Gamma(-{\rm i}\rho+3/2)}
\nonumber\\
\label{I-118} \phantom{C_{lm}^\varepsilon(\nu,\rho)=}{}
\times\int_{H_+^4}\frac{d^4x}{x_0}\psi(x)(\cosh a)^{-3/2}P_{-{\rm
i}\nu-1/2}^{{\rm i}\rho}(-\varepsilon\tanh a) \overline{V_{\nu
lm}(b,\theta,\phi)},
\end{gather}
where
\begin{equation}\label{I-119}
V_{\nu lm}(b,\theta,\varphi)=\frac{\Gamma({\rm i}\nu)}{\Gamma({\rm
i}\nu-l)} (\sinh b)^{-1/2}P_{{\rm i}\nu-1/2}^{-l-1/2}(\cosh b)
Y_{lm}(\theta,\phi).
\end{equation}
Here we have taken into account that
\begin{gather*}
\tau(\tau+1)d\tau=-\nu({\rm i}\nu-1)d\nu,\qquad
\cot\pi\sigma=-{\rm i}\tanh\pi\rho,
\\
(\sigma+2)(\sigma+1)\sigma d\sigma=-{\rm i}(\rho^2+1/4)({\rm
i}\rho-3/2)d\rho,
\\
\int_{-\infty}^\infty\,d\rho\, \Omega(\rho)
(\cdots)=\int_{-\infty}^0\,d\rho\,  \Omega(\rho) (\cdots)
+\int_0^\infty\,d\rho\, \Omega(\rho)
(\cdots)\\
\phantom{\int_{-\infty}^\infty\,d\rho\, \Omega(\rho)(\cdots)}{}
=2{\rm i}\,\int_0^\infty\,d\rho\rho(\rho^2+1/4)\tanh\pi\rho\,
(\cdots),
\end{gather*}
where $\Omega(\rho) =(\rho^2+1/4)({\rm i}\rho-3/2)\tanh\pi\rho$
and $(\cdots)$ is the expression under the sign of the integral
over $\rho$ in \eqref{I-117}.

The formula \eqref{I-117} is an expansion of the function
$\psi(x)$ in basis functions \eqref{I-44}. It follows from the
expansions \eqref{I-117}--\eqref{I-119} that the functions
\eqref{I-44} have the following normed form:
\begin{gather}
\Phi_{\rho\nu lm}^\varepsilon(a,b,\theta,\varphi)=
\frac{|\Gamma({\rm i}\rho+{\rm i}\nu+1/2) \Gamma({\rm i}\rho-{\rm
i}\nu+1/2)\Gamma({\rm i}\nu)|} {\sqrt{2\pi}\,|\Gamma({\rm
i}\rho+3/2)\Gamma({\rm i}\nu-l)|} (\cosh a)^{-3/2}
\nonumber\\
\label{I-120} \phantom{\Phi_{\rho\nu
lm}^\varepsilon(a,b,\theta,\varphi)=}{} \times P_{{\rm
i}\nu-1/2}^{-{\rm i}\rho} (\varepsilon \tanh a) (\sinh
b)^{-1/2}P_{{\rm i}\nu-1/2}^{-l-1/2}(\cosh b) Y_{lm}(\theta,\phi).
\end{gather}
This functions are normed by the formula
\begin{gather}
\langle\Phi_{\rho'\nu' l'm'}^{\varepsilon'},\Phi_{\rho\nu
lm}^\varepsilon\rangle=\,\int_{H_+^4}\frac{d^4x}{x_0}\Phi_{\rho\nu
lm}^\varepsilon(a,b,\theta,\varphi)\overline{\Phi_{\rho'\nu'
l'm'}^{\varepsilon'}(a,b,\theta,\varphi)}
\nonumber\\
\label{I-121} \phantom{\langle\Phi_{\rho'\nu'
l'm'}^{\varepsilon'},\Phi_{\rho\nu lm}^\varepsilon\rangle}{}
=\frac{\delta(\rho-\rho')}{\rho(\rho^2+1/4)\tanh\pi\rho}
\frac{\delta(\nu-\nu')}{\nu^2}\delta_{ll'}\delta_{mm'}\delta_{\varepsilon\varepsilon'}.
\end{gather}
The Plancherel formula for the transforms
\eqref{I-117}--\eqref{I-118} is of the form
\begin{gather}
\int_{H_+^4}\frac{d^4x}{x_0}|\psi(x)|^2=\frac{1}{(2\pi)^4}\,\sum_{l=0}^\infty\,
\sum_{m=-l}^l\,\int_0^\infty\,d\rho(\rho^2+1/4)\rho\tanh\pi\rho
\nonumber\\
\label{I-122} \phantom{\int_{H_+^4}\frac{d^4x}{x_0}|\psi(x)|^2=}{}
\times
\int_0^\infty\,d\nu\nu^2\left[|C_{lm}^+(\nu,\rho)|^2+|C_{lm}^-(\nu,\rho)|^2\right]
.
\end{gather}

\subsection[Expansion in $O$-coordinate system]{Expansion in $\boldsymbol{O}$-coordinate system}
\label{expan-hyp-o}

In this case, the cut of $C^4_+$ by the plane $k_0-k_4=1$ is
considered as a contour $\Gamma$. We parametrize this contour
$\Gamma\equiv \Gamma_O$ by the coordinates $\zeta,\Theta,\Phi$:
\begin{gather}
k'_0=(1+\zeta^2)/2,\ \ k'_1=\zeta\sin\Theta\cos\Phi,\qquad
k'_2=\zeta\sin\Theta\sin\Phi,
\nonumber\\
\label{I-123}
 k'_3=\zeta\cos\Theta,\qquad
k'_4=(-1+\zeta^2)/2,
\\
 0\leqslant\zeta<\infty,\qquad
0\leqslant\Theta<\pi,\qquad 0\leqslant\Phi<2\pi.\nonumber
\end{gather}
We expand the function $\Psi(k',\sigma)$, considered on this
contour $\Gamma_0$, in the spherical Bessel functions
\[
j_l(\kappa\zeta)=\left(\frac{\pi}{2\kappa\zeta}\right)^{1/2}J_{l+1/2}(\kappa\zeta)
\]
and spherical harmonics $Y_{lm}(\Theta,\Phi)$. Then
\begin{equation}\label{I-124}
\Psi(k',\sigma)=\sum_{l=0}^\infty\,
\sum_{m=-l}^l\,\int_0^\infty\,d\kappa\kappa^2A_{lm}(\kappa,\sigma)\Psi_{\kappa
lm}(k'),
\end{equation}
where $\Psi_{\kappa lm}(k')=j_l(\kappa\zeta)Y_{lm}(\Theta,\Phi)$.
The inverse transform is of the form
\begin{equation}\label{I-125}
A_{lm}(\kappa,\sigma)=\frac{2}{\pi}\,\int_{\Gamma_O}\,dk'\Psi(k',\sigma)
\overline{\Psi_{\kappa lm}(k')},
\end{equation}
where $dk'=\zeta^2\sin\Theta\, d \zeta\, d\Theta d\, \Phi$. Now we
have
\begin{gather}
\psi(x)=\frac{\rm i}{2(2\pi)^4}\,\sum_{l=0}^\infty\,
\sum_{m=-l}^l\,\int_{\delta-{\rm i}\infty}^{\delta+{\rm
i}\infty}\,d\sigma
\frac{\Gamma(\sigma+3)}{\Gamma(\sigma)}\cot\pi\sigma\int_0^\infty\,
d\kappa\kappa^2A_{lm}(\kappa,\sigma)
\nonumber\\
\label{I-126} \phantom{\psi(x)=}{}
\times\int_{\Gamma_O}\,dk'[x,k']^{-\sigma-3} \Psi_{\kappa lm}(k').
\end{gather}
We have to calculate the integral
\[
J_{lm}^{\kappa\sigma}(x\equiv
=\int_{\Gamma_O}\,dk'[x,k']^{-\sigma-3}
\left(\frac{\pi}{2\kappa\zeta}\right)^{1/2}\,J_{l+1/2}(\kappa\zeta)
Y_{lm}(\Theta,\Phi).
\]
Due to orthogonality of the Legendre functions $P_l^m(\cos\Theta)$
and orthogonality of the exponential functions $\exp {\rm
i}m\Phi$, for the point $x^0=x^0(a)=(\cosh a,0,0,0,-\sinh a)$ we
have $J_{lm}^{\kappa\sigma}(x^0)=0$ for $l\neq m$,  $m\neq 0$, and
\[
J_{00}^{\kappa\sigma}(x_0)=\pi(2e^{-a})^{\sigma+3}\int_0^\infty\,
d\zeta\, \zeta^2(e^{-2a}+\zeta)^{-\sigma-3}
\left(\frac{2}{\kappa\zeta}\right)^{1/2}\,J_{1/2}(\kappa\zeta).
\]
Setting $e^{-a}=b$ and using the formula 6.565(4) in \cite{RG}, we
get
\[
J_{00}^{\kappa\sigma}(x^0)= (2b)^{3/2}\pi\kappa^{\sigma+3/2}
\frac{1}{\Gamma(\sigma+3)}K_{\sigma+3/2}(\kappa b).
\]

For arbitrary $x$ we have (see \cite{VSm})
\[
J_{lm}^{\kappa\sigma}(x)=\sqrt{4\pi} \,
J_{00}^{\kappa\sigma}(x^0(a))\Psi_{\kappa lm}(r,\theta,\varphi).
\]
Thus,
\begin{gather}
\psi(x)=\frac{{\rm i}b^{3/2}}{(2\pi)^{5/2}}\,\sum_{l=0}^\infty\,
\sum_{m=-l}^l\,\int_{\delta-{\rm i}\infty}^{\delta+{\rm
i}\infty}\,d\sigma
\frac{\cot\pi\sigma}{\Gamma(\sigma)}\int_0^\infty\, d\kappa
A_{lm}(\kappa,\sigma)\nonumber\\
\label{I-127}    \phantom{\psi(x)=}{} \times
\kappa^{\sigma+7/2}K_{\sigma+3/2}(\kappa b)\Psi_{\kappa
lm}(r,\theta,\varphi),\qquad  b=e^{-a},
\end{gather}
where $\Psi_{\kappa lm}$ is such function as in \eqref{I-124} but
determined for other variables.

In order to express $A_{lm}(\kappa,\sigma)$ in terms of
$\Psi(k',\sigma)$, we substitute the expression \eqref{I-102} into
\eqref{I-125}:
\[
A_{lm}(\kappa,\sigma)=\frac{2}{\pi}\int_{H_+^4}\,\frac{d^4x}{x_0}\psi(x)
\int_{\Gamma_0}\,dk'[x,k']^\sigma\overline{\Psi_{\kappa lm}(k')}.
\]
Integrating over $k'$ we obtain
\begin{equation}\label{I-128}
A_{lm}(\kappa,\sigma)=\frac{8(2\pi)^{1/2}\kappa^{-\sigma-3/2}}{\Gamma(-\sigma)}
\int_{H_+^4}\,\frac{d^4x}{x_0}\psi(x) b^{3/2}K_{\sigma+3/2}(\kappa
b)\overline{\Psi_{\kappa lm}(r,\theta,\varphi)},
\end{equation}
where
\[
d^4x/x_0=e^{3a}r^2\sin\theta\, da\, dr\, d\theta\, d\varphi.
\]

For the unitary case, when $\sigma={\rm i}\rho-3/2$, $\rho\in
\mathbb{R}$, $\delta=-3/2$, the relations \eqref{I-127} and
\eqref{I-128} take the form
\begin{gather}
\psi(x)=\frac{b^{3/2}}{\pi(2\pi)^{3/2}}\,\sum_{l=0}^\infty\,
\sum_{m=-l}^l\,\int_0^\infty d\rho\, (\rho^2+1/4)\rho
\frac{\tanh\pi\rho}{\Gamma({\rm i}\rho+3/2)}
\nonumber\\
\label{I-129}    \phantom{\psi(x)=}{}
\times\int_0^\infty\,d\kappa\, \kappa^2
A_{lm}(\kappa,\rho)\kappa^{{\rm i}\rho}K_{{\rm i}\rho}(\kappa
b)\Psi_{\kappa lm}(r,\theta,\varphi),
\\
\label{I-130} A_{lm}(\kappa,\rho)=\frac{8(2\pi)^{1/2}\kappa^{-{\rm
i}\rho}} {\Gamma(-{\rm i}\rho+3/2)}
\int_{H_+^4}\,\frac{d^4x}{x_0}\psi(x) b^{3/2}K_{{\rm
i}\rho}(\kappa b)\overline{\Psi_{\kappa lm}(r,\theta,\varphi)}.
\end{gather}
The formula \eqref{I-129} is an expansion of the function
$\psi(x)$ in basis elements \eqref{I-57}. The formulas
\eqref{I-129} and \eqref{I-130} show that the functions
\eqref{I-57} in a normed form have the form
\begin{equation}\label{I-131}
\Phi_{lm}^{\rho\kappa}(b,r,\theta,\varphi)= \frac{(2/\pi\kappa
r)^{1/2}b^{3/2}}{|\Gamma(i\rho+3/2)|}\, K_{{\rm i}\rho}(\kappa
b)J_{l+1/2}(\kappa r) Y_{lm}(\theta,\varphi).
\end{equation}
The normalization condition is
\begin{equation}\label{I-132}
\langle\Phi_{l'm'}^{\rho'\kappa'},\Phi_{lm}^{\rho\kappa}\rangle=
\frac{\delta(\rho-\rho')}{\rho(\rho^2+1/4)\tanh\pi\rho}
\frac{\delta(\kappa-\kappa')}{\kappa^2}\,\delta_{ll'}\delta_{mm'}.
\end{equation}
The Plancherel formula for the transforms
\eqref{I-129}--\eqref{I-130} is
\begin{equation}\label{I-133}
\int_{H_+^4} \frac{d^4x}{x_0}|\psi(x)|^2 {=} \frac{1}{4(2\pi)^3}
\sum_{l=0}^\infty\, \sum_{m=-l}^l\,\int_0^\infty
d\rho\,\left(\rho^2{+}\frac14 \right) \rho\tanh\pi\rho
 \int_0^\infty d\kappa\, \kappa^2|A_{lm}(\kappa,\rho)|^2.
\end{equation}

\subsection[Expansion in $OC$-coordinate system]{Expansion in $\boldsymbol{OC}$-coordinate system}
\label{expan-hyp-oc}
We leave in this case the contour $\Gamma_O$ of the previous
subsection and introduce on it the coordinates $\zeta$, $s$,
$\Phi$:
\begin{gather}
k'_0=(\zeta^2+s^2+1)/2,\qquad k'_1=\zeta\cos\Phi,\qquad
k'_2=\zeta\sin\Phi,\qquad k'_3=s,
\nonumber\\
\label{I-134} k'_4=(\zeta^2+s^2-1)/2,\qquad
0\leqslant\zeta<\infty,\qquad -\infty<s<\infty,\qquad
0\leqslant\Phi<2\pi.
\end{gather}
We expand the function $\Psi(k',\sigma)$ in basis functions on
$\Gamma_O$,
\begin{equation}\label{I-135}
\Psi_{\eta q}^m(k')\equiv\Psi_{\eta q}^m(\zeta,s,\Phi)=
\frac{1}{2\pi}J_m(\eta\zeta)e^{{\rm i}qs}e^{{\rm i}m\Phi}.
\end{equation}
This expansion is of the form
\begin{gather}\label{I-136}
\Psi(k',\sigma)=\sum_{m=-\infty}^\infty\,\int_0^\infty\,d\eta\eta
\int_{-\infty}^\infty\,dqA_m(\eta,q,\sigma)\Psi_{\eta q}^m(k'),
\\
\label{I-137}
A_m(\eta,q,\sigma)=\int_{\Gamma_{O}}\,dk'\Psi(k',\sigma)
\overline{\Psi_{\eta q}^m(k')},
\end{gather}
where $dk'=\zeta\, d\zeta\, ds\, d\Phi$. Substituting the
expression \eqref{I-136} into \eqref{I-101}, we get
\begin{gather}
\psi(x)=\frac{\rm i}{2(2\pi)^4}\,\sum_{m=-\infty}^\infty\,
\int_{\delta-{\rm i}\infty}^{\delta+{\rm i}\infty}\,d\sigma
\frac{\Gamma(\sigma+3)}{\Gamma(\sigma)}\cot\pi\sigma
\int_0^\infty\,d\eta\eta\,\int_{-\infty}^\infty dq\,
A_m(\eta,q,\sigma)
\nonumber\\
\label{I-138}    \phantom{\psi(x)=}{} \times
\int_{\Gamma_{O}}\,dk'[x,k']^{-\sigma-3} \Psi_{\eta q}^m(k').
\end{gather}
The integral
\[
J_m^{\eta q\sigma}(x)=\int_{\Gamma_{O}}\,dk'[x,k']^{-\sigma-3}
\Psi_{\eta q}^m(k')
\]
for $x=x^0(a)=(\cosh a,0,0,0,-\sinh a)$ and $m=0$ takes the form
\[
J_{0}^{\eta q\sigma}(x^0(a))=(2b)^{\sigma+3} \int_0^\infty
d\zeta\, \zeta J_0(\eta\zeta)\, \int_{-\infty}^\infty\,ds\,e^{{\rm
i}qs}(\zeta^2+s^2+b^2)^{-\sigma-3},
\]
where $e^{-a}\equiv b$. Using the formula
\[
\int_{-\infty}^\infty\,dy(1+y^2)^\lambda e^{{\rm i}yt}=
\frac{2\sqrt{\pi}}{\Gamma(-\lambda)}\left(\frac{|t|}{2}\right)^{-\lambda-1/2}
K_{-\lambda-1/2}(|t|),
\]
by means of the relation 6.596(7) in \cite{RG} we get
\[
J_0^{\eta q\sigma}(a)=2(2\pi)^{1/2}b^{3/2}
\frac{\kappa^{\sigma+3/2}}{\Gamma(\sigma+3)}K_{\sigma+3/2}(\kappa
b),
\]
where $\kappa^2=\eta^2+q^2$ and $b=e^{-a}$. Then
\[
J_m^{\eta q\sigma}(x)=2(2\pi)^{3/2}b^{3/2}
\frac{\kappa^{\sigma+3/2}}{\Gamma(\sigma+3)}K_{\sigma+3/2}(\kappa
b)\Psi_{\eta q}^m(\xi,z,\varphi).
\]
It follows from \eqref{I-138} that
\begin{gather}
\psi(x)=\frac{{\rm
i}b^{3/2}}{(2\pi)^{5/2}}\,\sum_{m=-\infty}^\infty\,
\int_{\delta-{\rm i}\infty}^{\delta+{\rm i}\infty}\,d\sigma
\frac{\cot\pi\sigma}{\Gamma(\sigma)}
\int_0^\infty\,d\eta\eta\,\int_{-\infty}^\infty\,dq
A_m(\eta,q,\sigma)
\nonumber\\
\label{I-139}   \phantom{\psi(x)=}{} \times
\kappa^{\sigma+3/2}K_{\sigma+3/2}(\kappa b)\Psi_{\eta
q}^m(\xi,z,\varphi),
\end{gather}
where $\kappa^2=\eta^2+q^2$, $e^{-a}=b$, and $\Psi_{\eta
q}^m(\xi,z,\varphi)$ is such  function as in \eqref{I-135} but
determined for other variables.

We have from \eqref{I-137} and \eqref{I-102} that
\[
A_m(\eta,q,\sigma)=\int_{H_+^4}\frac{d^4x}{x_0}\psi(x)
\int_{\Gamma_{O}}\,dk'[x,k']^\sigma \overline{\Psi_{\eta
q}^m(k')}.
\]
Integrating over $k'$ we get
\begin{equation}\label{I-140}
A_m(\eta,q,\sigma)=2(2\pi)^{3/2}\frac{\kappa^{-\sigma-3/2}}{\Gamma(-\sigma)}
\int_{H_+^4}\frac{d^4x}{x_0}\psi(x)b^{3/2}K_{\sigma+3/2}(\kappa
b)\overline{\Psi_{\eta q}^m(\xi,z,\varphi)},
\end{equation}
where
\[
dx^4/x_0=e^{3a}da\, \xi d\xi\, dz\, d\varphi.
\]

For the unitary case, when $\sigma=i\rho-3/2$ and $\delta=-3/2$,
formulas \eqref{I-139} and \eqref{I-140} take the form
\begin{gather}
\psi(x)=\frac{1}{\pi} \left( \frac{b}{2\pi}\right)^{3/2}
\sum_{m=-\infty}^\infty\, \int_0^\infty\,d\rho \, (\rho^2+1/4)
\frac{\rho \tanh\pi\rho}{\Gamma({\rm i}\rho+3/2)}
\nonumber\\
\label{I-141}      \phantom{\psi(x)=}{} \times
\int_0^\infty\,d\eta\eta
\int_{-\infty}^\infty\,dqA_m(\eta,q,\rho)\kappa^{{\rm i}\rho}
K_{{\rm i}\rho}(\kappa b)\Psi_{\eta q}^m(\xi,z,\varphi),
\\
\label{I-142} A_m(\eta,q,\rho)=\frac{2(2\pi)^{3/2}\kappa^{-{\rm
i}\rho}}{\Gamma(-{\rm i}\rho+3/2)}
\int_{H_+^4}\frac{d^4x}{x_0}\psi(x)b^{3/2}K_{{\rm i}\rho}(\kappa
b)\overline{\Psi_{\eta q}^m(\xi,z,\Phi)}.
\end{gather}

The formula \eqref{I-141} is an expansion of the function
$\psi(x)$ in basis elements \eqref{I-69}. The
formu\-las~\eqref{I-141} and \eqref{I-142} show that the functions
\eqref{I-69} in a normed form have the form
\begin{equation}\label{I-143}
\Phi_{\rho\eta q}^m(b,\xi,z,\varphi)=\frac{1}{\pi\sqrt{2\pi}}
\frac{1}{|\Gamma({\rm i}\rho+3/2)|}b^{3/2}K_{{\rm i}\rho}(\kappa
b) J_m(\eta\xi)e^{{\rm i}qz}e^{{\rm i}m\varphi}.
\end{equation}
The normalization condition is
\begin{equation}\label{I-144}
\langle\Phi_{\rho'\eta'q'}^{m'},\Phi_{\rho\eta q}^m\rangle=
\frac{\delta(\rho-\rho')}{\rho(\rho^2+1/4)\tanh\pi\rho}
\frac{\delta(\eta-\eta')}{\eta}\,\delta(q-q')\delta_{mm'}.
\end{equation}
The Plancherel formula for the transforms
\eqref{I-141}--\eqref{I-142} is
\begin{gather}
\int_{H_+^4}\!\frac{d^4x}{x_0}|\psi(x)|^2=\frac{1}{(2\pi)^4}\sum_{m=-\infty}^\infty
\int_0^\infty\! d\rho\, (\rho^2{+}{{\tfrac14}})\rho\tanh\pi\rho
 \int_0^\infty\! d\eta \eta\int_{-\infty}^\infty\! dq|A_m(\eta,q,\rho)|^2.\!\!\label{I-145}
\end{gather}

\subsection[Expansion in $OT$-coordinate system]{Expansion in $\boldsymbol{OT}$-coordinate system}
\label{expan-hyp-ot}
We leave in this case the contour $\Gamma_O$ of two previous
subsections and introduce on it the coordinates $\chi_1$,
$\chi_2$, $\chi_3$:
\begin{gather*}
k'_0=(1+\chi^2)/2,\qquad k'_1=\chi_1,\quad k'_2=\chi_2,\qquad
k'_3=\chi_3,\qquad k'_4=(-1+\chi^2)/2,
\\
\chi^2=\chi_1^2+\chi_2^2+\chi_3^2,\qquad
-\infty<\chi_i<\infty,\qquad i=1,2,3.
\end{gather*}
We consider the expansion of the function $\Psi(k',\sigma)$ on
this contour in functions
\[
\Psi_{\boldsymbol{\kappa}}(k')\equiv\Psi_{\boldsymbol{\kappa}}
(\boldsymbol{\chi})=(2\pi)^{-3/2}
e^{i\boldsymbol{\kappa}\boldsymbol{\chi}}, \qquad
\boldsymbol{\kappa}=(\kappa_1,\kappa_2,\kappa_3),\qquad
\boldsymbol{\chi}=(\chi_1,\chi_2,\chi_3),
\]
and have
\begin{equation}\label{I-146}
\Psi(k',\sigma)=\int_{\mathbb{R}^3}\,d\boldsymbol{\kappa}\,
A(\boldsymbol{\kappa},\sigma)\Psi_{\boldsymbol{\kappa}}(k').
\end{equation}
The inverse transform is
\begin{equation}\label{I-147}
A(\boldsymbol{\kappa},\sigma)=\int_{\Gamma_{O}}\,dk'\Psi(k',\sigma)
\overline{\Psi_{\boldsymbol{\kappa}}(k')},
\end{equation}
where $dk'=d\boldsymbol{\chi}$. Substituting the expression
\eqref{I-146} into \eqref{I-101} we get
\[
\psi(x)=\frac{\rm i}{2(2\pi)^4}\, \int_{\delta-{\rm
i}\infty}^{\delta+{\rm i}\infty}\,d\sigma
\frac{\Gamma(\sigma+3)}{\Gamma(\sigma)}\cot\pi\sigma
\int_{\mathbb{R}^3}\,d\boldsymbol{\kappa}
A(\boldsymbol{\kappa},\sigma)\,\int_{\Gamma_{O}}\,dk'[x,k']^{-\sigma-3}
\Psi_{\boldsymbol{\kappa}}(k').
\]

We have to calculate the integral
\[
J^{\boldsymbol{\kappa}\sigma}(x)=\int_{\Gamma_{O}}\,dk'[x,k']^{-\sigma-3}
\Psi_{\boldsymbol{\kappa}}(k').
\]
First we consider the integral
\[
J^{\boldsymbol{\kappa}\sigma}(a)\equiv
J^{\boldsymbol{\kappa}\sigma}(x_0(a))=
(2\pi)^{-3/2}(2e^{-a})^{\sigma+3}\int_{\mathbb{R}^3}
d\boldsymbol{\chi}\, (e^{-2a}+\chi^2)^{-\sigma-3}
e^{i\boldsymbol{\kappa}\boldsymbol{\chi}}.
\]
We use the spherical coordinates. Taking into account that
\[
\int_0^\pi\,d\theta\sin\theta e^{{\rm i}\kappa \chi\cos\theta}=
\left(\frac{2\pi}{\kappa \chi}\right)^{1/2}\,J_{1/2}(\kappa \chi),
\]
where $\theta$ is an angle between $\boldsymbol{\kappa}$ and
$\boldsymbol{\chi}$, we have
\[
J^{\boldsymbol{\kappa}\sigma}(a)=(2e^{-a})^{\sigma+3}
\kappa^{-1/2}\int_0^\infty\,
d\chi\chi^{3/2}(e^{-2a}+\chi^2)^{-\sigma-3}J_{1/2}(\kappa \chi),
\]
where $\kappa=(\kappa_1^2+\kappa_2^2+\kappa_3^2)^{1/2}$. Using
formula 6.565(4) in \cite{RG}, one receives
\[
J^{\boldsymbol{\kappa}\sigma}(a)=2\kappa^{\sigma+3/2}b^{3/2}
\frac{1}{\Gamma(\sigma+3)} K_{\sigma+3/2}(\kappa b),\qquad
b=e^{-a}.
\]
Since $x=x(a,{\bf y})$ is obtained from $x^0(a)$ by means of the
shift from the subgroup $T(3)$, we get
\[
J^{\boldsymbol{\kappa}\sigma}({\bf
x})=2(2\pi)^{3/2}b^{3/2}\kappa^{\sigma+3/2}
\frac{1}{\Gamma(\sigma+3)}K_{\sigma+3/2}(\kappa
b)\Psi_{\boldsymbol{\kappa}}({\bf y}).
\]

Thus, we have the following formula for an expansion of the
function $\psi(x)$:
\begin{equation}\label{I-148}
\psi(x)=\frac{{\rm i}b^{3/2}}{(2\pi)^{5/2}}\,
\int_{\delta-i\infty}^{\delta+{\rm i}\infty}\,d\sigma
\frac{\cot\pi\sigma}{\Gamma(\sigma)}\int_{\mathbb{R}^3}
d\boldsymbol{\kappa}\,
A(\boldsymbol{\kappa},\sigma)\kappa^{\sigma+3/2}K_{\sigma+3/2}(\kappa
b)\Psi_{\boldsymbol{\kappa}}({\bf y}),
\end{equation}
where $b=e^{-a}$. In order to express
$A(\boldsymbol{\kappa},\sigma)$ in terms of $\psi(x)$ we
substitute the expression \eqref{I-102} for $\Psi(k',\sigma)$ into
\eqref{I-147}. Then
\[
A(\boldsymbol{\kappa},\sigma)=\,
\int_{H_+^4}\,\frac{d^4x}{x_0}\psi(x) \int_{\Gamma_{O}}
dk'\,[x,k']^\sigma\overline{\Psi_{\boldsymbol{\kappa}}(k')}.
\]
Performing integration in $k'$ over the contour $\Gamma_O$, we
f\/ind
\begin{equation}\label{I-149}
A(\boldsymbol{\kappa},\sigma)=\frac{2(2\pi)^{3/2}}{\Gamma(-\sigma)}
\kappa^{-\sigma-3/2}
\int_{H_+^4}\,\frac{d^4x}{x_0}\psi(x)b^{3/2}K_{\sigma+3/2} (\kappa
b)\overline{\Psi_{\boldsymbol{\kappa}}({\bf y})},
\end{equation}
where
\[
d^4x/x_0=e^{3a}da\, d{\bf y}=e^{3a}da\, dy_1\, dy_2\, dy_3.
\]

For the unitary case (when $\sigma={\rm i}\rho-3/2$ and
$\delta=-3/2$) one gets
\begin{gather}\label{I-150}
\psi(x)=\frac{1}{\pi}\left(\frac{b}{2\pi}\right)^{3/2}
\int_{0}^{\infty}\,d\rho\left(\rho^2+\frac{1}{4}\right)
\frac{\rho\tanh\pi\rho}{\Gamma({\rm
i}\rho+\frac32)}\,\int_{\mathbb{R}^3} d\boldsymbol{\kappa}\,
A(\boldsymbol{\kappa},\rho) \kappa^{{\rm i}\rho}K_{{\rm
i}\rho}(\kappa b) \Psi_{\boldsymbol{\kappa}}({\bf y}),
\\
\label{I-151} A(\boldsymbol{\kappa},\rho)=\frac{2(2\pi)^{3/2}}
{\Gamma(-{\rm i}\rho+3/2)}\kappa^{-{\rm i}\rho}
\int_{H_+^4}\frac{d^4x}{x_0}\psi(x)b^{3/2}K_{{\rm i}\rho}(\kappa
b)\overline{\Psi_{\boldsymbol{\kappa}}({\bf y})}.
\end{gather}

The formula \eqref{I-150} gives an expansion of the function
$\psi(x)$ in the basis functions \eqref{I-75}. A~normed form of
these basis functions is
\begin{equation}\label{I-152}
\Phi^{\rho\boldsymbol{\kappa}}(b,{\bf y})=\frac{1}{2\pi^2}
\frac{1}{|\Gamma({\rm i}\rho+3/2)|}\,b^{3/2} K_{{\rm
i}\rho}(\kappa b)e^{{\rm i}\boldsymbol{\kappa}{\bf y}}.
\end{equation}
The normalization condition is
\begin{equation}\label{I-153}
\langle\Phi^{\rho'\boldsymbol{\kappa}'},\Phi^{\rho\boldsymbol{\kappa}}\rangle=
\frac{\delta(\rho-\rho')}{(\rho^2+1/4)\rho\tanh\pi\rho}
\delta(\boldsymbol{\kappa}-\boldsymbol{\kappa}').
\end{equation}
The Plancherel formula for the transforms
\eqref{I-150}--\eqref{I-151} is
\begin{equation}\label{I-154}
\int_{H_+^4}\,\frac{d^4x}{x_0}|\psi(x)|^2=\frac{1}{(2\pi)^4}
\int_{0}^{\infty}\,d\rho\left(\rho^2{+}{\textstyle{\frac{1}{4}}}\right)
\rho\tanh\pi\rho\,\int_{\mathbb{R}^3}\,d\boldsymbol{\kappa}\,
|A(\boldsymbol{\kappa},\rho)|^2.
\end{equation}

\subsection[Expansion in $C$-coordinate system]{Expansion in $\boldsymbol{C}$-coordinate system}
\label{expan-hyp-c}

In this case we take a cut of the cone $C^4_+$ by the cylinder
$k_0^2-k_4^2=1$. This contour is denote by $\Gamma_C$. We
parametrize this contour by the coordinates $c$, $\Theta$, $\Phi$
such that
\begin{gather}
k'_0=\cosh c,\qquad k'_1=\sin\Theta\cos\Phi,\qquad
k'_2=\sin\Theta\sin\Phi,\qquad k'_3=\cos\Theta,\qquad k'_4=\sinh
c,\nonumber
\\
-\infty<c<\infty,\qquad 0<\Theta\leqslant\pi,\qquad
0\leqslant\Phi<2\pi.\label{I-155}
\end{gather}
We expand the function $\Psi(k',\sigma)$ in the spherical
functions $Y_{lm}(\Theta,\Phi)$ and fulf\/il the Fourier transform
in the parameter $c$. As a result, we get
\begin{equation}\label{I-156}
\Psi(k',\sigma)=\sum_{l=0}^\infty\,\sum_{m=-l}^l\,\int_{-\infty}^\infty\,
d\tau C_{lm}(\tau,\sigma)e^{{\rm i}\tau c}Y_{lm}(\Theta,\Phi).
\end{equation}
The coef\/f\/icients $C_{lm}(\tau,\sigma)$ are given by
\begin{equation}\label{I-157}
C_{lm}(\tau,\sigma)=\frac{1}{2\pi}\,\int_{\Gamma_C}\,dk'
\Psi(k',\sigma)e^{-{\rm i}\tau c}\overline{Y_{lm}(\Theta,\Phi)},
\end{equation}
where $dk'=\sin\Theta\, dc\, d\Theta\, d\Phi$. Substituting this
expression for the coef\/f\/icients $C_{lm}(\tau,\sigma)$ into
\eqref{I-101} one has
\begin{gather}
\psi(x)=\frac{\rm
i}{2(2\pi)^4}\,\sum_{l=0}^\infty\,\sum_{m=-l}^l\,
\int_{\delta-{\rm i}\infty}^{\delta+{\rm i}\infty}\,d\sigma\,
\frac{\Gamma(\sigma+3)}{\Gamma(\sigma)}\cot\pi\sigma
\int_{-\infty}^\infty d\tau\, C_{lm}(\tau,\sigma)
\nonumber\\
\label{I-158}     \phantom{\psi(x)=}{} \times
\int_{\Gamma_{C}}\,dk'\,[x,k']^{-\sigma-3} e^{{\rm i}\tau
c}Y_{lm}(\Theta,\Phi).
\end{gather}

We have to calculate the integral
\begin{equation}\label{I-159}
J_{lm}^{\sigma\tau}(x)=\int_{-\infty}^\infty dc\,e^{{\rm i}\tau c}
\int_0^\pi\,d\Theta\, \sin\Theta \int_0^{2\pi}\,d\Phi\,
[x,k']^{-\sigma-3}Y_{lm}(\Theta,\Phi).
\end{equation}
We do this f\/irst for the point $x^0=(\cosh a,0,0,\sinh a,0)$.
The problem is reduced to calculation of the integral
\begin{equation}\label{I-160}
J_l^{\sigma\tau}(a)=\lambda\int_{-\infty}^\infty dc\, e^{{\rm
i}\tau c} \int_0^\pi\,d\Theta\sin\Theta(\cosh a\cosh
c-\cos\Theta\sinh a)^{-\sigma-3}P_l(\cos\Theta),
\end{equation}
where $\lambda=\sqrt{(2l+1)\pi}$. Using the relation
\[
P_l(x)=\frac{(-1)^l}{2^ll!}\frac{d^l}{dx^l}[(1-x^2)^l],\quad
x=\cos\theta,
\]
and integrating $l$ times by parts, one gets
\begin{gather*}
J_l^{\sigma\tau}(a)=\lambda\frac{(-1)^l\Gamma(-\sigma-2)\sinh^la}
{\Gamma(-\sigma-l-2)2^ll!} \int_{-\infty}^\infty dc\,e^{{\rm
i}\tau c} \int_0^\pi\,d\Theta(\sin\Theta)^{2l+1}
\\
\phantom{J_l^{\sigma\tau}(a)=}{} \times (\cosh a\cosh
c-\cos\Theta\sinh a)^{-\sigma-l-3}.
\end{gather*}
Decomposing $(\cosh a\cosh c-\cos\Theta\sinh a)^{-\sigma-l-3}$
into series by using the formula for the Newton binomial and
performing termwise integration we f\/ind
\begin{equation}\label{I-163}
J_l^{\sigma\tau}(a)=\frac{\pi(2l+1)^{1/2}2^{\sigma+2}\Gamma(A)\Gamma(B)\tanh^l
a}
{\Gamma(\sigma+3)\Gamma(D)\cosh^{\sigma+3}a}\,_2F_1(A,B;D;\tanh^2a),
\end{equation}
where
\[
A=(\sigma+l+{\rm i}\tau+3)/2,\qquad B=(\sigma+l-{\rm
i}\tau+3)/2,\qquad D= l+3/2.
\]
For arbitrary value of $x$ we get
\begin{equation}\label{I-166}
J_{lm}^{\sigma\tau}(x)=\frac{2^{\sigma+3}\pi^{3/2}}{\Gamma(\sigma+3)}
\frac{\Gamma(A)\Gamma(B)\tanh^la}
{\Gamma(D)\cosh^{\sigma+3}a}\,_2F_1(A,B;D;\tanh^2a)
Y_{lm}(\theta\varphi)e^{{\rm i}\tau b}.
\end{equation}

Now we obtain the following expansion of the function $\psi(x)$:
\begin{gather}
\psi(x)=\frac{\rm
i}{(4\pi)^{5/2}}\,\sum_{l=0}^\infty\,\sum_{m=-l}^l\,
\tanh^la\,\int_{-\infty}^\infty d\tau\, e^{{\rm i}\tau b}
\int_{\delta-{\rm i}\infty}^{\delta+{\rm i}\infty}\,d\sigma
\frac{\cot\pi\sigma}{\Gamma(\sigma)}\left(\frac{2}{\cosh
a}\right)^{\sigma+3}
\nonumber\\
\label{I-167}    \phantom{\psi(x)=}{} \times
C_{lm}(\tau,\sigma)\frac{\Gamma(A)\Gamma(B)}{\Gamma(D)}\,
_2F_1(A,B;D;\tanh^2a)Y_{lm}(\theta,\varphi).
\end{gather}

In order to express the coef\/f\/icients $C_{lm}(\tau,\sigma)$ in
terms of $\psi(x)$, we substitute the expression~\eqref{I-102} for
$\Phi(k',\sigma)$ into \eqref{I-157}:
\begin{equation}\label{I-168}
C_{lm}(\tau,\sigma)=\frac{1}{2\pi}\,\int_{H_+^4}\,\frac{d^4x}{x_0}\psi(x)\,
\int_{\Gamma_C}dk'\,[x,k']^\sigma e^{-{\rm i}\tau
c}\overline{Y_{lm}(\Theta,\Phi)}.
\end{equation}
Integrating in $c$, $\Theta$ and $\Phi$ one obtains
\begin{gather}
C_{lm}(\tau,\sigma)=\frac{\pi^{1/2}\Gamma(A')\Gamma(B')}{2\Gamma(D')\Gamma(-\sigma)}\,
\int_{H_+^4}\,\frac{d^4x}{x_0}\psi(x)\tanh^la\left(\frac{2}{\cosh
a}\right)^{-\sigma}
\nonumber\\
\label{I-169}  \phantom{C_{lm}(\tau,\sigma)=}{} \times e^{-{\rm
i}\tau b}\, _2F_1(A', B'; D';
\tanh^2a)\overline{Y_{lm}(\theta,\varphi)},
\end{gather}
where
\begin{gather*}
A'=(l-\sigma+{\rm i}\tau)/2,\qquad B'=(l-\sigma-{\rm
i}\tau)/2,\qquad D'=D= l+3/2,
\\
 d^4x/x_0=\sinh^3a\sin\theta\, da\, db\, d\theta\, d\varphi.
\end{gather*}
For the unitary case (when $\sigma={\rm i}\rho-3/2$ and
$\delta=-3/2$) we have
\begin{gather}
\psi(x)=\frac{1}{(2\pi)^{5/2}}\,\sum_{l=0}^\infty\,\sum_{m=-l}^l
\tanh^la\,\int_0^\infty d\rho\, (\rho^2+1/4)\rho\tanh\pi\rho\,
2^{i\rho} (\cosh a)^{-{\rm i}\rho-3/2}
\nonumber\\
\phantom{\psi(x)=}{}\times\int_{-\infty}^\infty\,d\tau e^{{\rm
i}\tau b} C_{lm}(\tau,\rho)\frac{\Gamma\left(\frac{{\rm
i}\rho+{\rm i}\tau+l+3/2}{2}\right) \Gamma\left(\frac{{\rm
i}\rho-{\rm i}\tau+l+3/2}{2}\right)} {\Gamma({\rm
i}\rho+3/2)\Gamma(l+3/2)}
\nonumber\\
\phantom{\psi(x)=}{} \label{I-170} \times {}_2F_1\left(\frac{{\rm
i}\rho+{\rm i}\tau+l+3/2}{2}, \frac{{\rm i}\rho-{\rm
i}\tau+l+3/2}{2};
\,l+\frac{3}{2};\,\tanh^2a\right)Y_{lm}(\theta,\varphi),
\end{gather}
and
\begin{gather}
C_{lm}(\tau,\rho)=(2\pi)^{1/2}\frac{\Gamma\left( \frac{-{\rm
i}\rho+{\rm i}\tau+l+3/2}{2}\right) \Gamma\left(\frac{-{\rm
i}\rho-{\rm i}\tau+l+3/2}{2}\right)} {\Gamma(l+3/2)\Gamma(-{\rm
i}\rho+3/2)} 2^{-{\rm i}\rho}
\nonumber\\
\phantom{C_{lm}(\tau,\rho)=}{} \times
\int_{H_+^4}\,\frac{d^4x}{x_0}\, \psi(x)\tanh^la\, (\cosh a)^{{\rm
i}\rho-3/2}e^{-{\rm i}\tau b}\overline{Y_{lm}(\theta,\varphi)}
\nonumber\\
\phantom{C_{lm}(\tau,\rho)=}{} \label{I-171} \times
{}_2F_1\left(\frac{-{\rm i}\rho+{\rm i}\tau+l+3/2}{2}, \frac{-{\rm
i}\rho-{\rm i}\tau+l+3/2}{2}; \,l+\frac{3}{2};\,\tanh^2a\right).
\end{gather}

It follows from these formulas that the function $\psi(x)$ expands
in the basis \eqref{I-83} and the normed basis functions are of
the form
\begin{gather}
\Psi_{lm}^{\rho\tau}(a,b,\theta,\varphi)=\frac{1}{2\pi}
\frac{\left|\Gamma\left(\frac{{\rm i}\rho+{\rm
i}\tau+l+3/2}{2}\right) \Gamma\left(\frac{{\rm i}\rho-{\rm
i}\tau+l+3/2}{2}\right)\right|} {\Gamma({\rm
i}\rho+3/2)\Gamma(l+3/2)}(\cosh a)^{-3/2-{\rm i}\rho}\,\tanh^la
\label{I-170a} \\
\nonumber \phantom{\Psi_{lm}^{\rho\tau}(a,b,\theta,\varphi)=}{}
\times {}_2F_1\left(\frac{{\rm i}\rho+{\rm i}\tau+l+3/2}{2},
\frac{{\rm i}\rho-{\rm i}\tau+l+3/2}{2};
l+\frac{3}{2};\tanh^2a\right)e^{{\rm i}\tau
b}Y_{lm}(\theta,\varphi).
\end{gather}
The normalization condition is
\[
\langle\Psi_{l'm'}^{\rho'\tau'},\Psi_{lm}^{\rho\tau}\rangle=
\frac{\delta(\rho-\rho')}{\rho(\rho^2+1/4)\tanh\pi\rho}\,\delta(\tau-\tau')
\delta_{ll'}\delta_{mm'}.
\]
The Plancherel formula holds:
\[
\int_{H_+^4}\frac{d^4x}{x_0}|\psi(x)|^2=\frac{1}{(2\pi)^{3}}
\sum_{l=0}^\infty\,\sum_{m=-l}^l\,\int_0^\infty d\rho\,
(\rho^2+1/4)\rho\tanh\pi\rho\,
\int_{-\infty}^\infty\,d\tau|C_{lm}(\tau,\rho)|^2.
\]

\subsection[Expansion in $SH$-coordinate system]{Expansion in $\boldsymbol{SH}$-coordinate system}
\label{expan-hyp-sh}

In this case we take the cut $\Gamma_{SH}$ of $C^4_+$ by the
cylinder $k^2_3+k^2_4=1$ as the contour $\Gamma$ . This contour is
a product of a circle and the higher sheet of the two-sheeted
hyperboloid $H_+^2$. We choose on $\Gamma_{SH}$ the coordinates
$c,\alpha,\beta$, where
\begin{gather}
k'_0=\cosh c,\qquad k'_1=\sinh c \cos\alpha,\qquad k'_2=\sinh
c\sin\alpha,\qquad k'_3=\cos\beta,\qquad k'_4=\sin\beta,\nonumber
\\
0\leqslant c<\infty,\qquad 0\leqslant
\alpha,\beta<2\pi.\label{I-173}
\end{gather}

Let us perform the Fourier transform of the function
$\psi(k',\sigma)$ in the parameter $\beta$. One can consider
coef\/f\/icients of this expansion as functions on $H_+^2$,
parametrized by spherical coordinates $c$ and $\alpha$. We expand
these coef\/f\/icients in basis functions on $H_+^2$ in spherical
coordinates (they correspond to the reduction $SO_0(1,2)\supset
SO(2)$) using the formulas
\begin{gather*}
\phi(c,\alpha)=\frac{1}{4\pi{\rm i}}\sum_{m=-\infty}^\infty
\int_{\varepsilon -{\rm i}\infty}^{\varepsilon+{\rm i}\infty}
d\lambda\, \lambda \cot \pi\lambda \frac{\Gamma(-\lambda)}{\Gamma
(-\lambda+m)} B_m(\lambda)P^m_{-\lambda-1}(\cosh c)e^{{\rm
i}m\alpha},
\\
B_m(\lambda)=\frac{\Gamma(\lambda+1)}{\Gamma (\lambda-m+1)}
\int_0^\infty dc \int_0^{2\pi} d\alpha\, \phi(c,\alpha)
P^{-m}_\lambda(\cosh c) e^{-{\rm i}m\alpha}.
\end{gather*}
As a result, the formulas \eqref{I-101}--\eqref{I-102} take the
form
\begin{gather*}
\psi(x)=\frac{1}{4(2\pi)^5}\sum_{m,m'=-\infty}^\infty
\int_{\delta-{\rm i}\infty}^{\delta+{\rm i}\infty} d\sigma\,
\frac{\Gamma(\sigma+3)}{\Gamma(\sigma)} \cot \pi\sigma
\int_{\varepsilon -{\rm i}\infty}^{\varepsilon+{\rm i}\infty}
d\lambda\, \lambda \cot \pi\lambda \frac{\Gamma(-\lambda)}{\Gamma
(-\lambda+m)}
\\
\phantom{\psi(x)=}{}\times B_{mm'}(\lambda,\sigma)
\int_{\Gamma_{SH}} dk'\, [x,k']^{-\sigma-3} P^m_{-\lambda-1}(\cosh
c)e^{{\rm i}m\alpha}e^{{\rm i}m'\beta},
\\
B_{mm'}(\lambda,\sigma)=\frac{1}{2\pi}\frac{\Gamma(\lambda+1)}{\Gamma
(\lambda+m+1)} \int_{H_+^4} \frac{d^4 x}{x_0} \int_{\Gamma_{SH}}
dk' [x,k']^\sigma
 P^{m}_\lambda(\cosh c) e^{-{\rm i}m\alpha}e^{-{\rm i}m'\beta},
\end{gather*}
where $dk'=\sinh c\, dc\, d\alpha\, d\beta$. One integrates over
$\Gamma_{SH}$ in the same way as in the previous cases. We give a
f\/inal formula for expansion of the function $\psi(x)$ in basis
functions related to unitary representations ($\sigma={\rm
i}\rho-\frac 32$, $\lambda={\rm i}\omega-\frac12$, $0\leqslant
\rho,\omega<\infty$):
\begin{gather}
\psi(x)=\frac{1}{\pi(4\pi)^{5/2}}
\sum^\infty_{m'=-\infty}\sum^\infty_{m=-\infty}
\int^{\infty}_{0}d\rho \left(\rho^2{+}
{\textstyle{\frac{1}{4}}}\right)\rho\tanh
\pi\rho\left(\frac{2}{\cosh a}\right)^{{\rm i}\rho+3/2}
\nonumber\\
\phantom{\psi(x)=}{} \times \int^{\infty}_{0}d\omega\,
\omega\tanh\pi\omega \frac{\Gamma\left(-{\rm
i}\omega+\frac{1}{2}\right) \Gamma\left(\frac{{\rm i}\rho-{\rm
i}\omega+m'+1}{2}\right) \Gamma\left(\frac{{\rm i}\omega+{\rm
i}\rho+m'+1}{2}\right)} {m'!\Gamma\left({\rm
i}\rho+\frac{3}{2}\right) \Gamma\left(-{\rm
i}\omega+m+\frac{1}{2}\right)} B_{m'm}(\omega,\rho)
\label{I-190}\\
\phantom{\psi(x)=}{}
 \times
P^m_{-{\rm i}\omega-1/2}(\cosh b)e^{{\rm i}m\varphi}e^{{\rm
i}m'\Phi} {}_2F_1 \!\left(\!\frac{{\rm i}\rho{+}{\rm
i}\omega{+}m'{+}1}{2}, \frac{{\rm i}\rho{-}{\rm
i}\omega{+}m'{+}1}{2}; m'{+}1; \tanh^2 a\!\right)\!, \nonumber
\end{gather}
where
\begin{gather}
B_{m'm}(\omega,\rho)=\frac{\sqrt{\pi}\Gamma \left({\rm
i}\omega+\frac{1}{2}\right) \Gamma\left(\frac{-{\rm i}\rho+{\rm
i}\omega+m'+1}{2} \right)\Gamma\left(\frac{-{\rm i}\rho-{\rm
i}\omega+m'+1}{2}\right)} {2\Gamma\left(-{\rm
i}\rho+\frac{3}{2}\right)\Gamma(m'+1) \Gamma\left({\rm
i}\omega+m+\frac{1}{2}\right)}
\nonumber\\
\phantom{B_{m'm}(\omega,\rho)=}{}
 \times
\int_{H^4_+}\frac{d^4x}{x_0}\psi(x)\left(\frac{2}{\cosh
a}\right)^{-{\rm i}\rho+3/2}{\tanh^{m'}}a\, P^m_{{\rm
i}\omega-1/2}(\cosh b)e^{-{\rm i}m\varphi}e^{-{\rm i}m'\Phi}
\nonumber\\
\label{I-191} \phantom{B_{m'm}(\omega,\rho)=}{}\times {}_2F_1
\left(\frac{-{\rm i}\rho-{\rm i}\omega+m'+1}{2}, \frac{{\rm
i}\omega-{\rm i}\rho+m'+1}{2}; \,m'+1;\,\tanh^2 a\right)   ,
\end{gather}
and
\[
d^4x/x_0=\sinh^3 a\, \sinh b\, da\, db\, d\varphi\, d\Phi.
\]

It follows from these formulas that the function $\psi(x)$ expands
in the basis \eqref{I-90} and the normed basis functions are of
the form
\begin{gather}
\Psi_{\tilde{\rho\omega mm}}(a,b,\varphi,\Phi)=\frac{1}
{(2\pi)^{3/2}}\frac{\left|\Gamma\left(\frac {{\rm i}\rho+{\rm
i}\omega+\tilde{m}+1}{2}\right)\Gamma\left(\frac {{\rm i}\rho-{\rm
i}\omega+\tilde{m}+1}{2}\right)\right|}
{\Gamma(\tilde{m}+1)|\Gamma({\rm i}\rho+3/2)|}
\frac{|\Gamma(i\omega+1/2)|}{|\Gamma({\rm i}\omega+m+1/2)|}
\nonumber\\
\phantom{\Psi_{\tilde{\rho\omega mm}}(a,b,\varphi,\Phi)=}{}\times
\tanh^{\tilde{m}}a(\cosh a)^{-{\rm i}\rho-3/2}e^{{\rm
i}(m\varphi+\tilde{m}\Phi)} P^m_{{\rm i}\omega-1/2}(\cosh b)
\nonumber\\
\label{I-192} \phantom{\Psi_{\tilde{\rho\omega
mm}}(a,b,\varphi,\Phi)=}{} \times {}_2F_1\left(\frac{{\rm
i}\rho+{\rm i}\omega+\tilde{m}+1}{2},\, \frac{{\rm i}\rho-{\rm
i}\omega+\tilde{m}+1}{2};\,\tilde{m}+1;\,\tanh^2 a\right).
\end{gather}
The normalization condition is
\begin{equation}\label{I-193}
\big\langle
\Psi_{\tilde{\rho'\omega'm'm'}},\Psi_{\tilde{\rho\omega
mm}}\big\rangle=\frac{\delta(\rho-\rho')}{\rho(\rho^2+1/4)\tanh\pi\rho}
\frac{\delta(\omega-\omega')}{\omega\tanh\pi\omega}\delta_{\tilde{m}\tilde{m'}}
\delta_{mm'}.
\end{equation}

The Plancherel formula for the expansion
\eqref{I-190}--\eqref{I-191} is of the form
\begin{gather*}
\int_{H^4_+}\frac{d^4x}{x_0}|\psi(x)|^2=\frac{1}{(2\pi)^4}
\sum_{m,m'=-\infty}^\infty \int^{\infty}_{0}d\rho \left(\rho^2{+}
{\textstyle{\frac{1}{4}}}\right)\rho\tanh \pi\rho\\
\phantom{\int_{H^4_+}\frac{d^4x}{x_0}|\psi(x)|^2=}{} \times
\int^{\infty}_{0}d\omega\, \omega\tanh\pi\omega\,
|B_{mm'}(\omega,\rho)|^2.
\end{gather*}

\subsection[Expansion in $S$-coordinate system]{Expansion in $\boldsymbol{S}$-coordinate system}
\label{expan-hyp-s}
In this case, we take the cut $\Gamma_{S}$ of $C^4_+$ by the plane
$k_0=1$ as a contour $\Gamma$. This contour is the
three-dimensional sphere $S^3$. We choose on $\Gamma_{S}$ the
spherical coordinates determined by the formulas
\begin{gather*}
k'_1=\sin\gamma\sin\Theta\cos \Phi,\qquad
k'_2=\sin\gamma\sin\Theta\sin \Phi,\qquad
k'_3=\sin\gamma\cos\Theta,\quad k'_4=\cos\gamma,
\\
0\leqslant\gamma,\Theta<\pi,\qquad 0\leqslant\Phi<\pi,\qquad
dk'=\sin^2\gamma\sin\Theta\, d\gamma\, d\Theta\, d\Phi.
\end{gather*}

This case is well-studied (see, for example, \cite{VKII}). We
formulate only the result. The expansion of the function $\psi(x)$
is of the form
\begin{gather}
\psi(x)=\frac{1}{(2\pi)^2}\sum^\infty_{j=0}\sum^j_{l=0}\sum^l_{m=-l}
\int^\infty_0d\rho\left(\rho^2+\frac{1}{4}\right)\rho\tanh\pi\rho\,
A_{jlm}(\rho) \frac{(-1)^j\Gamma(-{\rm i}\rho-\frac{1}{2})}
{\Gamma(-{\rm i}\rho-j-\frac{1}{2})}
\nonumber\\
\label{I-204}   \phantom{\psi(x)=}{} \times(\sinh
a)^{-1}P^{-j-1}_{{\rm i}\rho-\frac{1}{2}}(\cosh
a)Y_{jlm}(\beta,\theta,\varphi),
\end{gather}
where
\[
A_{jlm}(\rho)=\frac{(2\pi)^2(-1)^j\Gamma\left({\rm
i}\rho-\frac{1}{2}\right)} {\Gamma\left({\rm
i}\rho-j-\frac{1}{2}\right)}
\int_{H^4_+}\frac{d^4x}{x_0}\psi(x)(\sinh a)^{-1}P^{-j-1}_{{\rm
i}\rho-1/2}(\cosh a)\overline{Y_{jlm}(\beta,\theta,\varphi)}.
\]
Here $Y_{jlm}(\beta,\theta,\varphi)$ is the spherical function on
$S^3$,
\[
Y_{jlm}(\beta,\theta,\varphi)=\left[ \frac{(j+1)\Gamma(j+l+2)}{
\Gamma(j-l+1)}\right]^{1/2} (\sin \beta)^{-1/2}
P^{-l-1/2}_{j+1/2}(\cos \beta) Y_{lm}(\theta,\varphi)
\]
and
\[
d^4x/x_0=\sinh^3a\, \sin^2\beta\, \sin \theta\, da\, d\beta\,
d\theta\, d\varphi.
\]

Therefore, the functions
\begin{equation}\label{I-206}
\Phi_{\rho jlm}(\alpha,\beta,\theta,\varphi)=\frac{\mid\Gamma({\rm
i}\rho-\frac{1}{2}) \mid} {\mid\Gamma({\rm
i}\rho-j-\frac{1}{2})\mid} \sinh ^{-1}a\, P^{-j-1}_{{\rm
i}\rho-1/2}(\cosh a)Y_{jlm}(\beta,\theta,\varphi)
\end{equation}
constitute a normed basis in the space $L^2(H^4_+)$ and
normalization condition is
\[
\langle\Phi_{\rho' j'l'm'},\Phi_{\rho
jlm}\rangle=\frac{\delta(\rho-\rho')}{\rho(\rho^2+1/4)\tanh
\pi\rho}\delta_{jj'}\delta_{ll'}\delta_{mm'}.
\]
The Plancherel formula is of the form
\[
\int_{H^4_+}\frac{d^4x}{x_0}\mid\psi(x)\mid^2=(2\pi)^{-4}
\sum^\infty_{j=0}\sum^j_{l=0}\sum^l_{m=-l}
\int^\infty_0d\rho\left(\rho^2+1/4\right)\rho\tanh\pi\rho
|A_{jlm}(\rho)|^2 .
\]

\section[Coordinate systems and generators
of $SO_0(1,4)$ on the con]{Coordinate systems and generators of
$\boldsymbol{SO_0(1,4)}$ on the cone} \label{Coord-sys-C^4}

In this section we consider coordinate systems on the upper sheet
$C^4_+$ of the cone $C^4$ in the 4-dimensional Minkowski space,
which is determined by the equations
\[
[x,x]=x_0^2-x_1^2-x_2^2-x_3^2-x_4^2=0,\qquad x_0>0.
\]
The group $SO_0(1,4)$ is a transitive group of transformations of
$C^4_+$. We take the point $x^0=(1,0,0,0,1)$ of $C^4_+$. A maximal
subgroup of $SO_0(1,4)$, whose elements leave this point
invariant, is the subgroup $SO(3)\times T(3)$, where $T(3)$ is
generated by $E_i=P_i+N_i$, $i=1,2,3$ (see \eqref{I-21}). Thus,
the cone $C^4_+$ is homeomorphic to the coset space
$SO_0(1,4)/(SO(3)\times T(3))$.

As in the case of the hyperboloid $H^4_+$, coordinate systems will
be determined by expressions of homogeneous coordinates $x_\mu$,
$\mu=0,1,2,3,4$, in terms of corresponding angles of a coordinate
system under consideration. We restrict ourselves by 7 coordinate
systems. They correspond to the same subgroup chains of
$SO_0(1,4)$ as in the case of the hyperboloid. Coordinate systems
on $C^4_+$ will be denoted by the same symbols as for the
hyperboloid $H^4_+$.

{\bf Spherical coordinate system $S$} (coordinates $a$, $\beta$,
$\theta$, $\varphi$):
\begin{gather}
x_0=e^a/2,\qquad x_1=(e^a/2)\sin\beta\sin\theta\cos\varphi,\qquad
x_2=(e^a/2)\sin\beta\sin\theta\sin\varphi,
\nonumber\\
\label{I-208} x_3=(e^a/2)\sin\beta\cos\theta,\quad
x_4=(e^a/2)\cos\beta,
\\
-\infty<a<\infty,\qquad  0\leqslant \beta, \theta<\pi,\qquad
0\leqslant \varphi<2\pi  .\nonumber
\end{gather}

{\bf Hyperbolic coordinate system $H$} (coordinates $a$, $b$,
$\theta$, $\varphi$):
\begin{gather}
x_0=(e^a/2)\cosh b,\qquad x_1=(e^a/2)\sinh
b\sin\theta\cos\varphi,\qquad x_2=(e^a/2)\sinh
b\sin\theta\sin\varphi,
\nonumber\\
\label{I-209}
 x_3=(e^a/2)\sinh b\cos\theta,\qquad
x_4=\varepsilon (e^a/2),
\\
 {-}\infty<a<\infty,\qquad 0\leqslant
b<\infty,\qquad 0\leqslant \theta<\pi,\qquad 0\leqslant
\varphi<2\pi,\qquad \varepsilon={\rm sign}\;  x_4=\pm.\nonumber
\end{gather}

{\bf Orispherical coordinate system $O$} (coordinates $a$, $r$,
$\theta$, $\varphi$):
\begin{gather}
x_0=e^a(r^2+1)/2,\qquad x_1=e^a r\sin\theta\cos\varphi,\qquad
x_2=e^a r\sin\theta\sin\varphi,
\nonumber\\
\label{I-210}
 x_3=e^a r\cos\theta,\qquad
x_4=e^a(r^2{-}1)/2,
\\
 -\infty<a<\infty,\qquad 0\leqslant
r<\infty,\qquad 0\leqslant \theta<\pi,\qquad 0\leqslant
\varphi<2\pi.\nonumber
\end{gather}

{\bf Cylindric coordinate system $C$} (coordinates $a$, $b$,
$\theta$, $\varphi$):
\begin{gather}
x_0=(e^a/2)\cosh b,\qquad  x_1=(e^a/2)\sin\theta\cos\varphi,\qquad
x_2=(e^a/2)\sin\theta\sin\varphi,
\nonumber\\
\label{I-211}
 x_3=(e^a/2)\cos\theta,\qquad
x_4=(e^a/2)\sinh b,
\\
-\infty<a<\infty,\qquad -\infty<b<\infty,\qquad
0\leqslant\theta<\pi,\qquad 0\leqslant \varphi<2\pi.\nonumber
\end{gather}

{\bf Spherically-hyperbolic coordinate system $SH$} (coordinates
$a$, $b$, $\Phi$, $\varphi$):
\begin{gather}
x_0=(e^a/2)\cosh b,\qquad x_1=(e^a/2)\sinh b\cos\varphi, \qquad
x_2=(e^a/2)\sinh b\sin\varphi,
\nonumber\\
\label{I-212}
 x_3=(e^a/2)\cos\Phi,\qquad
x_4=(e^a/2)\sin\Phi,
\\
 -\infty<a<\infty,\qquad 0\leqslant
b<\infty,\qquad 0\leqslant\varphi,\Phi<2\pi.\nonumber
\end{gather}

{\bf Orispherically-cylindric coordinate system $OC$} (coordinates
$a$, $\xi$, $z$, $\varphi$):
\begin{gather}
x_0=e^a(\xi^2{+}z^2{+}1)/2,\qquad x_1=e^a\xi\cos\varphi,\qquad
x_2=e^a\xi\sin\varphi,
\nonumber\\
\label{I-213}
 x_3=e^a\xi,\qquad x_4=e^a(\xi^2{+}z^2{-}1)/2,
\\
  {-}\infty<a<\infty,\qquad
0\leqslant \xi<\infty,\qquad {-}\infty<z<\infty,\qquad 0\leqslant
\varphi<2\pi.\nonumber
\end{gather}

{\bf Orispherically-translational coordinate system $OT$}
(coordinates $a$, $y_1$, $y_2$, $y_3$):
\begin{gather}
x_0=e^a(y^2+1)/2,\qquad x_1=e^ay_1,\qquad x_2=e^ay_2,\qquad
x_3=e^ay_3,
\nonumber\\
\label{I-214} x_4=e^a(y^2-1)/2,\qquad y^2=y_1^2+y_2^2+y_3^2,
\\
-\infty<a<\infty,\qquad -\infty<y_i<\infty,\qquad
i=1,2,3.\nonumber
\end{gather}

Since the cone $C^4_+$ is an asymptotical surface for the
hyperboloid $H^4_+$, the coordinate systems on~$C^4_+$ are
asymptotically (at $a\to \infty$) obtained from the corresponding
coordinate systems on~$H^4_+$.

We can write down a dif\/ferential form of inf\/initesimal
operators $J_{\mu\nu}$, $\mu,\nu=0,1,2,3,4$, $\mu<\nu$, on the
cone $C^4_+$. For this we use the formulas
\[
J_{rs}=-{\rm i}\left( x_r\frac{\partial}{\partial
x_s}{-}x_s\frac{\partial}{\partial x_r} \right),\qquad
J_{0s}=-{\rm i}\left( x_0\frac{\partial}{\partial
x_s}{+}x_s\frac{\partial}{\partial x_0} \right),
\]
where $r,s=1,2,3,4$. We substitute into these formulas the
expressions for $x_\mu$, $\mu=0,1,2,3,4$, in terms of coordinates
of the corresponding system. The results of such calculations can
be found in \cite{KK86}, section 2.6. As an example, we give here
a dif\/ferential form for generators of $SO_0(1,4)$ in the
hyperbolic coordinate system.

{\bf $H$-system}:
\begin{gather*}
M_1={-}{\rm i}\left( {-}\sin\varphi\frac{\partial}{\partial
\theta}{-}\cot\theta\cos\varphi\frac{\partial}{\partial \varphi}
\right),\qquad M_2={-}{\rm i}\left(
\cos\varphi\frac{\partial}{\partial
\theta}{-}\cot\theta\sin\varphi\frac{\partial}{\partial \varphi}
\right),
\\
 M_3={-}{\rm i}\frac{\partial}{\partial \varphi},\qquad
P_0={-}{\rm i}\varepsilon\left( \cosh b\frac{\partial}{\partial
a}{-}\sinh b\frac{\partial}{\partial b} \right),
\\
P_1={-}{\rm i}\varepsilon\left( \sinh
b\sin\theta\cos\varphi\frac{\partial}{\partial a}{-}\cosh
b\sin\theta\cos\varphi\frac{\partial}{\partial
b}{-}\frac{\cos\theta\cos\varphi}{\sinh b}\frac{\partial}{\partial
\theta}{+}\frac{\sin\varphi}{\sinh
b\sin\theta}\frac{\partial}{\partial \varphi} \right),
\\
P_2={-}{\rm i}\varepsilon\left( \sinh
b\sin\theta\sin\varphi\frac{\partial}{\partial a}{-}\cosh
b\sin\theta\sin\varphi\frac{\partial}{\partial
b}{-}\frac{\cos\theta\sin\varphi}{\sinh b}\frac{\partial}{\partial
\theta}{-}\frac{\cos\varphi}{\sinh
b\sin\theta}\frac{\partial}{\partial \varphi} \right),
\\
P_3={-}{\rm i}\varepsilon\left( \sinh
b\cos\theta\frac{\partial}{\partial a}{-}\cosh
b\cos\theta\frac{\partial}{\partial b}{+}\frac{\sin\theta}{\sinh
b}\frac{\partial}{\partial \theta} \right),
\\
 N_1={-}{\rm i}\left(
\sin\theta\cos\varphi\frac{\partial}{\partial b}{+}\coth
b\cos\theta\cos\varphi\frac{\partial}{\partial \theta}{-}\coth
b\frac{\sin\varphi}{\sin\theta}\frac{\partial}{\partial \varphi}
\right),
\\
 N_2={-}{\rm i}\left(
\sin\theta\sin\varphi\frac{\partial}{\partial b}{+}\coth
b\cos\theta\sin\varphi\frac{\partial}{\partial \theta}{+}\coth
b\frac{\cos\varphi}{\sin\theta}\frac{\partial}{\partial \varphi}
\right),
\\
 N_3={-}{\rm i}\left( \cos\theta\frac{\partial}{\partial
b}{-}\coth b\sin\theta\frac{\partial}{\partial\theta} \right).
\end{gather*}

\section{Invariant operators and their eigenfunctions on the cone}
\label{Invar-oper-cone}
Our aim in this section is to f\/ind basis functions on the cone
$C^4_+$ corresponding to dif\/ferent coordinate systems. As in the
case of the hyperboloid $H^4_+$, these functions are constructed
as eigenfunctions of a full system of self-adjoint dif\/ferential
operators on the cone $C^4_+$, including the Casimir operator $F$
(see formula \eqref{I-7}) of $SO_0(1,4)$ and Casimir operators of
the corresponding chain of subgroups.

In all coordinate systems on the cone $C^4_+$ the Casimir operator
$F$ is of the form
\begin{equation}\label{I-215}
F=-\frac{\partial^2}{\partial a^2}-3\frac{\partial}{\partial a}.
\end{equation}
This expression can be obtained from the dif\/ferential form of
the operator $F$ in any coordinate system on $H^4_+$ by the
asymptotic limit $a\to \infty$.

Let us give the rest of self-adjoint operators of a full system in
each coordinate system on~$C^4_+$.

{\bf $S$-system.}
\begin{gather*}
{\bf M}^2+{\bf
P}^2=-\frac{1}{\sin^2\beta}\left(\frac{\partial}{\partial\beta}
\sin^2\beta\frac{\partial}{\partial\beta}+\frac{1}{\sin\theta}
\frac{\partial}{\partial\theta}\sin\theta\frac{\partial}{\partial\theta}
+\frac{1}{\sin^2\theta}\frac{\partial^2}{\partial\varphi^2}\right),
\\
{\bf
M}^2=-\left(\frac{1}{\sin\theta}\frac{\partial}{\partial\theta}\sin\theta
\frac{\partial}{\partial\theta}+\frac{1}{\sin^2\theta}
\frac{\partial^2}{\partial\varphi^2}\right),\qquad
M_3=-i\frac{\partial}{\partial\varphi}.
\end{gather*}
These operators correspond to the chain of subgroups
\[
SO_0(1,4)\supset SO(4)\supset SO(3)\supset SO(2).
\]

{\bf $H$-system.}
\begin{gather*}
{\bf N}^2-{\bf
M}^2=-\frac{1}{\sinh^2\beta}\left(\frac{\partial}{\partial b}
\sinh^2 b\frac{\partial}{\partial b}+\frac{1}{\sin\theta}
\frac{\partial}{\partial\theta}\sin\theta\frac{\partial}{\partial\theta}
+\frac{1}{\sin^2\theta}\frac{\partial^2}{\partial\varphi^2}\right),
\\
{\bf
M}^2=-\left(\frac{1}{\sin\theta}\frac{\partial}{\partial\theta}\sin\theta
\frac{\partial}{\partial\theta}+\frac{1}{\sin^2\theta}
\frac{\partial^2}{\partial\varphi^2}\right),\qquad
M_3=-i\frac{\partial}{\partial\varphi}.
\end{gather*}
These operators correspond to the chain of subgroups
\[
SO_0(1,4)\supset SO_0(1,3)\supset SO(3)\supset SO(2).
\]

{\bf $O$-system.}
\begin{gather*}
{\bf E}^2=-\left[\frac{\partial^2}{\partial
r^2}+\frac{2}{r}\frac{\partial}{\partial
r}+\frac{1}{r^2}\left(\frac{\partial^2}{\partial \theta^2}+\cot
\theta\frac{\partial}{\partial\theta}+\frac{1}{\sin^2\theta}
\frac{\partial^2}{\partial\varphi^2}\right)\right],
\\
{\bf
M}^2=-\left(\frac{1}{\sin\theta}\frac{\partial}{\partial\theta}
\sin\theta\frac{\partial}{\partial \theta}+\frac{1}{\sin^2\theta}
\frac{\partial^2}{\partial^2\varphi}\right),\qquad
M_3=-i\frac{\partial}{\partial\varphi}.
\end{gather*}
These operators correspond to the chain of subgroups
\[
SO_0(1,4)\supset ISO(3)\supset SO(3)\supset SO(2).
\]

{\bf $OC$-system.}
\begin{gather*}
{\bf E}^2=-\left(\frac{\partial^2}{\partial
\xi^2}+\frac{1}{\xi}\frac{\partial}{\partial
\xi}+\frac{1}{\xi^2}\frac{\partial^2}{\partial \varphi^2}
+\frac{\partial^2}{\partial z^2}\right),\qquad
E_3=-i\frac{\partial}{\partial z},
\\
M_3=-i\frac{\partial}{\partial\varphi},\qquad \tilde{{\bf
E}^2}\equiv E^2_1+E^2_2=-\left(\frac{\partial^2}{\partial
\xi^2}+\frac{1}{\xi}\frac{\partial}{\partial
\xi}+\frac{1}{\xi^2}\frac{\partial^2}{\partial \varphi^2}\right).
\end{gather*}
These operators correspond to the chain of subgroups
\[
SO_0(1,4)\supset ISO(3)\supset ISO(2)\otimes T_\bot\supset
SO(2)\otimes T_\bot.
\]

{\bf $OT$-system.}
\[
{\bf E}^2=-\left(\frac{\partial^2}{\partial
y^2_1}+\frac{\partial^2}{\partial
y^2_2}+\frac{\partial^2}{\partial y^2_3}\right),\qquad
E_j=-i\frac{\partial}{\partial y_j},\qquad j=1,2,3.
\]
The corresponding chain of subgroups is
\[
SO_0(1,4)\supset ISO(3)\supset T(1)\otimes T(2)\otimes T(3).
\]

{\bf $C$-system.}
\[
{\bf
M}^2=-\left(\frac{1}{\sin\theta}\frac{\partial}{\partial\theta}
\sin\theta\frac{\partial}{\partial \theta}+\frac{1}{\sin^2\theta}
\frac{\partial^2}{\partial^2\varphi}\right),\qquad
M_3=-i\frac{\partial}{\partial\varphi},\qquad
P_0=-i\frac{\partial}{\partial b}.
\]
These operators correspond to the chain of subgroups
\[
SO_0(1,4)\supset SO_0(1,1)\otimes SO(3)\supset SO(2).
\]

{\bf $SH$-system.}
\begin{gather*}
N^2_1+N^2_2-M^2_3=-\left(\frac{\partial^2}{\partial b^2} +\coth b
\frac{\partial}{\partial b}+\frac{1}{\sinh^2 b}
\frac{\partial^2}{\partial^2\varphi}\right),
\\
M_3=-i\frac{\partial}{\partial\varphi},\qquad
P_3=-i\frac{\partial}{\partial\Phi}.
\end{gather*}
The corresponding chain of subgroups is{\samepage
\[
SO_0(1,4)\supset SO_0(1,2)\otimes SO'(2)\supset SO(2),\qquad
SO(2)\sim SO'(2),
\]
(the subgroup $SO(2)$ is embedded into $SO_0(1,2)$).}

As we see from the expressions for invariant operators in each
coordinate system  given above their dif\/ferential form (except
for the operator $F$) coincides with the corresponding
dif\/ferential form in the corresponding coordinate system on the
hyperboloid $H^4_+$.

It is known (see~\cite{VKII}, Chapter 10) that the quasi-regular
representation of $SO_0(1,4)$ on the Hilbert space $L^2(C^4_+)$
decomposes into a direct integral of irreducible unitary
representations of this group, and each of these irreducible
representations is contained in the decomposition twice. (Let us
recall that the quasi-regular representation of $SO_0(1,4)$ on the
Hilbert space $L^2(H^4_+)$ decomposes into a direct integral of
irreducible unitary representations of this group, and each of
them is contained in the decomposition once.) Therefore, the
spectrum of each full collection of self-adjoint operators in
$L^2(C^4_+)$ given above dif\/fers from the corresponding spectrum
on~$L^2(H^4_+)$ only by the fact that multiplicity of each
eigenvalue is doubled.

As in the case of functions on the hyperboloid, we try to f\/ind
eigenfunctions of the collections of self-adjoint dif\/ferential
operators on $C^4_+$ in the form of separated variables. As we
have said, dif\/ferential forms of invariant operators (except for
the operator $F$) for $H^4_+$ and for $C^4_+$ coincide in the same
type of coordinate systems. For this reason, the corresponding
solutions for them are the same. Thus, in order to f\/ind
eigenfunctions of full collections of self-adjoint operators we
have to solve the equation
\[
\left(\frac{\partial^2}{\partial a^2}+3\frac{\partial}{\partial
a}\right)\langle a\mid\sigma \rangle'=\sigma(\sigma+3) \langle
a\mid\sigma\rangle',
\]
corresponding to the operator $F$. For each f\/ixed complex value
of $\sigma$, there are two corresponding linearly independent
solutions of this equation:
\[
\langle a\mid\sigma\rangle'_1=\exp\sigma a,\qquad \langle
a\mid\sigma\rangle'_2=\exp(-\sigma-3)a.
\]
In particular, for values of $\sigma$, corresponding to the
principal unitary series representations of $SO_0(1,4)$ (that is,
for $\sigma={\rm i}\rho-\frac{3}{2}$, $\rho\in \mathbb{R}$), we
have the solutions
\begin{equation}\label{I-216}
\langle a\mid\rho \rangle\equiv\left<a
\left|-\tfrac{3}{2}+i\rho\right.\right>' =
\exp\left(-\tfrac{3}{2}\pm i\rho\right)a.
\end{equation}

Thus, a full system of eigenfunctions of a collection of
self-adjoint operators, corresponding to a f\/ixed coordinate
system, is given by the formula
\begin{equation}\label{I-217}
\Phi_{\rho\gamma}(a,\alpha)\equiv \langle
a,\alpha\mid\rho,\gamma\rangle= \langle
a\mid\rho\rangle\langle\alpha\mid\gamma\rangle,
\end{equation}
where $\alpha$ is a collection of all coordinates except for the
coordinate $a$, $\gamma$ are eigenvalues of all operators except
for the operator $F$, and $\langle\alpha\mid\gamma\rangle$ are
eigenfunctions of these operators.

It is evident that the function \eqref{I-216} is homogeneous in
$e^a$ of homogeneity degree $(-3/2\pm {\rm i}\rho)$. Therefore,
the basis functions \eqref{I-217} are homogeneous in $e^a$.

The orispherical transform \eqref{I-97} maps $L^2(H^4_+)$ not upon
the whole space $L^2(C^4_+)$, but rather upon its subspace (we
denote it by $L_0^2(C^4_+)$). Indeed, if this transform would map
$L^2(H^4_+)$ upon the whole space $L^2(C^4_+)$, then
multiplicities of irreducible representations of $SO_0(1,4)$ in
$L^2(C^4_+)$ would exceed those in $L^2(H^4_+)$; this cannot be
true.

Basis functions of the space $L^2(C^4_+)$, corresponding to a
f\/ixed coordinate system, can be found in dif\/ferent ways. One
of these ways is making orispherical transform of the
corresponding basis functions of $L^2(H^4_+)$. However, for most
coordinate systems this transform leads to divergent integrals
(let us remind that our basis functions are normed to the
delta-function). For this reason, we choose another way. As we
have said, in the homogeneous coordinates on $H^4_+$ and on
$C^4_+$ coordinates and dif\/ferential operators of the operator
$F$ and other operators coincide in the asymptotics $a\to \infty$.
The corresponding basis functions on $H^4_+$ and on $C^4_+$ also
coincide in this asymptotics. The part $\Phi_\rho(a)$ of a basis
function of $L_0^2(C^4_+)$, which depends on $a$, is a linear
combination of the functions $\exp\sigma a$ and
$\exp(-\sigma-3)a$, $\sigma=i\rho-3/2$:
\begin{equation}\label{I-218}
\Phi_\rho(a)=C_1\exp\sigma a+C_2\exp(-\sigma-3) a,
\end{equation}
where $C_1$ and $C_2$ are independent of $a$. The coef\/f\/icients
$C_1$ and $C_2$ can be found using the fact that the function
$\Phi_\rho(a)$ and the corresponding basis function on the
hyperboloid $H^4_+$ coincide in the asymptotics $a\to \infty$. As
an example, let us consider the case of the spherical coordinate
system.

We take the function $(\sinh a)^{-1}P^{-j-1}_{{\rm
i}\rho-1/2}(\cosh a)$ of $a$, entering into the expression for the
basis function $\Phi_{\rho jlm}(a,\beta,\theta,\varphi)$ of the
space $L^2(H^4_+)$ (see formula \eqref{I-206}). Expressing the
function $P^{-j-1}_{{\rm i}\rho-1/2}(\cosh a)$ in terms of the
Gauss hypergeometric function and using the formula (15.3.6)
in~\cite{Hand}, we f\/ind that for large $a$ the following
asymptotics holds:
\[
(\sinh a)^{-1}P^{-j-1}_{{\rm i}\rho-1/2}(\cosh
a)\simeq\frac{2}{\sqrt{\pi}}\left[\frac{\Gamma({\rm i}\rho)
\exp(-3/2+{\rm i}\rho)a} {\Gamma({\rm
i}\rho+j+3/2)}+\frac{\Gamma(-{\rm i}\rho) \exp(-3/2-{\rm i}\rho)a}
{\Gamma(-{\rm i}\rho+j+3/2)}\right].
\]
Therefore, for large $a$ the function $\Phi_{\rho
jlm}(a,\beta,\theta,\varphi)$ has the form
\begin{gather}
\Phi_{\rho
jlm}(a,\beta,\theta,\varphi)\simeq\frac{2\mid\Gamma({\rm
i}\rho-1/2)\mid} {\sqrt{\pi}\mid\Gamma({\rm
i}\rho-j-1/2)\mid}Y_{jlm}(\beta,\theta,\varphi)
\nonumber\\
\label{I-219}    \phantom{\Phi_{\rho
jlm}(a,\beta,\theta,\varphi)\simeq}{}
\times\left[\frac{\Gamma({\rm i}\rho)\exp(-3/2+{\rm i}\rho)a}
{\Gamma({\rm i}\rho+j+3/2)} +\frac{\Gamma(-i\rho)\exp(-3/2-{\rm
i}\rho)a} {\Gamma(-{\rm i}\rho+j+3/2)}\right].
\end{gather}
Then
\[
C_{1,2}\equiv C_{\pm}=\frac{2}{\sqrt{\pi}}\frac{\left|
\Gamma\left({\rm i}\rho-\frac{1}{2}\right)\right|}
{\left|\Gamma\left({\rm
i}\rho-j-\frac{1}{2}\right)\right|}\frac{\Gamma(\pm {\rm i}\rho)}
{\Gamma\left(\pm {\rm i}\rho+j+\frac{3}{2}\right)}.
\]
Thus, in the $S$-coordinate system the basis functions of the
space $L_0^2(C^4_+)$, corresponding to the basis functions on
$L^2(H^4_+)$, are of the form
\begin{gather*}
\Phi'_{\rho jlm}(a,\beta,\theta,\varphi)=\frac{2}{\sqrt{\pi}}
\frac{\mid\Gamma({\rm i}\rho-1/2)\mid} {\mid\Gamma({\rm
i}\rho-j-1/2)\mid}Y_{jlm}(\beta,\theta,\varphi)
\\
\phantom{\Phi'_{\rho jlm}(a,\beta,\theta,\varphi)=}{}
\times\left[\frac{\Gamma({\rm i}\rho)} {\Gamma({\rm
i}\rho+j+3/2)}e^{(-3/2+{\rm i}\rho)a}+\frac{\Gamma(-{\rm i}\rho)}
{\Gamma(-{\rm i}\rho+j+3/2)}e^{(-3/2-{\rm i}\rho)a}\right].
\end{gather*}
The domains of def\/inition for the indices $\rho$, $j$, $l$, $m$
are the same as in the case of the space $L^2(H^4_+)$. This remark
is true for other coordinate systems considered below.

An explicit form of basis functions in other coordinate systems
can be found analogously. Let us write down basis functions on the
cone $C^4_+$ in all the coordinate systems, normed by a~delta-function.
\medskip

{\bf $S$-system.}
\begin{equation}\label{I-220}
\Phi_{\rho jlm}(a,\beta,\theta,\varphi)\equiv \langle
a,\beta,\theta,\varphi\mid\rho,j,l,m\rangle =(C^+_Se^{({\rm
i}\rho-3/2)a}+ C^-_S e^{(-{\rm
i}\rho-3/2)a})Y_{jlm}(\beta,\theta,\varphi),
\end{equation}
where
\[
C^\pm_S=\frac{\sqrt{2}\, |\Gamma({\rm i}\rho+j+3/2)|\,\Gamma(\pm
{\rm i}\rho)}{\sqrt{\pi}\, |\Gamma({\rm i}\rho+3/2)|\,\Gamma(\pm
{\rm i}\rho+j+3/2)}.
\]
The orthogonality relation has the form{\samepage
\[
\langle\Phi_{\rho'j'l'm'},\Phi_{\rho
jlm}\rangle=\int_{C^4_+}\frac{d^4x}{x_0}\,
\Phi_{\rho'j'l'm'}(a,\beta,\theta,\varphi)\overline{\Phi_{\rho
jlm} (a,\beta,\theta,\varphi)}
=\frac{\delta(\rho-\rho')\delta_{jj'}\delta_{ll'}\delta_{mm'}}
{\rho(\rho^2+1/4)\tanh\pi\rho},
\]
where $d^4x/x_0=\frac18 e^{3a}\sin^2\beta\sin\theta\, da\,
d\beta\, d\theta\, d\varphi$.}

{\bf $H$-system.}
\begin{gather}
\Phi^\pm_{\rho \nu lm}(a,\beta,\theta,\varphi)\equiv\langle
a,\beta,\theta,\varphi\mid\rho, \nu,l,m\rangle
\nonumber\\
\label{I-221}   \phantom{\Phi^\pm_{\rho \nu
lm}(a,\beta,\theta,\varphi)}{} =(C^+_{H_\pm}e^{(-3/2+{\rm
i}\rho)a}+ C^-_{H_\pm}e^{(-3/2-{\rm i}\rho)a})V_{\nu
lm}(\beta,\theta,\varphi),
\end{gather}
where
\begin{gather*}
V_{\nu lm}(\beta,\theta,\varphi)=\frac{\mid\Gamma({\rm i}\nu)\mid}
{\mid\Gamma({\rm i}\nu-l)\mid}(\sinh b)^{-1/2}P^{-l-1/2}_{{\rm
i}\nu-1/2}(\cosh b)Y_{lm}(\theta,\varphi),
\\
C^+_{H_\pm}=\frac{\sqrt{2}\mid\Gamma({\rm i}\rho+{\rm i}\nu+1/2)
\Gamma({\rm i}\rho-{\rm i}\nu+1/2) \mid\Gamma({\rm
i}\rho)}{\sqrt{\pi}\mid\Gamma({\rm i}\rho+3/2)\mid \Gamma({\rm
i}\rho+{\rm i}\nu+1/2)\Gamma({\rm i}\rho-{\rm i}\nu+1/2)},
\\
C^-_{H_\pm}=\frac{\sqrt{2}\mid\Gamma({\rm i}\rho+{\rm i}\nu+1/2)
\Gamma({\rm i}\rho-{\rm i}\nu+1/2) \mid}{\sqrt{\pi}\mid\Gamma({\rm
i}\rho+3/2)\mid} \left(\frac{\cosh\pi\rho}{\pi}\Gamma(-{\rm
i}\rho)\pm\frac{1} {\Gamma({\rm i}\rho+1)} \right).
\end{gather*}
The orthogonality relation for the functions \eqref{I-221} is
\begin{gather*}
\langle\Phi^{\pm}_{\rho'\nu'l'm'},\Phi^{\pm}_{\rho\nu lm}\rangle
=\int_{C^4_+}\frac{d^4x}{x_0}
\Phi^{\pm}_{\rho'\nu'l'm'}(a,\beta,\theta,
\varphi)\overline{\Phi^{\pm}_{\rho\nu lm}(a,\beta,\theta,
\varphi)}
\\
\phantom{\langle\Phi^{\pm}_{\rho'\nu'l'm'},\Phi^{\pm}_{\rho\nu
lm}\rangle}{}
=\frac{\delta(\rho-\rho')}{\rho(\rho^2+1/4)\tanh\pi\rho}
\frac{\delta(\nu-\nu')}{\nu^2}\delta_{ll'}\delta_{mm'},
\\
\langle\Phi^{\pm}_{\rho'\nu'l'm'},\Phi^{\mp}_{\rho\nu lm}\rangle
=0,
\end{gather*}
where $d^4x/x_0=(e^{3a}/8)\sinh^2b\sin\theta\, da\, db\, d\theta\,
d\varphi$.
\medskip

{\bf $O$-system.}
\begin{gather}\label{I-223}
\Phi^{\rho \kappa}_{lm}(a,r,\theta,\varphi)\equiv \langle
a,r,\theta,\varphi\mid\rho,\kappa,l,m\rangle =(C^+_O e^{(-3/2+{\rm
i}\rho)a}+ C^-_O e^{(-3/2-{\rm i}\rho)a})J_{\kappa
lm}(r,\theta,\varphi),\!\!
\end{gather}
where
\[
J_{\kappa lm}(r,\theta,\varphi)=(\kappa r)^{-1/2}J_{l+1/2}(\kappa
r) Y_{lm}(\theta,\varphi), \qquad C^{\pm}_O=\frac{(\kappa/2)^{\mp
{\rm i}\rho}\Gamma(\pm {\rm i}\rho)}{2\sqrt{\pi}\mid \Gamma({\rm
i}\rho-3/2)\mid}.
\]
The orthogonality relation for the functions \eqref{I-223} is
\begin{gather*}
\langle\Phi^{\rho'\kappa'}_{l'm'},\Phi^{\rho \kappa}_{lm}\rangle
=\int_{C^4_+}\frac{d^4x}{x_0}\Phi^{\rho'\kappa'}_{l'm'}(a,r,\theta,
\varphi)\overline{\Phi^{\rho \kappa}_{lm}(a,r,\theta, \varphi)}
\\
\phantom{\langle\Phi^{\rho'\kappa'}_{l'm'},\Phi^{\rho
\kappa}_{lm}\rangle}{}
=\frac{\delta(\rho-\rho')}{\rho(\rho^2+1/4)\tanh\pi\rho}
\frac{\delta(\kappa-\kappa')}{\kappa^2}\delta_{ll'}\delta_{mm'},
\end{gather*}
where $d^4x/x_0=e^{3a}r^2\sin\theta\, da\, dr\, d\theta\,
d\varphi$.

{\bf $OC$-system.}
\begin{gather}\label{I-224}
\Phi^{m}_{\rho\eta q }(a,\xi,z,\varphi){\equiv} \langle
a,\xi,z,\varphi\mid\rho,\eta,q,m\rangle {=}(C^+_{OC}e^{(-3/2+{\rm
i}\rho)a}+C^-_{OC}e^{(-3/2-{\rm i}\rho)a})\Psi^m_{\eta q
}(\xi,z,\varphi),\!\!
\end{gather}
where $q^2+\eta^2=\kappa^2$ and
\[
\Psi^m_{\eta q
}(\xi,z,\varphi)=\frac{1}{2\pi}J_m(\eta\xi)e^{iqz}e^{im\varphi},
\qquad C^{\pm}_{OC}=C^{\pm}_{O}= \frac{(\kappa/2)^{\mp {\rm
i}\rho}\Gamma(\pm {\rm i}\rho)}{2\sqrt{\pi}|\Gamma({\rm
i}\rho+3/2)|}.
\]
The orthogonality relation for the functions \eqref{I-224} is
\begin{gather*}
\langle\Phi^{m'}_{\rho'\eta'q'},\Phi^{m}_{\rho\eta q}\rangle
=\int_{C^4_+}\frac{d^4x}{x_0}\Phi^{m'}_{\rho'\eta'q'}(a,\xi,z,
\varphi)\overline{\Phi^{m}_{\rho\eta q}(a,\xi,z, \varphi)}
\\
\phantom{\langle\Phi^{m'}_{\rho'\eta'q'},\Phi^{m}_{\rho\eta
q}\rangle}{}
=\frac{\delta(\rho-\rho')}{\rho(\rho^2+1/4)\tanh\pi\rho}
\frac{\delta(\eta-\eta')}{\eta}\delta(q-q')\delta_{mm'},
\end{gather*}
where $d^4x/x_0=e^{3a}\xi\, da\, d\xi\, dz\, d\varphi$.
\medskip

{\bf $OT$-system.}
\begin{equation}\label{I-225}
\Phi^{\rho \boldsymbol{\kappa}}(a,{\bf y}){\equiv} \langle a,{\bf
y},\mid\rho, \boldsymbol{\kappa}\rangle {=}(C^+_{OT}e^{(-3/2+{\rm
i}\rho)a}+C^-_{OT}e^{(-3/2-{\rm i}\rho)a})
\Psi_{\boldsymbol{\kappa}}({\bf y}),
\end{equation}
where
\begin{gather*}
\Psi_{\boldsymbol{\kappa}}({\bf y})=(2\pi)^{-3/2}\exp({\rm
i}\boldsymbol{\kappa} {\bf y}),
\\
C^{\pm}_{OT}=C^{\pm}_{O}=\frac{({\kappa}/2)^{\mp {\rm
i}\rho}\Gamma(\pm i\rho)}{2\sqrt{\pi}\mid \Gamma({\rm
i}\rho+3/2)\mid},\qquad \kappa=|\boldsymbol{\kappa}|.
\end{gather*}
The orthogonality relation for the functions \eqref{I-225} is
\[
\langle\Phi^{\rho'\boldsymbol{\kappa}'},\Phi^{\rho
\boldsymbol{\kappa}}\rangle
=\int_{C^4_+}\frac{d^4x}{x_0}\Phi^{\rho'\boldsymbol{\kappa}'}(a,{\bf
y}) \overline{\Phi^{\rho \boldsymbol{\kappa}}(a,{\bf y})}=
\frac{\delta(\rho-\rho')\delta(\boldsymbol{\kappa}-
\boldsymbol{\kappa}')}{\rho(\rho^2+1/4)\tanh\pi\rho},
\]
where $d^4x/x_0=e^{3a}da\, d{\bf y}$.

{\bf $C$-system.}
\begin{equation}\label{I-226}
\Psi^{\rho \tau}_{l,m}(a,b,\theta,\varphi){\equiv} \langle
a,b,\theta,\varphi\mid\rho,\tau,l,m\rangle
{=}(C^+_{C}e^{(-3/2+i\rho)a}+C^-_C
e^{(-3/2-i\rho)a})Y^{\tau}_{lm}(b,\theta, \varphi),
\end{equation}
where
\begin{gather*}
Y^{\tau}_{lm}(b,\theta,\varphi)=(2\pi)^{-1/2}\exp({\rm i}\tau
b)Y_{lm}(\theta,\varphi),
\\
C^{\pm}_{C}=\frac{\sqrt{2}\mid\Gamma[({\rm i}\rho+{\rm
i}\tau+l+3/2)/2] \Gamma[({\rm i}\rho-{\rm i}\tau+l+3/2)/2]\mid
2^{\mp {\rm i}\rho}\Gamma(\pm {\rm
i}\rho)}{\sqrt{\pi}\mid\Gamma({\rm i}\rho+3/2)\mid\Gamma [(\pm
{\rm i}\rho+{\rm i}\tau+l+3/2)/2] \Gamma[(\pm {\rm i}\rho-{\rm
i}\tau+l+3/2)/2]}.
\end{gather*}
The orthogonality relation for the functions \eqref{I-226} is
\begin{gather*}
\langle\Psi^{\rho'\tau'}_{l'm'},\Psi^{\rho\tau}_{lm}\rangle
=\int_{C^4_+}
\frac{d^4x}{x_0}\Psi^{\rho'\tau'}_{l'm'}(a,b,\theta,\varphi)
\overline{\Psi^{\rho\tau}_{lm}(a,b,\theta,\varphi)}
\\
\phantom{\langle\Psi^{\rho'\tau'}_{l'm'},\Psi^{\rho\tau}_{lm}\rangle
}{} =\frac{\delta(\rho-\rho')} {\rho(\rho^2+1/4)\tanh\pi\rho}
\delta(\tau-\tau')\delta_{ll'}\delta_{mm'},
\end{gather*}
where $d^4x/x_0=\frac18 e^{3a}\sin\theta\, da\, db\, d\theta\,
d\varphi$.

{\bf $SH$-system.}
\begin{gather*}
\Psi_{\rho\omega m\tilde{m}}(a,b,\varphi,\Phi) \equiv \langle
a,b,\varphi,\Phi\mid\rho,\omega,m,
\tilde{m}\rangle\\
\phantom{\Psi_{\rho\omega m\tilde{m}}(a,b,\varphi,\Phi)}{}
=(C^+_{SH} e^{(-3/2+{\rm i}\rho)a} +C^-_{SH} e^{(-3/2-{\rm
i}\rho)a})W^{\omega}_{m\tilde{m}} (b, \varphi,\Phi),
\end{gather*}
where
\begin{gather*}
W^{\omega}_{m\tilde{m}}(b,\varphi,\Phi)=\frac{1}{2\pi}\frac{\mid\Gamma
({\rm i}\omega+1/2)\mid}{\mid\Gamma({\rm i}\omega+m+1/2)\mid}
P^m_{{\rm i}\omega-1/2}(\cosh b )e^{{\rm
i}(m\varphi+\tilde{m}\Phi)},
\\
C^{\pm}_{SH}=\frac{\sqrt{2}\mid\Gamma[({\rm i}\rho+{\rm
i}\omega+\tilde{m}+1)/2] \Gamma[({\rm i}\rho-{\rm
i}\omega+\tilde{m}+1)/2]\mid 2^{\mp {\rm i}\rho}\Gamma(\pm {\rm
i}\rho)}{\sqrt{\pi}\mid\Gamma({\rm i}\rho+3/2)\mid\Gamma [(\pm
{\rm i}\rho+{\rm i}\omega+\tilde{m}+1)/2] \Gamma[(\pm {\rm
i}\rho-{\rm i}\omega+\tilde{m}+1)/2]}.
\end{gather*}
The orthogonality relation for these basis functions is
\begin{gather*}
\langle\Psi_{\rho'\omega'm'\tilde{m'}},\Psi_{\rho\omega
m\tilde{m}}\rangle =\int_{C^4_+}
\frac{d^4x}{x_0}\Psi_{\rho'\omega'm'\tilde{m'}}
(a,b,\varphi,\Phi)\overline{\Psi_{\rho\omega m\tilde{m}}
(a,b,\varphi,\Phi)}
\\
\phantom{\langle\Psi_{\rho'\omega'm'\tilde{m'}},\Psi_{\rho\omega
m\tilde{m}}\rangle}{} =\frac{\delta(\rho-\rho')}
{\rho(\rho^2+1/4)\tanh\pi\rho}
\frac{\delta(\omega-\omega')}{\omega\tanh\pi\omega}\delta_{mm'}
\delta_{\tilde{m}\tilde{m'}},
\end{gather*}
where $d^4x/x_0=\frac18 e^{3a}\,\sinh b\, da\, db\, d\varphi\,
d\Phi$.

{\bf Transition coef\/f\/icients.} In applications of harmonic
analysis on the hyperboloid and on the cones, it is necessary
sometimes to go from expansion of functions $\psi(x)$, $x\in
H^4_+$ (or $x\in C^4_+$), in basis elements in a certain
coordinate system to expansion of this function in basis elements
in another coordinate system. This transition is fulf\/illed by
means of matrix elements (kernels) of the corresponding transition
operator. These matrix elements (kernels) are called {\it
coefficients of the transition}.

If we denote a basis function in a coordinate system $A$ by
$\langle x|\gamma\rangle$ and in a coordinate system $B$ by
$\langle x|\mu\rangle$, where $x\in H_+^4$ (or $x\in C_+^4$), then
$\langle x|\gamma\rangle$ and $\langle x|\mu\rangle$ are connected
by the formula
\[
\langle x|\mu\rangle=\int\,d\nu(\gamma)\langle x|\gamma\rangle
\,\langle\gamma|\mu\rangle,
\]
where $d\nu(\gamma)$ is the measure in continuous parameters and a
sum in discrete parameters in $\gamma$ (this measure coincides
with the measure with respect to which the expansion of the
function $\psi(x)$ in basis functions $\langle x|\gamma\rangle$ is
made. If $\langle\gamma|\mu\rangle$ is continuous function in
variables $\gamma$, then according to orthogonality properties of
the functions $\langle x|\gamma\rangle$ we have
\begin{equation}\label{I-227}
\langle\gamma|\mu\rangle=\int\,\frac{d^4x}{x_0}\langle
\gamma|x\rangle\,\langle x|\mu\rangle,\qquad \langle
\gamma|x\rangle=\overline{\langle x|\gamma\rangle},
\end{equation}
where integration is over the hyperboloid or the cone.

The quasi-regular representations of the group $SO_0(1,4)$ are
realized on the spaces $L^2(H_+^4)$ and $L_0^2(C_+^4)$. These
representations are unitary equivalent and the equivalence is
given by the orispherical transform \eqref{I-97}. Besides, for
each coordinate system there exist basis functions of these
spaces, which are also connected by the orispherical transform
\eqref{I-97}. Therefore, coef\/f\/icients of transition from one
basis to another in $L^2(H_+^4)$ coincide with the corresponding
coef\/f\/icients of transition in the space $L_0^2(C_+^4)$,
\[
\int_{H_+^4}\,\frac{d^4x}{x_0}\langle\gamma|x\rangle\,\langle
x|\mu\rangle=\int_{C_+^4}\,\frac{d^4x}{x_0}\,
(O\langle\gamma|x\rangle)(O\langle x|\mu\rangle),
\]
where $O$ is the operator of the orispherical transform. Thus, we
do not need to derive coef\/f\/icients of transition in
$L^2(H_+^4)$ and in $L_0^2(C_+^4)$. Since basis functions on
$L_0^2(C_+^4)$ are simpler than basis functions on $L^2(H_+^4)$,
these coef\/f\/icients usually derive for the space
$L_0^2(C_+^4)$.

\section{Information on semisimple Lie groups and Lie algebras}
\label{Lie groups}

Let $G$ be a connected linear (matrix) noncompact real semisimple
Lie group and let $\mathfrak{g}$ be its Lie algebra. Let $K$ be a
maximal compact subgroup in $G$ and let $\mathfrak{k}$ be the Lie
subalgebra in~$\mathfrak{g}$ corresponding to $K$. Then one has a
{\it Cartan decomposition}
$\mathfrak{g}=\mathfrak{k}+\mathfrak{p}$, where $\mathfrak{p}$ is
a linear subspace of $\mathfrak{g}$. This sum is direct. The
Cartan decomposition is characterized by the inclusions
\begin{equation}\label{1}
[\mathfrak{k},\mathfrak{k}]\subset \mathfrak{k},\qquad
[\mathfrak{p},\mathfrak{p}]\subset \mathfrak{k},\qquad
[\mathfrak{k},\mathfrak{p}]\subset \mathfrak{p}.
\end{equation}

The formula
\begin{equation}
\label{2} B(X,Y)={\rm Tr}\; (({\rm ad}\; X)({\rm ad}\; Y)),\qquad
X,Y\in \mathfrak{g},
\end{equation}
determines a symmetric bilinear form on $\mathfrak{g}$, which is
called a {\it Killing--Cartan form}. Here ${\rm ad}\; X$ is the
operator on $\mathfrak{g}$ acting as $({\rm ad}\; X)Y=[X,Y]$,
$Y\in \mathfrak{g}$. The fact that a Lie algebra $\mathfrak{g}$ is
semisimple means that the bilinear form $B(X,Y)$ is
non-degenerate. If $\mathfrak{g}=\mathfrak{k}+\mathfrak{p}$ is a
Cartan decomposition of $\mathfrak{g}$, then $B(X,X)<0$ for $X\in
\mathfrak{k}$ and $B(X,X)>0$ for $X\in \mathfrak{p}$.

Let $\theta$ be the involutive Cartan automorphism on
$\mathfrak{g}$. Then $\theta$ leaves elements of $\mathfrak{k}$
invariant and multiplies elements of $\mathfrak{p}$ by $-1$. We
introduce on $\mathfrak{g}$ the bilinear form
\begin{equation}\label{3}
\langle X,Y\rangle=-B(X,\theta Y).
\end{equation}
This form determines a positive scalar product on $\mathfrak{g}$.
The algebra $\mathfrak{g}$ with this scalar product turns into a
f\/inite dimensional Hilbert space.

Let $\mathfrak{a}$ be a maximal commutative subalgebra in
$\mathfrak{p}$. A dimension of $\mathfrak{a}$ is called a {\it
real rank} of the Lie algebra $\mathfrak{g}$. We consider the set
of operators ${\rm ad}\; H$, $H\in \mathfrak{a}$, acting on the
space $\mathfrak{g}$. It is easy to see that
\[
\langle ({\rm ad}\; H)X,Y\rangle =\langle X,({\rm ad}\;
H)Y\rangle, \qquad X,Y\in \mathfrak{g},
\]
that is, ${\rm ad}\, H$ is a self-adjoint operator on
$\mathfrak{g}$. Thus, if $\mathfrak{g}$ is supplied by the scalar
product \eqref{3}, then the operators ${\rm ad}\; H$, $H\in
\mathfrak{a}$, constitute a commuting collection of self-adjoint
operators. For this reason, $\mathfrak{g}$ decomposes into an
orthogonal sum of eigenspaces of these operators:
\begin{equation}
\label{4} \mathfrak{g}=\mathfrak{g}_0+\sum_\gamma
\mathfrak{g}_\gamma ,
\end{equation}
where $\mathfrak{g}_0$ is an eigenspace with zero eigenvalue for
all operators ${\rm ad}\; H$, $H\in \mathfrak{a}$, and
$\mathfrak{g}_\gamma$ are eigenspaces with eigenvalues
$\gamma(H)$, $H\in \mathfrak{a}$ ($\gamma$ are linear forms on
$\mathfrak{a}$).

It is evident that elements of dif\/ferent eigenspaces
$\mathfrak{g}_\gamma$ are orthogonal with respect to the scalar
product \eqref{3}. Linear forms $\gamma$ on $\mathfrak{a}$ in
\eqref{4} are called {\it restricted roots} of the algebra
$\mathfrak{g}$ with respect to $\mathfrak{a}$ (or restricted roots
of the pair $(\mathfrak{g}, \mathfrak{a})$). The subspaces
$\mathfrak{g}_\gamma$ are called {\it root subspaces}.

Restricted roots are split into two sets: a set of positive
restricted roots and a set of nega\-ti\-ve restricted roots. In
order to determine a sign of a restricted root we have to f\/ix a
basis $H_1,H_2,\ldots,H_l$ of the subalgebra $\mathfrak{a}$. If
the f\/irst non-zero number in the set
$\gamma(H_1),\gamma(H_2),\ldots$, $\gamma(H_l)$ is positive
(negative), then the root $\gamma$ is positive (negative). The set
$\alpha_1,\alpha_2,\dots,\alpha_l$ of restricted positive roots is
called a system of {\it simple roots} if each restricted positive
root is a linear combination of $\alpha_1,\alpha_2,\dots,\alpha_l$
with non-negative integral coef\/f\/icients.

Restricted roots and root subspaces possess the following
properties:
 \begin{enumerate}\itemsep=0pt

\item[(a)] if $\gamma_1$, $\gamma_2$ and $\gamma_1+\gamma_2$ are
restricted roots of the pair $(\mathfrak{g},\mathfrak{a})$, then
$[\mathfrak{g}_{\gamma_1},\mathfrak{g}_{\gamma_2}]\subset
\mathfrak{g}_{\gamma_1+\gamma_2}$; if $\gamma_1+\gamma_2$ is not a
restricted root, then
$[\mathfrak{g}_{\gamma_1},\mathfrak{g}_{\gamma_2}]=0$;

\item[(b)] if $\gamma$ is a restricted root, then $-\gamma$ is
also a restricted root, that is, to each positive root $\gamma$
there corresponds a negative root $-\gamma$;

\item[(c)] root subspaces $\mathfrak{g}_\gamma$ may be more than
one-dimensional;

\item[(d)] only the roots $2\gamma$, $\gamma$, $-\gamma$,
$-2\gamma$ or the roots $\gamma$, $\gamma/2$, $-\gamma/2$,
$\gamma$ can be roots multiple to a~restricted root $\gamma$; in
particular, if a semisimple Lie algebra is of rank 1, then the
subalgebra~$\mathfrak{a}$ is one-dimensional and all restricted
roots are linearly dependent, that is, in this case there are only
two restricted roots $\gamma$, $-\gamma$ or 4 restricted roots
$2\gamma$, $\gamma$, $-\gamma$, $-2\gamma$.
\end{enumerate}

A dimension of a root subspace $\mathfrak{g}_\gamma$ is called a
{\it multiplicity} of the restricted root $\gamma$, which is
denoted by $m(\gamma)$.

Let $\gamma_1,\gamma_2,\ldots,\gamma_n$ be a set of all positive
restricted roots of the pair $(\mathfrak{g},\mathfrak{a})$. The
linear form
\begin{equation}
\label{5} \rho=\frac12 \sum_{i=1}^n m(\gamma_i)\gamma_i
\end{equation}
is called a {\it half-sum} of positive restricted roots of the
pair $(\mathfrak{g},\mathfrak{a})$. This form is important in the
representation theory of semisimple Lie groups.

Let $\mathfrak{n}=\sum\limits_{\gamma>0} \mathfrak{g}_\gamma$,
where summation is over positive restricted roots of the pair
$(\mathfrak{g},\mathfrak{a})$. Then $\mathfrak{n}$ is a maximal
nilpotent subalgebra in $\mathfrak{g}$ and the Iwasawa
decomposition
\[
\mathfrak{g}=\mathfrak{k}+\mathfrak{a}+\mathfrak{n}
\]
holds, where the sum is direct.

Let $\mathfrak{m}$ be the centralizer of the subalgebra
$\mathfrak{a}$ in $\mathfrak{k}$ (that is, the set of all elements
of $\mathfrak{k}$ which commute with all elements of
$\mathfrak{a}$). Then $\mathfrak{m}$ is a subalgebra in
$\mathfrak{g}$. Moreover, $\mathfrak{m}$ is a reductive Lie
algebra (that is, it is a direct sum of a semisimple subalgebra of
$\mathfrak{g}$ and a center of $\mathfrak{m}$). It is clear that
$\mathfrak{m}$ belongs to the subalgebra $\mathfrak{g}_0$ from
\eqref{4}. Moreover, the subalgebra $\mathfrak{g}_0$ coincides
with the sum $\mathfrak{a}+\mathfrak{m}$.

Let $N$ be a closed subgroup in $G$ with a Lie algebra
$\mathfrak{n}$ and let $A$ be a subgroup in $G$ with the Lie
algebra $\mathfrak{a}$. Then $G=ANK$. Moreover, each element $g\in
G$ decomposes uniquely as a product $g=hnk$, $h\in A$, $n\in N$,
$k\in K$. The mapping $(h,n,k)\to hnk$ is an analytic
dif\/feomorphism of the manifold $A\times N\times K$ onto $G$. The
decomposition of elements of the group~$G$ into a product of
elements of the subgroups~$A$, $N$ and $K$ is called its {\it
Iwasawa decomposition}. It is a global analogue of the Iwasawa
decomposition
$\mathfrak{g}=\mathfrak{k}+\mathfrak{a}+\mathfrak{n}$ of the Lie
algebra $\mathfrak{g}$. This decomposition is extensively used in
the representation theory.

The Iwasawa decomposition of $G$ can be written in another form.
Since $G=G^{-1}=(ANK)^{-1}=K^{-1}N^{-1}A^{-1}=KNA$, then $G=KNA$.
Each element $g\in G$ decomposes uniquely as a product $g=knh$,
$k\in K$, $n\in N$, $h\in A$.

Let $M$ be the centralizer of the subgroup $A$ in $K$, that is,
$M$ consists of all elements of~$K$ commuting with all elements of
$A$. Then the subalgebra $\mathfrak{m}$ is a Lie algebra for $M$.
The subgroup~$M$ may be not connected. It can be represented in
the form $M=ZM_0$, where $M_0$ is a~connected component of $M$,
containing the unit element, and $Z$ is a f\/inite group. The
subgroup $P=ANM=MNA$ of the group $G$ is called a {\it minimal
parabolic subgroup} of~$G$. A~subgroup of~$G$, containing a
minimal parabolic subgroup, is called a {\it parabolic subgroup}
of~$G$. Below we shall consider representations of the group~$G$,
induced by irreducible representations of a minimal parabolic
subgroup.

\looseness=1 Let us consider a connection of compact and
noncompact Lie groups and a connection of their Lie algebras. Let
$G$ and $\mathfrak{g}$ be as above. Since $G$ is a linear group,
it has a complexif\/ication; let us denote it by $[G]$. Let
$[\mathfrak{g}]$ be a complexif\/ication of the Lie algebra
$\mathfrak{g}$. Let $G_k$ be a compact real form of the complex
group $[G]$, and let $\mathfrak{g}_k$ be a compact real form of
the Lie algebra $[\mathfrak{g}]$. If $\mathfrak{g}=\mathfrak{k}+
\mathfrak{p}$ is a Cartan decomposition of the Lie algebra
$\mathfrak{g}$, then for the Lie algebra $\mathfrak{g}_k$ we have
the decomposition
\begin{equation}
\label{10} \mathfrak{g}_k=\mathfrak{k}+{\rm i}\mathfrak{p},\qquad
{\rm i}=\sqrt{-1},
\end{equation}
(see, for example, \cite{Hel}).\looseness=1 There exists also a
connection between decompositions of the groups~$G$ and $G_k$. For
$G$, the decomposition $G=KAK$ (the {\it Cartan decomposition})
holds. The subgroup~$A$ is commutative and noncompact. Moreover,
it is a direct product of~$l$ copies of the group~$\mathbb{R}^+$
of positive real numbers with the usual multiplication as a group
operation. Under complexif\/ication of $G$ the subgroup $A\subset
G$ turns into the commutative subgroup \mbox{$[A]\subset [G]$}. Then
$A_k=[A]\bigcap G_k$ is a commutative subgroup in $G_k$. Thus, the
subgroup $A_k\subset G_k$ is obtained by analytic continuation of
the parameters of the subgroup $A\subset G$ to the corresponding
compact domain. To the decomposition $G=KAK$ there corresponds the
decomposition $G_k=KA_kA$ of the compact group
$G_k$. This fact is used in consideration of f\/inite dimensional
representations of semisimple Lie groups. It is known that
f\/inite dimensional representations of the groups $G$ and $G_k$
are obtained by analytic continuation in group parameters
\cite{Zh}. It follows from the decompositions $G=KAK$ and
$G_k=KA_kK$ that it is enough to make an analytical continuation
only for parameters of the subgroups~$A$ and~$A_k$.

More details on the structure of semisimple Lie groups and Lie
algebra can be found in \cite{War,Hel} and \cite{Knapp}.

\newpage

\section{Representations of semisimple Lie groups and Lie algebras}
\label{Rep}
We consider representations of the {\it principal nonunitary
series} of a noncompact real semisimple Lie group $G$. These
representations are constructed by means of irreducible f\/inite
dimensional representations of a minimal parabolic subgroup
$P=MNA$. Let $\delta$ be a unitary irreducible representation of
the compact subgroup $M$ on a f\/inite dimensional Hilbert space
$\mathcal{V}$, and let $\lambda$ be a complex linear form on
$\mathfrak{a}$. Then the mapping $h\to \exp (\lambda (\log h))$,
$h\in A$, is a representation of the commutative subgroup $A$. Here
$\log$ denotes the mapping of the group $A$ onto its Lie
algebra~$\mathfrak{a}$, which is inverse to the exponential
mapping $\exp: \mathfrak{a}\to A$. The correspondence
\begin{equation}
\label{11} mnh\to \exp (\lambda (\log h)) \delta(m),\qquad h\in
A,\quad n\in N,\quad m\in M,
\end{equation}
is an irreducible representation of the group $P=MNA$.

By means of the representations \eqref{11} of $P$ one constructs
(induces) representations of the group $G$. Let us f\/irst
construct Hilbert spaces on which these representations of $G$
act. Let $f$ be a function from $G$ to the f\/inite dimensional
Hilbert space $\mathcal{V}$ satisfying the relation
\begin{equation}
\label{12} f(gmnh)=\delta(m^{-1})\exp (-\lambda (\log
h))f(g),\qquad m\in M,\quad n\in N,\quad h\in A.
\end{equation}
These functions are uniquely determined by their values on
representatives of cosets in $G/P=G/MNA$. Indeed, each element
$g\in G$ is uniquely decomposed into a product $g=xp=xmnh$,
$p=mnh\in P=MNA$, where $x$ denotes representatives of $G/P$. Then
\begin{equation}
\label{13} f(g)=f(xmnh)=\delta(m^{-1})\exp (-\lambda (\log
h))f(x).
\end{equation}
Thus, if we have $f(x)$, then $f(g)$, $g\in G$, are uniquely
determined.

Since $G/P=KNA/MNA\sim K/M$, then instead of representatives of
$G/P$ we can take representatives of cosets in $K/M$. If we extend
the set of the latter representatives to the whole subgroup $K$,
then instead of functions $f(g)$ on $G$, satisfying the condition
\eqref{12}, we obtain functions on $K$, satisfying the condition
\begin{equation}
\label{14} f(km)=\delta(m^{-1})f(k),\qquad m\in M.
\end{equation}
Now we construct a Hilbert space of functions $f(k)$ on $K$ with
values in $\mathcal{V}$, satisfying the condition \eqref{14} and
the conditio
\[
\int _K \Vert f(k)\Vert ^2_{\mathcal{V}}\, dk<\infty,
\]
where $dk$ is an invariant measure on $K$ and $\Vert
f(k)\Vert_{\mathcal{V}}$ is a norm on ${\mathcal{V}}$. Namely, we
def\/ine a scalar product in the space of such functions by the
formula
\begin{equation}
\label{15} \langle f_1,f_2\rangle=\int _K \langle f_1(k),f_2(k)
\rangle_{\mathcal{V}}\, dk
\end{equation}
and close this space by means of this scalar product. Note that
this scalar product can be introduced both in the space of
functions on $G$, satisfying the condition \eqref{12}, and in the
space of functions on $K$, satisfying the condition \eqref{14}. In
the f\/irst case the Hilbert space is denoted by
$\mathcal{H}_{\delta\lambda}$ and in the second case by
$L^2_\delta(K,{\mathcal{V}})$. Now the formula
\begin{equation}
\label{16} \pi_{\delta\lambda}(g_0) f(g)=f(g^{-1}_0g), \qquad
g_0\in G,
\end{equation}
determines a linear representation of the group $G$ on the space
$\mathcal{H}_{\delta\lambda}$ which is denoted by
$\pi_{\delta\lambda}$. Going from functions given on $G$ to
functions given on $K$ we obtain
\begin{equation}
\label{17} \pi_{\delta\lambda}(g) f(k)=\exp (-\lambda (\log h))
f(k_g),\qquad g\in G,
\end{equation}
where $h\in A$ and $k_g\in K$ are determined by the Iwasawa
decomposition of the element $g^{-1}k$:
\[
g^{-1}k=k_gnh,\qquad n\in N.
\]

Thus, in $L^2_\delta(K,{\mathcal{V}})$ the representation
$\pi_{\delta\lambda}$ is given by formula \eqref{17}. The
representations \eqref{16} and \eqref{17}, realized on the spaces
$\mathcal{H}_{\delta\lambda}$ and $L^2_\delta(K,{\mathcal{V}})$,
respectively, are in fact dif\/ferent realizations of the same
representation. One says that this representation is induced by
the representation \eqref{11} of the subgroup $P$.

If $\delta$ runs over all non-equivalent irreducible unitary
representations of the subgroup $M$ and~$\lambda$ runs over all
complex linear forms on $\mathfrak{a}$, then $\pi_{\delta\lambda}$
constitute the {\it principal nonunitary series} of
representations of $G$.

The restriction of the representation $\pi_{\delta\lambda}$ to the
subgroup $K$ acts on $L^2_\delta(K,{\mathcal{V}})$ by the formula
\[
\pi_{\delta\lambda} (k_0)f(k)=f(k_0^{-1}k).
\]
Since the functions $f$ satisfy the condition
$f(km)=\delta(m^{-1})f(k)$, $m\in M$, the space
$L^2_\delta(K,{\mathcal{V}})$ decomposes into orthogonal sum of
linear f\/inite dimensional subspaces, on which irreducible
unitary representations (we denote them by $\nu$) of the subgroup
$K$ are realized. Moreover, each such representation $\nu$ is
contained in the decomposition of the representation
$\pi_{\delta\lambda}\downarrow _K$ of $K$ with multiplicity
$b^\nu_\delta$, where $b^\nu_\delta$ is the multiplicity of the
representation $\nu$ of $M$ in the representation $\delta$ of $K$.

The representation $\pi_{\delta\lambda}$ of the group $G$
determines the corresponding representation of the Lie algebra
$\mathfrak{g}$, which is denotes also by $\pi_{\delta\lambda}$.
To  noncompact elements of the Lie algebra $\mathfrak{g}$ there
correspond unbounded operators of this representation.
For this reason, these operators are not def\/ined on the whole
Hilbert space $L^2_\delta(K,{\mathcal{V}})$. However, each of
these operators is determined on dif\/ferentiable functions from
$L^2_\delta(K,{\mathcal{V}})$. Moreover, on the set of
inf\/initely dif\/ferentiable functions of
$L^2_\delta(K,{\mathcal{V}})$ an action of products of such
operators is determined. In particular, an action of Casimir
operators is determined on such functions. Note that the linear
space of inf\/initely dif\/ferentiable functions of
$L^2_\delta(K,{\mathcal{V}})$ is dense in this Hilbert space.

\section{Hyperboloids and cones for semisimple noncompact\\
Lie groups and coordinate systems on them} \label{Hyp-Cone}

\subsection{Hyperboloids and cones}
\label{H-C}
Let us construct analogues of upper sheet of the hyperboloid
$H^4_+$ of the two-sheeted hyperboloid and the upper sheet $C^4_+$
of the cone constructed in the previous sections. The motion group
of~$H^4_+$ and of~$C^4_+$ is the de Sitter group $SO_0(1,4)$. The
hyperboloid $H^4_+$ is identif\/ied with the homogeneous space
$SO_0(1,4)/SO(4)$, where $SO(4)$ is a maximal compact subgroup in
$SO_0(1,4)$, and the cone $C^4_+$ with the homogeneous space
$SO_0(1,4)/ISO(3)$, where $ISO(3)$ is the subgroup~$MN$ (see the
previous section) for $SO_0(1,4)$. Taking into account these
facts, we def\/ine a hyperboloid and a cone for an arbitrary
connected linear semisimple noncompact Lie group $G$ as follows.
An analogue of the upper sheet $H^4_+$ of the two-sheeted
hyperboloid is the homogeneous space $G/K$, where $K$ is a maximal
compact subgroup of $G$. The space $G/K$ is a noncompact
Riemannian symmetric space (see \cite{Hel}). In order to have an
analogy with the case of the group $SO_0(1,4)$, we call this space
a {\it hyperboloid}. A {\it cone} (an analogue of the upper sheet
$C^4_+$ of the cone) with a motion group $G$ is the homogeneous
space $G/MN$, where $N$ is a maximal nilpotent subgroup (see the
previous section) and $M$ is a subgroup of $K$, which is a
centralizer of the subgroup $A=\exp \mathfrak{a}$ in $K$. A list
of classical simple noncompact real Lie groups $G$ and their
subgroups $K$ and $M$ is given in Table~1.

\begin{center}
{\bf Table 1}

\vspace{1mm}

\begin{tabular}{| c | c | c | c |}
\hline $G$ & $G_k$ & $K$ & $M$ \\ \hline $SL(n,\mathbb{C})$ &
$SU(n)\times SU(n)$ & $SU(n)$ & $D$ \\ \hline $SO(n,\mathbb{C})$
& $SO(n)\times SO(n)$ & $SO(n)$ & $D$ \\ \hline $Sp(n,\mathbb{C})$
& $Sp(n)\times Sp(n)$ & $Sp(n)$ & $D$ \\ \hline $SL(n,\mathbb{R})$
& $SU(n)$ & $SO(n)$ & $(Z_2)_{n-1}$ \\ \hline $SU^*(2n)$ &
$SU(2n)$ & $Sp(n)$ & $(Z_2)_{n-1}$ \\ \hline $SU(p,q)$  &
$SU(p+q)$ & $S(U(p)\times U(q))$ &
                      $S(U(p-q)\times U(1)\times\cdots\times U(1))$ \\ \hline
$U(p,q)$  & $U(p+q)$ & $U(p)\times U(q)$ &
                      $U(p-q)\times U(1)\times\cdots\times U(1)$ \\ \hline
$SO_0(p,q)$ & $SO(p+q)$ & $SO(p)\times SO(q)$ & $SO(p-q)\times
(Z_2)_q$ \\ \hline
$SO^*(2n)$ & $SO(2n)$ & $U(n)$  & $SU(2)\times\cdots\times SU(2)$ \\
           &          &          & if $n=2k$ \\
           &          &          & $SU(2)\times\cdots\times SU(2)\times U(1)$ \\
           &          &          & if $n=2k+1$ \\  \hline
$Sp(n,\mathbb{R})$ & $Sp(n)$ & $U(n)$ & $(Z_2)_{n-1}$ \\ \hline
$Sp(p,q)$  & $Sp(p+q)$ & $Sp(p)\times Sp(q)$ &
                             $Sp(p-q)\times Sp(1)\times\cdots\times Sp(1)$ \\ \hline
\end{tabular}
\end{center}

Complex groups $SL(n,\mathbb{C})$, $SO(n,\mathbb{C})$ and
$Sp(n,\mathbb{C})$ in Table 1 are understood as real groups with
double number of real parameters. The subgroups $D$ in these
groups coincide with maximal torus in $K$. $Z_2$ denotes a group
consisting of two elements and $(Z_2)_m$ is a direct product of
$m$ copies of the group $Z_2$.

The nilpotent subgroup $N$ of $G$ is obtained by exponential
mapping from the nilpotent subalgebra $\mathfrak{n}$ of the Lie
algebra $\mathfrak{g}$. The subalgebra $\mathfrak{n}$ is
constructed by means of the system of restricted roots of the pair
$(\mathfrak{g},\mathfrak{a})$. In \cite{7}, examples of
construction of subalgebras $\mathfrak{n}$ are given. The subgroup
$N$ for the group $SO_0(1,n)$ is constructed in \cite{VKII},
Chapter 9. The subgroup $N$ of the group $SU(1,n)$ is given in
\cite{7}.

\subsection[Cartan decomposition and $S$-coordinate system on $G/K$]{Cartan decomposition and $\boldsymbol{S}$-coordinate system on $\boldsymbol{G/K}$}
\label{S-syst}

The Cartan decomposition of the group $G$ is given by the formula
$G=KAK$. This means that each element $g\in G$ can be represented
in the form of a product $khk'$, $k,k'\in K$, $h\in A$. However
the decomposition $g=khk'$ is not unique since $kmhk'=khmk'$,
$m\in M$. Therefore, taking a~decomposition of elements of $K$ in
the form $k=\tilde k m$, $m\in M$, where $\tilde k$ are
representatives of cosets of the homogeneous space $K/M$, we
obtain from the Cartan decomposition $g=khk'$ of $g\in G$ the
decomposition
\begin{equation}
\label{251} g=\tilde k hk\qquad {\rm or}\qquad g=kh\tilde k .
\end{equation}
The set of elements $\tilde k$ of $K$, parametrizing the space
$K/M$, is denoted by $(K/M)$. An ambiguousness of the
decomposition \eqref{251} is determined by the relation
\[
\tilde k hmk=\tilde k mhk,\qquad m\in M^*/M\equiv W,
\]
where $M^*$ is a normalizer of the subgroup $A$ in $K$. The
quotient group $W$ is f\/inite and is called a {\it Weyl group} of
the pair $(\mathfrak{g}, \mathfrak{a})$. Thus, the decomposition
\eqref{251} will be unique almost for each $g\in G$ if $h$ is
taken from the part $\overline{A^+}$ of the subgroup $A$, where
$\overline{A^+}$ is a closure of the set ${A^+}$. The set
$\overline{A^+}$ is characterized by the property that there are
no two elements $h$ and $h'$ in $\overline{A^+}$ such that
$h=mh'm^{-1}$ for some $m\in M^*$. The set $\overline{A^+}$ can be
constructed in the following way. The set $A^+$ is the set of all
elements $h=\exp H$, $H\in \mathfrak{a}$, satisfying the
conditions $\alpha_i(H)>0$, $i=1,2,\dots, l$, where $\alpha_i$,
$i=1,2,\dots,l$, are simple restricted roots of the pair
$(\mathfrak{g}, \mathfrak{a})$. The set $\overline{A^+}$ coincides
with the closure of the set $A^+$ in $A$. Thus, starting from the
decomposition
\begin{equation}
\label{252} G=(K/M) \overline{A^+} K,
\end{equation}
we obtain  the decomposition $g=\tilde k hk$, $\tilde k \in
(K/M)$, $h\in \overline{A^+}$, $k\in K$, of elements of the group
$G$. According to the formula \eqref{252}, there exists a set
$(G/K)$ of representative of cosets of $G/K$,
\begin{equation}
\label{253} (G/K)=(K/M) \overline{A^+},
\end{equation}
such that to each element of
$(G/K)$ there corresponds only one coset in $G/K$ and vise versa.

If $l$ is a real rank of the group $G$, then for elements $h\in A$
we have
\begin{equation}
\label{254} h=\exp H=\exp (t_1H_1+t_2H_2+\cdots +t_lH_l)=
\prod_{i=1}^l \exp t_iH_i,
\end{equation}
where $H_1,H_2,\dots,H_l$ is a basis of the subalgebra
$\mathfrak{a}$. The part $\overline{A^+}$ of the subgroup $A$ is
characterized by the condition that the parameters
$t_1,t_2,\dots,t_l$ of elements $h=\exp H\in \overline{A^+}$
satisfy the conditions $\alpha_i(H)\equiv
\alpha_i(t_1H_1+t_2H_2+\cdots +t_lH_l)\ge 0$, $i=1,2,\dots,l$. It
follows from \eqref{253} that {\it for obtaining a
parametrization of the space $G/K$ we have to determine a
parametrization of the ``sphere''} $K/M$. It is clear that the
space $K/M$ is compact. The numbers $t_1,t_2,\dots,t_l$ together
with parameters, determining $K/M$, parametrize the set $(G/K)$.

A parametrization of the space $K/M$ is not unique. It can be
received as follows. Let~$G'_k$ and~$G'$ be compact and noncompact
connected semisimple real Lie groups with the same
complexif\/ication $[G]$. The pairs $(G',G'_k)$, corresponding to
simple Lie groups $G'$, are given in Table~1. Let $K'$ be a
maximal compact subgroup in $G'$ and let $G'=K'A'K'$ be a Cartan
decomposition of the group $G'$. The group $G'_k$ is obtained from
the group $G'$ by an analytic continuation (in the framework of
the complexif\/ication $[G']$ of $G'$) of noncompact parameters of
the group $G'$ to the corresponding compact parameters of the
group $G'_k$. It follows from the Cartan decomposition $G'=K'A'K'$
that we have to make only an analytical continuation of parameters
of the subgroup $A'$. As a result of this continuation, we obtain
the decomposition $G'_k=K'A_k'K'$ of the group $G'_k$ with the
commutative subgroup (torus) $A'_k$. If elements of the subgroup
$A'$ are of the form \eqref{254}, then elements of the subgroup
$A'_k$ are of the form
\begin{equation}
\label{255} \prod_{i=1}^l \exp {\rm i}\varphi_iH_i,\qquad
0\leqslant \varphi_j<2\pi.
\end{equation}
The decomposition $G'_k=K'A'_kK'$ is not unique. The reason is
that elements of the subgroup $M'\subset K'$ permute with elements
of $A'_k$. Therefore, as in the case of noncompact simple real Lie
groups, we have the decomposition $G'_k=(K'/M') A'_kK'$. This
decomposition is not unique, since elements from ${M'}^*/M'\equiv
W'$, as well as elements from the set $J=K'\bigcap A'_k$, permute
with~$A_k'$ (for a noncompact group $G'$ the set $J=K'\bigcap A'$
consists of one (unit) element). Restricting (in an appropriate
manner) values of the parameters $\varphi_1,\varphi_2,
\dots,\varphi_l$ in \eqref{255}, we obtain the decomposition
$G'_k=(K'/M')\overline{{A'_k}^+}K'$. This decomposition leads to
the decomposition of the set $(G'_k/K')$:
\begin{equation}
\label{256} (G'_k/K')=(K'/M')\overline{{A'_k}^+}.
\end{equation}
Now we set here $G'_k=K$ and $K'=M$ and f\/ind
\begin{equation}
\label{257} (K/M)=(M/M')\overline{{A_k}^+},
\end{equation}
where $A_k$ is a commutative subgroup (torus) in $K$ and $M'$ is a
subgroup in $K$ and in $M$. Note that the subgroup $M$ can be
nonconnected. Then it is a product of a connected subgroup $M_0$
with a f\/inite subgroup $Z$. In this case the decomposition
\eqref{256} must be applied to the space $K/M_0$ (not to the space
$K/M$). Then for transition from $K/M_0$ to $K/M$ we have to
perform an identif\/ication of the corresponding values of
parameters.

The relation \eqref{257} leads a parametrization of the space
$K/M$ to a parametrization of the space $M/M'$ such that $\dim
M/M'< \dim K/M$. We may apply decomposition \eqref{256} to the
set~$M/M'$ and reduce its parametrization to a parametrization of
a space which has smaller dimension than that of~$M/M'$.
Continuing this procedure, after a f\/inite number of steps we
obtain a parametrization of the whole space $K/M$, and therefore
of the space $G/K$. A parametrization, obtained in this way,
corresponds to the chain of subgroups $G\supset K\supset M\supset
M'\supset\cdots$. Such coordinate system is an analogue of the
spherical system ($S$-system) of coordinates for the hyperboloid
$H^+_4$. For this reason, we call such  set of coordinates on
$G/K$ a {\it spherical coordinate system} or an {\it $S$-system of
coordinates}.

Thus, a parametrization of the space $G/K$ was obtained by using
the Cartan decompositions of the semisimple Lie group $G$ and of
the corresponding compact Lie groups $G'_k$. By using these
decompositions we can receive a $G$-invariant measure on $G/K$,
expressed in terms of the corresponding parameters. If $f$ is a
continuous function on $G$ with compact support and $g=k_1(\exp
H)k_2$, $k_1,k_2\in K$, $h=\exp H\in A$, then the following
formula holds (see \cite{Hel}):
\begin{equation}
\label{258} \int_G f(g)dg= c\frac1{|W|} \int_\mathfrak{a}
\left\vert \prod_{\alpha>0} \sinh \alpha(H)\right\vert dH
 \int_K \int_K f(k_1(\exp H)k_2)dk_1dk_2,
\end{equation}
where $dg$ and $dk$ are invariant measures on $G$ and $K$,
respectively; the product is over the set of positive restricted
roots of the pair $(\mathfrak{g}, \mathfrak{a})$ (and each root
$\alpha$ appears in the product the number of times equal to
multiplicity of this root), $dH$ is the Lebesgue measure on the
space $\mathfrak{a}$, $|W|$ is an order of the Weyl group $W$ of
the pair $(\mathfrak{g},\mathfrak{a})$, and $c$ is a constant,
depending on a normalization of measures on $G$, $\mathfrak{a}$
and $K$.

The formula
\[
\int_G f(g)dg= \int_{G/K}\left( \int_K f(xk)dk\right) dx
\]
determines an invariant measure $dx$ on $G/K$. Applying this
formula to the relation \eqref{258} we obtain
\begin{equation}
\label{259} \int_{G/K} f(x)dx= c_1\frac1{|W|} \int_\mathfrak{a}
\left\vert \prod_{\alpha>0} \sinh \alpha(H)\right\vert dH
 \int_{(K/M)}  f(y(\exp H))dy,
\end{equation}
where $dy$ is an invariant (with respect to $K$) measure on $K/M$.
In order to obtain  integration over $\overline{A^+}$ instead of
integration over $A=\exp \mathfrak{a}$, it is necessary to
restrict the integral over $\mathfrak{a}$ to an integral over
$\mathfrak{a}^+$. The set $\mathfrak{a}^+$ consists of elements
$H\in \mathfrak{a}$ such that $\alpha_i(H)>0$ for any restricted
simple root $\alpha_i$ of the pair $(\mathfrak{g}, \mathfrak{a})$.
The complete algebra $\mathfrak{a}$ is obtained from
$\mathfrak{a}^+$ by action by elements of the Weyl group $W$ (with
a subsequent closure). (Note that for any $w_1,w_2\in W$ the sets
$w_1\mathfrak{a}^+$ and $w_2\mathfrak{a}^+$ do not intersect if
$w_1\ne w_2$.) Then
\begin{equation}
\label{260} \int_{G/K} f(x)dx= c_1 \int_{\mathfrak{a}^+}
\left\vert \prod_{\alpha>0} \sinh \alpha(H)\right\vert dH
 \int_{(K/M)}  f(y(\exp H))dy.
\end{equation}
This relation reduces a $G$-invariant measure on $G/K$ to a
$K$-invariant measure on $K/M$. In order to obtain the latter
measure we use the integral relation for $G'_k$, which is an
analogue of the formula \eqref{258} (see \cite{Hel}):
\begin{equation}
\label{261} \int_{G'_k} f(g)dg= c \int_{A'_k} \left\vert D(h)
\right\vert dh
 \int_{K'}\int_{K'}  f(k_1hk_2)dk_1dk_2.
\end{equation}
The function $D(h)\equiv D(\exp {\rm i}H)$, $H\in \mathfrak{a}$,
$h\in A'_k$, is determined by the relation
\[
D(\exp {\rm i}H)=\prod_{\alpha>0} \sin \alpha(H).
\]
The relation \eqref{261} can be easily reduced to the relation
similar to \eqref{260}. This gives a~possibility to determine a
$K$-invariant measure on $K/M$ if we set $G'_K=K$ and $K'=M$.

\smallskip

\noindent {\bf Remark.} The formulas \eqref{260} and \eqref{261}
contain products over restricted positive roots of the pair
$(\mathfrak{g}, \mathfrak{a})$. Systems of simple restricted roots
(including their multiplicities) of pairs
$(\mathfrak{g},\mathfrak{a})$ for all simple noncompact Lie
algebras $\mathfrak{g}$ are given in~\cite{War} (see also Table~3
in~\cite{7}). A system of simple roots determines uniquely a
system of the corresponding positive roots.

\subsection[Iwasawa decomposition and $T$-coordinate system on $G/K$]{Iwasawa decomposition and $\boldsymbol{T}$-coordinate system on $\boldsymbol{G/K}$}
\label{T-syst}

Let $G=NAK$ be an Iwasawa decomposition of the group $G$. It gives
a possibility to parametrize the space $G/K$ by means of elements
of the subgroup $NA$. The formula \eqref{254} gives a
para\-met\-ri\-za\-tion of the subgroup $A$ by numbers $t_1,
t_2,\dots,t_l$. In order to obtain a parametrization of the
subgroup $N$, it is necessary to represent $N$ in the form of a
product of the one-parameter subgroups $\exp s_\alpha X_\alpha$
corresponding to root elements $X_\alpha$ of the algebra
$\mathfrak{g}$ with positive restricted roots $\alpha$. In an
analogy with the group $SO_0(1,4)$, this parametrization of $G$ is
called a {\it translational system of coordinates} on $G/K$, which
is also called a {\it $T$-system} of coordinates. Note that for a
generic semisimple Lie group $G$ the subgroup $N$ is not
commutative. The subgroup $N=\exp \mathfrak{n}$ is commutative if
and only if the corresponding Lie algebra $\mathfrak{n}$ is
commutative. A~commutativity of the algebra $\mathfrak{n}$ is
determined by means of a root system of the pair $(\mathfrak{g},
\mathfrak{a})$. If along with roots $\alpha$ and $\beta$ there
exists a root $\alpha+\beta$ (or along with a root $\alpha$ there
exists a root $2\alpha$), then for the corresponding root
subspaces we have $[\mathfrak{g}_\alpha,
\mathfrak{g}_\beta]\subset \mathfrak{g}_{\alpha+\beta}$. This
leads to a violation of commutativity in $\mathfrak{n}$ and,
therefore, in $N$. Analyzing systems of simple restricted roots of
simple real Lie algebras (see, for example, \cite{7}), we conclude
that a subgroup $N$ is commutative only for the groups
$SO_0(1,n)$.

Using the Iwasawa decomposition $G=NAK$, it is possible to f\/ind
a $G$-invariant measure on $G/K$, expressed in the $T$-coordinate
system. If $f$ is a continuous function with a compact support and
$g=nhk$, $n\in N$, $h\in A$, $k\in K$, then (see, for example,
\cite{Hel})
\[
\int_G f(g)dg=\int_K \int_A \int_N f(nhk) e^{2\rho (\log h)} dn\,
dh\, dk,
\]
where $dg$, $dn$, $dh$ and $dk$ are invariant measures on $G$,
$N$, $A$ and $K$, respectively, and $\rho=\frac12
\sum\limits_{\alpha>0} \alpha$. Therefore, if  an element $x\in
G/K$ is represented by an element $nh\in NA$, then
\begin{equation}
\label{262} \int_{G/K} f(x)dx=  \int_{A}\int_N  f(nh)e^{2\rho
(\log h)} dn\, dh,
\end{equation}
where $dx$ is a $G$-invariant measure on $G/K$.

\subsection[Iwasawa decomposition and $O$-coordinate system on $G/K$]{Iwasawa decomposition and $\boldsymbol{O}$-coordinate system on $\boldsymbol{G/K}$}
\label{O-syst}

Using the Iwasawa decomposition $G=NAK$ of the group $G$, we
represent $G$ in the form
\begin{equation}
\label{263} G=  (NM)AK,
\end{equation}
where $NM$ is a closed subgroup of $G$. Since $MNM^{-1}\subset N$,
then the subgroup $NM$ is a~semidirect product of the subgroup $M$
and the normal subgroup $N$. The most interesting case is when
$NM$ is a semidirect product of the compact subgroup $M$ and the
commutative subgroup $N$. As we know this is the case when
$G=SO_0(1,n)$. For this group the subgroup $NM$ is isomorphic to
the group $ISO(n-1)$ (see, for example, \cite{VKII}, Chapter 9).

The decomposition \eqref{263} is not unique. Indeed, elements of
the subgroup $A$ commute with elements of $M$. In order to obtain
non-ambiguous decomposition we have to take the space $NM/M$
instead of the space $NM$. Then for $(G/K)$ we have the
decomposition
\begin{equation}
\label{264} (G/K)=  (NM/M)A,
\end{equation}
where $(NM/M)$ is the set of representatives of cosets of the
quotient space $NM/M$. We take a parametrization of the subgroup
$A$ by the numbers $t_1,t_2, \dots,t_l$. Choosing a
parametrization of the quotient space $NM/M$ we obtain, according
to the formula \eqref{254}, a parametrization of the space $G/K$.
This parametrization is called {\it orispherical}. It is also
called an {\it $O$-coordinate system} on $G/K$. Note that one of
possibilities for parametrization of $(NM/M)$ is given by the
$T$-system of coordinates.

If we choose (according to a parametrization of the space $NM/M$)
representatives $y\in NM$ of cosets of $NM/M$ and set $x=yh\in
(G/K)$, then we have the integral relation
\begin{equation}
\label{265} \int_{G/K} f(x)dx= \int_{NM/M} dy\int_{A}
f(yh)e^{2\rho (\log h)} dh,
\end{equation}
where $dh$ is an invariant measure on $A$, $dx$ is a $G$-invariant
measure on $G/K$, and $dy$ is a~$NM$-invariant measure on $NM/M$.

\subsection{Generalized Cartan decomposition}
\label{Gener-car}
Let $G$ be a semisimple noncompact connected linear real Lie group
and let $\mathfrak{g}$ be its Lie algebra. There exists an
involutive Cartan automorphism $\theta$ on $\mathfrak{g}$ such
that $\mathfrak{g}$ decomposes into a direct sum of eigenspaces of
$\theta$ as $\mathfrak{g}=\mathfrak{k}+\mathfrak{p}$ (the Cartan
decomposition), where $\mathfrak{k}$ corresponds to
eigenvalue~$+1$ and $\mathfrak{p}$ corresponds to eigenvalue $-1$.
Under transition from $\mathfrak{g}$ to $G$, to the automorphism
$\theta$ of~$\mathfrak{g}$ there corresponds the automorphism
$\Theta$ of $G$. The set of points of $G$, which are invariant
with respect to $\Theta$, coincides with the maximal compact
subgroup $K$ of $G$. The homogeneous space $G/K$ is a Riemannian
symmetric space, which is called a hyperboloid. Thus,
we associate a~Riemannian symmetric space with the involutive
Cartan automorphism $\theta$ of $\mathfrak{g}$.

However, the Lie algebra $\mathfrak{g}$ can have other involutive
automorphisms, that is, automorphisms with noncompact stationary
subgroup. Let $\tau$ be such an automorphism of $\mathfrak{g}$.
Then $\mathfrak{g}$ decomposes into a direct sum of eigenspaces of
$\tau$ with eigenvalues $\pm 1$ (since $\tau^2=1$, other
eigenvalues cannot exist). Let this decomposition be of the form
\begin{equation}
\label{266} \mathfrak{g}=\mathfrak{b}+\mathfrak{q},
\end{equation}
where $\mathfrak{b}$ belongs to eigenvalue $+1$ and $\mathfrak{q}$
to eigenvalue $-1$. Since $\tau$ is an automorphism, then the
eigenvalues show that
\[
[\mathfrak{b},\mathfrak{b}]\subset \mathfrak{b},
\]
that is, $\mathfrak{b}$ is a subalgebra of $\mathfrak{g}$. Let
$\mathcal{T}$ be an automorphism of the group $G$, corresponding
to the automorphism $\tau$ of $\mathfrak{g}$, and let $G_\tau$ be
a closed subgroup of $G$ consisting of all points of $G$ invariant
under $\mathcal{T}$. A connected component of the unit element in
$G_\tau$ will be denoted by $G^0_\tau$. If $H$ is a~subgroup of
$G$ such that $G^0_\tau\subset H\subset G_\tau$, then the
homogeneous space $G/H$ is called {\it symmetric}. Since the
subgroup $H$ is not compact, a quotient space $G/H$ is called {\it
pseudo-Riemannian} or {\it affine} symmetric space. The group $G$
is a motion group of the pseudo-Riemannian symmetric space $G/H$.
A classif\/ication of pseudo-Riemannian symmetric spaces is given
in \cite{84}.

With a pseudo-Riemannian symmetric space with the transitive
motion group $G$ a generalized Cartan decomposition of $G$ is
connected. For simplicity, we shall consider only those
pseudo-Riemannian symmetric spaces for which the automorphism
$\tau$ is determined by a signature of the set of restricted roots
of the pair $(\mathfrak{g},\mathfrak{a})$ (see \cite{85}).

Let $\Sigma$ be a set of all restricted roots of the pair
$(\mathfrak{g},\mathfrak{a})$. A mapping $\varepsilon$ of roots of
$\Sigma$ to the set $\{ +1,-1\}$ is called a {\it signature} of
$\Sigma$ if

\begin{enumerate}\itemsep=0pt

\item[(a)] $\varepsilon(\alpha)=\varepsilon(-\alpha)$,  $\alpha\in
\Sigma$;
 \medskip

\item[(b)] if $\alpha,\beta,\alpha+\beta$ are restricted roots,
then
$\varepsilon(\alpha+\beta)=\varepsilon(\alpha)\varepsilon(\beta)$.
\end{enumerate}

The Lie algebra $\mathfrak{g}$ can be represented in the form
\[
\mathfrak{g}=\mathfrak{m}+\mathfrak{a}+\sum_{\alpha\in \Sigma}
\mathfrak{g}_\alpha,
\]
where $\mathfrak{g}_\alpha$ is a root subspace corresponding to a
root $\alpha\in \Sigma$. Using a signature $\varepsilon$, we give
an automorphism $\theta_\varepsilon$ on $\mathfrak{g}$ such that

\begin{enumerate}\itemsep=0pt
\item[(a)] $\theta_\varepsilon (X)=\varepsilon(\alpha)\theta(X)$,
$X\in \mathfrak{g}_\alpha$, where $\theta$ is the involutive
Cartan automorphism, introduced above;

\item[(b)] $\theta_\varepsilon (X)=X$ for $X\in
\mathfrak{m}+\mathfrak{a}$.
\end{enumerate}

It is easy to show that $\theta_\varepsilon$ is an involutive
automorphism of $\mathfrak{g}$ such that $\theta_\varepsilon\ne
\theta$ for  a nontrivial signature $\varepsilon$ on $\Sigma$.

We write down the decomposition \eqref{266} for the automorphism
$\theta_\varepsilon$ in the form
\begin{equation}
\label{267}
\mathfrak{g}=\mathfrak{t}_\varepsilon+\mathfrak{p}_\varepsilon,
\end{equation}
where $\mathfrak{t}_\varepsilon$ is a Lie subalgebra (noncompact)
in $\mathfrak{g}$. We have $\mathfrak{m}\subset
\mathfrak{t}_\varepsilon$ and $\mathfrak{a}\subset
\mathfrak{p}_\varepsilon$.

If the automorphism $\theta_\varepsilon$ can be continued to an
automorphism of $G$, we denote this automorphism of $G$ by
$\Theta_\varepsilon$. Let $(K_\varepsilon)_0$ be an analytical
subgroup of $G$ with the Lie algebra $\mathfrak{t}_\varepsilon$.
We set $K_\varepsilon=(K_\varepsilon)_0M$. Since the connected
component $M_0$ of the unit element in the subgroup $M$ is
contained in $(K_\varepsilon)_0$ (since $\mathfrak{m}\subset
\mathfrak{k}_\varepsilon$), then $K_\varepsilon=(K_\varepsilon)_0
Z$, where $Z$ is a discrete subgroup such that $M=M_0Z$. The
subgroups $(K_\varepsilon)_0$ and $K_\varepsilon$ have the
following properties (see \cite{85}:

\begin{enumerate}\itemsep=0pt
\item[(a)] for each $m\in M$ we have
$m(K_\varepsilon)_0m^{-1}\subset (K_\varepsilon)_0$;

\item[(b)] $K_\varepsilon$ is a closed subgroup of $G$ and
$\Theta_\varepsilon k=k$ for each $k\in K_\varepsilon$.
\end{enumerate}

Let us give a classif\/ication of pairs $(G,K_\varepsilon)$ for
simple Lie groups $G$. A pair $(G,K_\varepsilon)$ is determined by
the corresponding pair $(\mathfrak{g},\mathfrak{k}_\varepsilon)$
for the Lie algebra $\mathfrak{g}$. In order to classify pairs
$(\mathfrak{g},\mathfrak{k}_\varepsilon)$ with a simple Lie
algebra $\mathfrak{g}$, we consider signatures of the
corresponding root system $\Sigma$. These signatures can be easily
described as follows. Let $\alpha_1,\alpha_2,\cdots, \alpha_l$ be
simple restricted roots of the pair $(\mathfrak{g},
\mathfrak{a})$. Each restricted root $\alpha\in \Sigma$ can be
represented uniquely in the form
\begin{equation}
\label{268} \alpha=\sum_{i=1}^l m_i\alpha_i.
\end{equation}
Let $\varepsilon_j$ ($j=1,2,\dots,l$) be a signature  of $\Sigma$
such that $\varepsilon_j(\alpha)=(-1)^{m_j}$, where $m_j$ is
determined by the decomposition \eqref{268}. Then any signature
$\varepsilon$ of the root system $\Sigma$ reduces to some
signature~$\varepsilon_j$, namely, for any signature $\varepsilon$
there exists an element $w$ of the Weyl group $W$ and an integer
$j$ ($1\leqslant j\leqslant l$) such that for each $\alpha\in
\Sigma$ we have $\varepsilon(\alpha)=\varepsilon_j(w\alpha)$ (see
\cite{86}). Thus, any signature $\varepsilon$ can be transformed
to some signature $\varepsilon_j$.

Let us describe signatures $\varepsilon_j$ and the corresponding
systems of roots $\Sigma_\varepsilon$, where $\Sigma_\varepsilon$
is a set of roots $\alpha$ from $\Sigma$ for which
$\varepsilon(\alpha)=1$.

{\it Let a root system $\Sigma$ coincide with the root system}
$A_l$, $l\geqslant 1$. We have signatures
\[
\varepsilon_j,\qquad 2j<l+1.
\]
The corresponding $\Sigma_{\varepsilon_j}$ coincide respectively
with the root systems
\[
A_{l-j}+A_{j-1}.
\]

{\it Let a root system $\Sigma$ coincide with the root system}
$B_l$, $l\geqslant 2$. We have signatures
\[
\varepsilon_j,\qquad j\leqslant l.
\]
The corresponding $\Sigma_{\varepsilon_j}$ coincide respectively
with the root systems
\[
B_{l-j}+D_{j}.
\]

{\it Let $\Sigma$ coincide with the root system} $BC_l$,
$l\geqslant 1$. We have signatures
\[
\varepsilon_j,\qquad j\leqslant l.
\]
The corresponding $\Sigma_{\varepsilon_j}$ coincide respectively
with the root systems
\[
BC_{l-j}+C_{j}.
\]

{\it Let $\Sigma$ coincide with the root system} $C_l$,
$l\geqslant 3$. We have signatures
\[
\varepsilon_j,\qquad  2j\leqslant l,\qquad \varepsilon_l.
\]
The corresponding $\Sigma_{\varepsilon_j}$ coincide respectively
with the root systems
\[
C_{l-j}+C_{j}\qquad {\rm if}\qquad 2j\leqslant l
\]
and
\[
A_l\qquad {\rm for}\qquad \varepsilon_l.
\]

{\it Let $\Sigma$ coincide with the root system} $D_l$,
$l\geqslant 4$. Then we have signatures
\[
\varepsilon_j,\qquad  2j\leqslant l, \qquad
\varepsilon_{l-1},\qquad \varepsilon_l.
\]
The corresponding $\Sigma_{\varepsilon_j}$ coincide respectively
with the root systems
\[
D_{l-j}+D_{j}\qquad {\rm if}\qquad 2j\leqslant l
\]
and
\[
A_l\qquad {\rm for}\qquad \varepsilon_{l-1}\qquad {\rm and}\qquad
\varepsilon_l.
\]

{\it Let $\Sigma$ coincide with the root system} $E_6$. We have
signatures
\[
\varepsilon_1 \qquad {\rm and}\qquad \varepsilon_2.
\]
Then the corresponding $\Sigma_{\varepsilon_j}$ coincide
respectively with the root systems
\[
D_5\qquad {\rm for}\qquad \varepsilon_1
\]
and
\[
A_1+A_5\qquad {\rm for}\qquad \varepsilon_2.
\]

{\it Let a root system $\Sigma$ coincide with the root system}
$E_7$. We have signatures
\[
\varepsilon_j, \qquad j=1,2,7.
\]
The corresponding $\Sigma_{\varepsilon_j}$ coincide respectively
with the root systems
\begin{gather*}
A_1+D_6\qquad {\rm for}\qquad \varepsilon_1,
\\
A_7\qquad {\rm for}\qquad \varepsilon_2,
\\
E_6\qquad {\rm for}\qquad \varepsilon_7.
\end{gather*}

{\it Let $\Sigma$ coincide with the root system} $E_8$. We have
signatures
\[
\varepsilon_j, \qquad j=1,8.
\]
The corresponding $\Sigma_{\varepsilon_j}$ coincide respectively
with the root systems
\[
D_8\qquad {\rm for}\qquad \varepsilon_1
\]
and
\[
A_1+E_7\qquad {\rm for}\qquad \varepsilon_8.
\]

{\it Let $\Sigma$ coincide with the root system} $F_4$. We have
signatures
\[
\varepsilon_j, \qquad j=1,4.
\]
The corresponding $\Sigma_{\varepsilon_j}$ coincide respectively
with the root system
\[
A_1+C_3\qquad {\rm for}\qquad \varepsilon_1
\]
and with root system
\[
B_4\qquad {\rm for}\qquad \varepsilon_4.
\]

{\it Let a root system $\Sigma$ coincide with the root system}
$G_2$. Then we have only one signature $\varepsilon=1$. The
corresponding $\Sigma_{\varepsilon_j}$ is the root system of
\[
A_1+A_1.
\]

For a f\/ixed $\Sigma$, there can be several corresponding pairs
$(\mathfrak{g}, \mathfrak{k}_\varepsilon)$. Let us give a list of
possible subalgeb\-ras~$\mathfrak{k}_\varepsilon$ for classical
simple real Lie algebras:
\begin{alignat*}{3}
& \mathfrak{g}={\rm sl}(l+1,\mathbb{C}):\quad &&
\mathfrak{k}_\varepsilon={\rm su}(l-j+1,j), \quad 0\leqslant 2j\leqslant l; &\\
& \mathfrak{g}={\rm sl}(l+1,\mathbb{R}): &&
\mathfrak{k}_\varepsilon={\rm so}(l-j+1,j), \quad 0\leqslant 2j\leqslant l; &\\
& \mathfrak{g}={\rm su}^*(2l+2):&&
\mathfrak{k}_\varepsilon={\rm sp}(l-j+1,j), \quad 0\leqslant 2j\leqslant l; &\\
& \mathfrak{g}={\rm su}(l+m,l): && \mathfrak{k}_\varepsilon={\rm
su}(l+m-j,j)+{\rm su}(l-j,j)+{\rm so}(2),\quad m\geqslant 1,
\quad 0 \leqslant j\leqslant l; &\\
& \mathfrak{g}={\rm su}(l,l): && \mathfrak{k}_\varepsilon={\rm
su}(l-j,j)+{\rm su}(l-j,j)+{\rm so}(2), \quad
0\leqslant 2j\leqslant l; &\\
& \mathfrak{g}={\rm su}(l,l): &&
\mathfrak{k}_\varepsilon={\rm sl}(l,\mathbb{C})+\mathbb{R}; &\\
&\mathfrak{g}={\rm so}(2l+1,\mathbb{C}):\quad  &&
\mathfrak{k}_\varepsilon={\rm so}(2l-2j+1,2j), \quad 0\leqslant j\leqslant l; &\\
&\mathfrak{g}={\rm so}(2l,\mathbb{C}):&&
\mathfrak{k}_\varepsilon={\rm so}(2l-2j,2j), \quad 0\leqslant j\leqslant l; &\\
&\mathfrak{g}={\rm so}(2l,\mathbb{C}): &&
\mathfrak{k}_\varepsilon={\rm so}^*(2l); &\\
&\mathfrak{g}={\rm so}(l+m,l):&& \mathfrak{k}_\varepsilon={\rm
so}(l-j+m,j)+{\rm so}(l-j,j), \quad m\geqslant 1,\quad
0\leqslant j\leqslant l; &\\
&\mathfrak{g}={\rm so}(l,l): && \mathfrak{k}_\varepsilon={\rm
so}(l-j,j)+{\rm so}(l-j,j), \quad
0\leqslant 2j\leqslant l;&\\
&\mathfrak{g}={\rm so}(l,l): &&
\mathfrak{k}_\varepsilon={\rm so}(l,\mathbb{C}); &\\
&\mathfrak{g}={\rm so}^*(4l+2):&&
\mathfrak{k}_\varepsilon={\rm u}(2l-2j+1,2j), \quad 0\leqslant 2j\leqslant l;&\\
&\mathfrak{g}={\rm so}^*(4l):&&
\mathfrak{k}_\varepsilon={\rm u}(2l-2j,2j), \quad 0\leqslant 2j\leqslant l;&\\
&\mathfrak{g}={\rm so}^*(4l):&&
\mathfrak{k}_\varepsilon={\rm su}^*(2l)+{\rm so}(1,1);&\\
&\mathfrak{g}={\rm sp}(l,\mathbb{C}):&&
\mathfrak{k}_\varepsilon={\rm sp}(l-j,j), \quad 0\leqslant 2j\leqslant l;&\\
&\mathfrak{g}={\rm sp}(l,\mathbb{C}):&&
\mathfrak{k}_\varepsilon={\rm sp}(l,\mathbb{R});&\\
&\mathfrak{g}={\rm sp}(l,\mathbb{R}):&&
\mathfrak{k}_\varepsilon={\rm u}(l-j,j), \quad 0\leqslant 2j\leqslant l;&\\
&\mathfrak{g}={\rm sp}(l,\mathbb{R}):&&
\mathfrak{k}_\varepsilon={\rm gl}(l,\mathbb{R});&\\
&\mathfrak{g}={\rm sp}(l+m,l):&& \mathfrak{k}_\varepsilon={\rm
sp}(l+m-j,j)+{\rm sp}(l-j,j),\quad
m\geqslant 1,\quad 0\leqslant j\leqslant l;&\\
&\mathfrak{g}={\rm sp}(l,l):&&
\mathfrak{k}_\varepsilon={\rm sp}(l-j,j)+{\rm sp}(l-j,j),\quad 0\leqslant 2j\leqslant l;&\\
&\mathfrak{g}={\rm sp}(l,l): && \mathfrak{k}_\varepsilon={\rm
sp}(l,\mathbb{C}).&
\end{alignat*}

A {\it generalized Cartan decomposition} of the Lie group $G$ with
respect to a pair of subgroups $(K_\varepsilon,K)$ has a form
\begin{equation}
\label{269} G=K_\varepsilon AK
\end{equation}
(see \cite{85}). If
\[
k_1h_1k_1'=k_2h_2k_2', \qquad k_1,k_2\in K_\varepsilon, \qquad
h_1,h_2\in A,\qquad k_1',k_2'\in K,
\]
then
\[
k_1'{k_2'}^{-1}=k_1^{-1}k_2\in K\bigcap K_\varepsilon,\qquad
h_1=(k_1^{-1}k_2)h_2(k_1^{-1}k_2)^{-1}.
\]

\subsection[Generalized Cartan decomposition and
$H$-coordinate systems on $G/K$]{Generalized Cartan decomposition
and $\boldsymbol{H}$-coordinate systems on $\boldsymbol{G/K}$}
\label{Gener-coord}

The decomposition \eqref{269} does not give a unique decomposition
$g=khk'$ of elements $g\in G$ into a product of elements of
$K_\varepsilon$, $A$ and $K$. In order to have a unique
decomposition we use the following procedure. Let
$\mathfrak{a}_\varepsilon^+$ denote the set of all elements $H\in
\mathfrak{a}$ for which $\alpha(H)>0$ for each $\alpha\in
\Sigma_\varepsilon^+$, where $\Sigma_\varepsilon^+$ is the set of
positive roots in $\Sigma_\varepsilon$. Let $A_\varepsilon^+=\exp
\mathfrak{a}_\varepsilon^+$. We denote a set of representatives of
cosets of $K_\varepsilon/M$ by $(K_\varepsilon/M)$. Then almost
each element $g\in G$ decomposes uniquely as a product
\begin{equation}
\label{270} g=yhk,\qquad y\in (K_\varepsilon/M),\qquad h\in
A_\varepsilon^+,\qquad k\in K
\end{equation}
(see \cite{85}). Note that the set $A_\varepsilon^+$ does not
coincide with the set $A^+$ in \eqref{252}. Moreover, the set
$A_\varepsilon^+$ is a union of the sets $wA^+$, where $w$ runs
over a part of elements (including the unit element) of the Weyl
group $W$ of the pair $(\mathfrak{g}, \mathfrak{a})$.

Taking into account the unique decomposition \eqref{270}, we can
state that the space $G/K$ is parametrized almost everywhere by
elements from $(K_\varepsilon/M)A_\varepsilon^+$. A
parametrization of $A_\varepsilon^+$ can be obtained from
\eqref{254}, if we introduce necessary restrictions upon values of
the parameters $t_1,t_2,\dots,t_l$.

Thus, a parametrization of the space $G/K$ (by using the
generalized Cartan decomposition) is reduced to a parametrization
of the space $K_\varepsilon/M$. Generally speaking, it is possible
to choose many coordinate systems on $K_\varepsilon/M$. To each
of such coordinate systems on $K_\varepsilon/M$  there 
corresponds a coordinate system on $G/K$. These coordinate systems
on $G/K$ are called {\it hyperbolic} or {\it H-systems}.

A group $G$ may have many subgroups $K_\varepsilon$. To each of
these subgroups there corresponds its coordinate systems
on $G/K$.

\medskip

\noindent {\bf Remark.} The generalized Cartan decomposition
\eqref{269} can be written in the form $G=KAK_\varepsilon$. This
form of the decomposition can be used for parametrization of the
pseudo-Riemannian symmetric space $G/K_\varepsilon$. This space is
an analogue of the one-sheeted hyperboloid $SO_0(1,4)/SO_0(1,3)\!$
in the 5-dimensional Minkowski space-time. The space
$G/K_\varepsilon$ can be parametrized by elements from
$(K/M)A^+_\varepsilon$; then the corresponding parametrization
gives an analogue of the spherical coordinate system on
$SO_0(1,4)/SO_0(1,3)$.

\subsection[Iwasawa decomposition and $S$-coordinate systems on $G/MN$]{Iwasawa decomposition and $\boldsymbol{S}$-coordinate systems on $\boldsymbol{G/MN}$}
\label{Iw-cone}

Let $G=KAN$ be an Iwasawa decomposition of the group $G$. We
represent elements of $K$ in the form of the product $k= \tilde k
m$, $m\in M$, where $\tilde k$ is an element of $K$ representing a
coset of $K/M$ containing the element $k$. The set of elements
$\tilde k$ is denoted as $(K/M)$. Then the decomposition $G=KAN$
can be written as $G=(K/M)MAN=(K/M)AMN$. Therefore,
\begin{equation}
\label{271} (G/MN)=(K/M)A,
\end{equation}
where $(G/MN)$ is a set of representatives of cosets in $G/MN$.
The relation \eqref{254} gives a~parametrization of elements $h\in
A$ by the parameters $t_1,t_2,\dots,t_l$. Thus, according to
\eqref{271} a parametrization of the space $G/MN$ is reduced to a
parametrization of the space $K/M$. A~procedure of the latter
parametrization was described above. In analogy with the
parametrization of the upper sheet of the cone $C^+_4$ (related to
the group $G=SO_0(1,4)$), we call the coordinate system given by
this parametrization of $G/MN$ {\it spherical} or the {\it
$S$-coordinate system}.

Comparing the relations \eqref{253} and \eqref{271}, we see that
$S$-coordinate systems on $G/K$ and on $G/MN$ almost coincide. A
dif\/ference consists in the fact that for the parametrization of
$G/MN$ the whole subgroup $A$ is used, whereas for the
parametrization of the space $G/K$ we used only its part
$\overline{A^+}$.

In order to obtain a $G$-invariant measure on $G/MN$ we use the
integral relation
\begin{equation}
\label{272} \int_G f(g)dg=\int_K \int_A \int_N f(khn)e^{2\rho
(\log h)} dn\, dh\, dk ,
\end{equation}
where $dg$, $dk$, $dh$ and $dn$ are invariant measures on $G$,
$K$, $A$ and $N$, respectively. If $k=\tilde k m$, $m\in M$,
$\tilde k \in (K/M)$, then
\[
\int_K f(k)dk=\int_{(K/M)} \int_M f(\tilde k m)dm\, d\tilde k,
\]
where $d\tilde k$ is a $K$-invariant measure on $(K/M)$. It
follows from \eqref{272} that for $(G/K)\ni x=\tilde kh\in (K/M)A$
we have
\begin{equation}
\label{273} \int_{G/K} f(x)dx=\int_{(K/M)} \int_A  f(\tilde
kh)e^{2\rho (\log h)} dh\, d\tilde k,
\end{equation}
where $dx$ is a $G$-invariant measure on $G/K$.

\subsection[Gelfand-Naimark-Bruhat decomposition and $T$-coordinate
systems on $G/MN$]{Gelfand--Naimark--Bruhat decomposition\\ and
$\boldsymbol{T}$-coordinate systems on $\boldsymbol{G/MN}$}
\label{GNB-cone}

The decomposition of the Lie algebra $\mathfrak{g}$ into  root
subspaces is of the form
\[
\mathfrak{g}=\mathfrak{m}+\mathfrak{a}+\sum_{\alpha\in \Sigma}
\mathfrak{g}_\alpha ,
\]
where $\Sigma$ is the set of restricted roots of the pair
$(\mathfrak{g},\mathfrak{a})$. We separate here the subspace
$\bar{\mathfrak{n}} =\sum\limits_{\alpha<0} \mathfrak{g}_\alpha$,
where the summation is over all negative restricted roots of the
pair $(\mathfrak{g},\mathfrak{a})$. It is easy to see that
$\bar{\mathfrak{n}}$ is a nilpotent subalgebra in $\mathfrak{g}$.
Besides, $\bar{\mathfrak{n}}$ is obtained from $\mathfrak{n}$ by
acting by the involutive Cartan automorphism $\theta$: $\theta
\mathfrak{n} =\bar{\mathfrak{n}}$. We denote by $\overline{N}$ the
analytical subgroup in $G$ with the Lie algebra
$\bar{\mathfrak{n}}$. The set $\overline{N} AMN$ is everywhere
dense  in $G$ and almost each element $g\in G$ decomposes uniquely
as a product $g=\tilde{n}hmn$, $\tilde{n}\in \bar{N}$, $h\in A$,
$m\in M$, $n\in N$ (see \cite{87}). This decomposition of $G$ is
called the {\it Gelfand--Naimark--Bruhat decomposition}. We have
\begin{equation}
\label{274} (G/MN)=\bar{N} A.
\end{equation}
The equality here is understood in the sense that $\bar{N} A$ is
everywhere dense in $(G/MN)$. The relation \eqref{254} gives a
parametrization of the subgroup $A$ by numbers $t_1,t_2,\dots,
t_l$.

Thus, a parametrization of the space $G/MN$ is reduced to a
parametrization of the subgroup~$\bar N$. This subgroup can be
parametrized by representing $\bar N$ in a form of a product of
one parameter subgroups $\exp \mathfrak{g}_\alpha$. This
parametrization is called {\it translational} ({\it T-system of
coordinates} on $G/MN$). The most interesting case is when the
subgroup $\bar N$ is commutative. It is a~case when $G=SO_0(1,n)$.

As we have seen, the $T$-coordinate system on $G/K$ is introduced
by means of the subgroup $NA$ which dif\/fers from the subgroup
$\bar NA$ in \eqref{274}. However, it can be done that
parametrizations of the corresponding subgroups $N$ and $\bar N$
(by means of which $T$-coordinate systems are introduced on $G/K$
and $G/MN$) will coincide with each other. For this aim we
parametrize the set $G/K$ starting from the Iwasawa decomposition
$G=\bar NAK$ of $G$. Then parametrizations of $G/K$ and $G/MN$ are
fulf\/illed by means of parameters of the same set $\bar N A$.

A $G$-invariant measure $dx$ on $G/MN$ in the $T$-coordinate
system is determined by the relation
\[
\int_{G/MN} f(x) dx=\int_{\bar N} \int_{A} f(\bar n h)e^{2\rho
(\log h)} dh\, d\bar n,
\]
where $x$ represents the element $\bar n h\in \bar N A$, and
$d\bar n$ and $dh$ are invariant measures on $\bar N$ and $A$,
respectively.

\subsection[Gelfand-Naimark-Bruhat decomposition and $O$-coordinate
systems on $G/MN$]{Gelfand--Naimark--Bruhat decomposition\\ and
$\boldsymbol{O}$-coordinate systems on $\boldsymbol{G/MN}$}
\label{GNB-O-cone}

Let $G=\bar N AMN$ be the Gelfand--Naimark--Bruhat decomposition
of the group $G$, where the equality is understood on a dense
subspace in $G$. It can be represented in the form
\begin{equation}
\label{275} G=(\bar{N} M) AMN,
\end{equation}
where $\bar N M$ is a closed subgroup in $G$. This decomposition
is not unique. Since elements of the subgroup $A$ commute with
elements of $M$, then for obtaining a unique decomposition we have
to take a set of representatives $y$ of cosets of $\bar N M/M$
instead of the subgroup $\bar N M$. We denote the set of these
representative by $(\bar N M/M)$.  It follows from \eqref{275}
that $G=(\bar N M/M)AMN$, where the equality is understood on a
dense subspace of $G$. Thus,
\begin{equation}
\label{276} (G/MN)=(\bar{N} M/M)A.
\end{equation}
The relation \eqref{254} gives a parametrization of the subgroup
$A$ by numbers $t_1,t_2,\dots,t_l$. The relation \eqref{276}
reduces a parametrization of the space $G/MN$ to a parametrization
of the space $(\bar{N} M/M)$. A parametrization of the cone
$G/MN$, obtained in this way, is called {\it orispherical} ({\it
$O$-coordinate system}). The $O$-coordinate system on the cone
$C^+_4$ for $G=SO_0(1,4)$ was considered in Section~8.

The most interesting case is when the subgroup $\bar N$ is
commutative, that is, when $G=SO_0(1,n)$. In this case, the
subgroup $\bar N M$ is isomorphic to the group $ISO(n-1)$
(see~\cite{VKII}, Chapter~9).

Comparing the parametrizations \eqref{264} and \eqref{276} of the
spaces $G/K$ and $G/MN$, we see that they dif\/fer by appearing
dif\/ferent nilpotent subgroups $N$ and $\bar N$. Replacing the
Iwasawa decomposition $G=NAK$ by the decomposition $G=\bar N AK$
we obtain the relation
\begin{equation}
\label{277} (G/K)=(\bar{N} M/M)A
\end{equation}
instead of the relation \eqref{264}. Then the spaces $G/K$ and
$G/MN$ are parametrized by the same sets.

A $G$-invariant measure $dx$ on $G/MN$ in the $O$-coordinate
system is determined by the integral relation
\[
\int_{G/MN} f(x)dx=\int_{(\bar N M/M)} \int_A f(yh)e^{2\rho (\log
h)} dh\, dy,
\]
where $dy$ is an $\bar N M$-invariant measure on $(\bar N M/M)$.

\subsection{Generalized Iwasawa decomposition}
\label{Iw-g-cone}
The generalized Iwasawa decompositions are constructed by means of
subgroups of $G$ used for construction of the generalized Cartan
decomposition of $G$.

Let $M^*$ be a normalizer of the subgroup $A$ in $K$. We set
$M^*_\varepsilon=K_\varepsilon \bigcap M^*$ and
$W_\varepsilon=M^*_\varepsilon/M$. Then $W_\varepsilon$ is a
symmetry group for the system of roots $\Sigma_\varepsilon$ (see
\cite{85}).

Elements of the quotient space $W_\varepsilon\backslash  W$ can be
represented by elements $w_1\equiv 1, w_2,\dots,w_r$ of the
subgroup $M^*$ which do not belong to $M^*_\varepsilon$. The
integer $r$ is equal to the order of the quotient space
$W_\varepsilon\backslash  W$.

We create the sets $K_\varepsilon w_i AN$, $i=1,2,\dots,r$. Then

\begin{enumerate}\itemsep=0pt

\item[(a)] if $kw_ihn=k'w_jh'n'$ with $k,k'\in K_\varepsilon$,
$h,h'\in A$, $n,n'\in N$, then $k=k'$, $i=j$, $h=h'$, $n=n'$;

\item[(b)] the mapping $(k,h,n)\to kw_ihn$ is an analytic
dif\/feomorphism of the manifold $K_\varepsilon\times A\times N$
onto $K_\varepsilon w_iAN$ ($i=1,2,\dots,r$);

\item[(c)] the set $\bigcup_{i=1}^r K_\varepsilon w_i AN$ is open
and everywhere dense in $G$ (see \cite{85}). Thus,
\begin{equation}
\label{278} G=\bigcup_{i=1}^r K_\varepsilon w_i AN .
\end{equation}
This equality is true on a dense manifold of $G$.
\end{enumerate}

According to these assertions, almost each element $g\in G$ can be
uniquely decomposed into a product
\[
g=kw_i hn,\qquad k\in K_\varepsilon,\qquad 1\leqslant i\leqslant
r,\qquad h\in A,\qquad n\in N.
\]
The equality \eqref{278} is called a {\it generalized Iwasawa
decomposition} of the group $G$. This decomposition is written for
those pseudo-Riemannian symmetric spaces $G/K_\varepsilon$, which
are associated with the corresponding signatures $\varepsilon$ of
the root system $\Sigma$. These symmetric spaces do not exhaust
all pseudo-Riemannian symmetric spaces (see \cite{84}). However,
the decomposition \eqref{278} can be generalized to any
pseudo-Riemannian symmetric space \cite{88}. We consider the
decomposition \eqref{278} only for spaces determined by signatures
$\varepsilon$ of the root system $\Sigma$.

Integral relations for the usual Iwasawa decomposition of the
group $G$ can be generalized to generalized Iwasawa
decompositions. We have
\begin{equation}
\label{279} \int_G f(g)dg=\sum_{i=1}^r \int_{K_\varepsilon}\int_A
\int_N f(kw_i h n)e^{2\rho(\log h)} dn\, dh\, dk,
\end{equation}
where $f$ is a continuous function on $G$ with a compact support,
and $dg$, $dk$, $dh$, $dn$ are invariant measures on $G$,
$K_\varepsilon$, $A$ and $N$, respectively. This relation is true
for an appropriate normalization of the measures. In another case,
the right hand side must be multiplied by a constant.

\subsection[Generalized Iwasawa decomposition and $H$-coordinate systems on
$G/MN$]{Generalized Iwasawa decomposition\\ and
$\boldsymbol{H}$-coordinate systems on $\boldsymbol{G/MN}$}

\label{Iw-coord-cone}
Let us write down the relation \eqref{278} in the form
\[
G=\bigcup_{i=1}^r K_\varepsilon w_i AMN ,
\]
where the equality is understood on a dense manifold in $G$. Since
$AM=MA$ and $w_iMw_i^{-1}=M$,  it can be represented as
\begin{equation}
\label{280} G=\bigcup_{i=1}^r (K_\varepsilon/M) w_i AMN ,
\end{equation}
where $(K_\varepsilon/M)$ is a set of representatives of cosets of
the quotient space $K_\varepsilon/M$. The decomposition
\eqref{280} obeys the following property: If
\begin{gather*}
yw_ihmn=y'w_jh'm'n', \qquad y,y'\in (K_\varepsilon/M), \qquad
h,h'\in A,\qquad m,m'\in M,\qquad n,n'\in N,
\end{gather*}
then $y=y'$, $i=j$, $h=h'$, $m=m'$, $n=n'$. By other words, almost
every element $g\in G$ can be uniquely decomposed as a product
\[
g=yw_ihmn,\qquad y\in (K_\varepsilon/M),\qquad 1\leqslant
i\leqslant r,\qquad h\in A,\qquad m\in M,\qquad n\in N.
\]
We obtain from the decomposition \eqref{280} that
\begin{equation}
\label{281} (G/MN)=\bigcup_{i=1}^r (K_\varepsilon/M) w_i A ,
\end{equation}
where the equality is understood on a dense (in $G/MN$) set. The
relation \eqref{254} gives a~para\-met\-rization of the subgroup
$A$ by the numbers $t_1,t_2,\dots, t_l$. Therefore, the relation
\eqref{281} reduces a parametrization of the space $G/MN$ to a
parametrization of the quotient space $K_\varepsilon/M$.
A~parametrization of the cone $G/MN$, obtained in this way, is
called {\it hyperbolic}. The correspon\-ding coordinate system is
called {\it H-system} of coordinates. Clearly, to
dif\/ferent subgroups~$K_\varepsilon$ there corresponds
dif\/ferent coordinate systems on~$G/MN$.

\medskip

\noindent {\bf Remark.} The generalized Iwasawa decomposition
\eqref{278} can be written as
\[
G=\bigcup_{i=1}^r NA w_i K_\varepsilon .
\]
This relation can be used for parametrization of the
pseudo-Riemannian symmetric space $G/K_\varepsilon$ by elements of
the set $NA\bigcup NAw_2\bigcup \cdots \bigcup NAw_r$. As a
result, one obtains a $T$-coordinate system on $G/K_\varepsilon$.

\section[Spectra of quasi-regular representations of $G$
on functions on $G/K$ and on $G/MN$]{Spectra of quasi-regular representations of $\boldsymbol{G}$\\
on functions on $\boldsymbol{G/K}$ and on $\boldsymbol{G/MN}$}
\label{Quasir}

\subsection[Quasi-regular representations of $G$]{Quasi-regular representations of $\boldsymbol{G}$}
\label{Quasi}

We consider the Hilbert space $L^2(G/K)$ of functions on the
hyperboloid $G/K$ and the Hilbert space $L^2(G/MN)$ of functions
on the cone $G/MN$ with respect to $G$-invariant measures on~$G/K$
and $G/MN$, respectively. The group $G$ acts on $G/K$ and $G/MN$
as a transitive motion group. To each element $g\in G$ there 
corresponds transformation $x\to g^{-1}x$ of points of $G/K$ and
of points of $G/MN$. If some coordinate system is introduced on
$G/K$ (on $G/MN$) then this transform can be written in terms of
the corresponding coordinates.

The transform $x\to g^{-1}x$ on $G/K$ and on $G/MN$ gives a
possibility to introduce unitary representations on $L^2(G/K)$ and
on $L^2(G/MN)$. These representations are given by the formula
\begin{gather}
\label{282} \pi(g)f(x)=f(g^{-1}x),\qquad x\in G/K \quad {\rm
or}\quad G/MN,
\end{gather}
and are called {\it quasi-regular}. The quasi-regular
representation on $G/K$ will be denoted by $\pi^H$ and on $G/MN$
by $\pi^C$.

One of the main problems of our paper is to construct orthogonal
bases on the Hilbert spaces $L^2(G/K)$ and $L^2(G/MN)$ for each
coordinate system introduced above. To each coordinate system
there corresponds a chain of subgroups of $G$:
\medskip

$S$-system: $G\supset K\supset \cdots$,
\medskip

$T$-system: $G\supset \bar{N}\supset \cdots$,
\medskip

$O$-system: $G\supset \bar{N}M\supset \cdots$,
\medskip

$H$-system: $G\supset K_\varepsilon\supset \cdots$.
\medskip

\noindent There are several choices for each of these chains. An
explicit form of other subgroups gives dif\/ferent varieties of a
f\/ixed type of coordinate systems. We do not consider them. As in
the case of the group $G=SO_0(1,4)$, most interesting bases of
$L^2(G/K)$ and $L^2(G/MN)$ are bases with separated variables.
Irreducible representations of the group $G$ and of its subgroups
are realized on parts of the corresponding bases. These basis
functions are eigenfunctions of operators invariant with respect
to the group $G$ and its subgroups associated with the
corresponding coordinate system (see the case $G=SO_0(1,4)$).
Eigenvalues of these operators (and also other indices if
invariant operators have multiple eigenvalues) characterize basis
functions. Therefore, we have to know a list of irreducible
representations which are contained in the quasi-regular
representations on $L^2(G/K)$ and on $L^2(G/MN)$ (these
representations determine eigenvalues of the corresponding
invariant operators). We also have to f\/ind decompositions of the
irreducible representations of $G$, which are contained in the
decomposition of the quasi-regular representations, into
irreducible representations of the corresponding subgroups. These
representations of subgroups determine a part of indices
characterizing basis functions.

\subsection[Decomposition of quasi-regular representation $\pi^H$]{Decomposition of quasi-regular representation $\boldsymbol{\pi^H}$}
\label{Quasi-H}

Let $\mathcal{F}$ be a space of real linear forms on the
subalgebra $\mathfrak{a}$ of the Lie algebra $\mathfrak{g}$. The
subset of~$\mathcal{F}$, consisting of linear form $\lambda$ for
which $(\lambda,\alpha_i)\geqslant 0$ for each simple restricted
root $\alpha_i$ of the pair $(\mathfrak{g},\mathfrak{a})$, is
denoted by $\mathcal{F}^+$. Let $\pi_{\delta\lambda}$ be a
principal nonunitary series representation of the group $G$ from
section 10. We are interested in representations
$\pi_{\delta\lambda}$ of class~1 with respect to the subgroup~$K$
(that is, in representations $\pi_{\delta\lambda}$, which contain
a trivial representation of~$K$). A restriction
$\pi_{\delta\lambda}\vert_K$ of the representation
$\pi_{\delta\lambda}$ upon the subgroup~$K$ contains an
irreducible representation $\omega$ of this subgroup if and only
if a restriction of $\omega$ upon the subgroup $M$ contains the
representation $\delta$ of $M$ (see \cite{War}). Therefore, the
representation $\pi_{\delta\lambda}$ is of class 1 with respect to
$K$ if and only if $\delta$ is the trivial representation of the
subgroup $M$. Such representations $\pi_{\delta\lambda}$ will be
denoted by $\pi_{\lambda}$. The representations $\pi_{\lambda}$,
for which $\lambda+\rho$ (where $\rho$ is the half-sum of positive
restricted roots) are pure imaginary linear forms, are unitary
(see, for example, \cite{War}). They constitute class 1
representations of the principal unitary series of the group $G$.
These representations are irreducible. In the class of the unitary
representations $\pi_{\lambda}$, $\lambda\in {\rm
i}\mathcal{F}-\rho$, equivalence relations exist. Namely,
representations $\pi_{\lambda}$ and $\pi_{\lambda'}$ are
equivalent if and only if $\lambda'=w(\lambda+\rho)-\rho$, $w\in
W$. In order to obtain non-equivalent representations we have to
restrict ourselves by the representations $\pi_{\lambda}$,
$\lambda\in {\rm i}\mathcal{F}^+-\rho$.

Let $g\in G$ and $k\in K$. Then for the element $g^{-1}k$ we have
a decomposition
\[
g^{-1}k=k_ghn,\qquad k_g\in K,\quad h\in A,\quad n\in N
\]
(the Iwasawa decomposition). We represent the element $h$ in this
decomposition in the form
\[
h=\exp H(g^{-1}k),\qquad H(g^{-1}k)\in \mathfrak{a}.
\]
By means of this element we introduce a Fourier transform of
functions on $G/K$. If $f\in C^\infty_c(G/K)$, where $
C^\infty_c(G/K)$ is the set of inf\/initely dif\/ferentiable
functions on $G/K$ with compact support, then $f$ can be
considered as a function on $G$ which is constant on cosets from
$G/K$. A~Fourier transform of the function $f$ is a function on
$K/M\times \mathcal{F}$ determined by
\begin{equation}
\label{283} \tilde f (\tilde k, \nu)=\int_G f(g)\exp \{ (-{\rm
i}\nu-\rho) [H(g^{-1}k)]\} dg,\qquad \nu\in \mathcal{F},
\end{equation}
where $\tilde k$ is the coset in $K/M$ which contains the element
$k\in K$. The function $\tilde f (\tilde k, \nu)$ can be
represented as a function on $K$ constant on cosets of $K/M$. Let
us prove that if $f(g)$ transforms under action of $G$ by
quasi-regular representation $\pi^H$ of $G$, then $\tilde f
(\tilde k, \nu)$ transforms under the representation
$\pi_\lambda\equiv \pi_{-{\rm i}\nu-\rho}$ of the principal
unitary series. Indeed, if $f(g)$ is replaced by
$F(g):=\pi^H(g_0)f(g)= f(g_0^{-1}g)$, then the function
\eqref{283} is replaced by
\begin{equation} \label{283'}
\int_G f(g_0^{-1}g)\exp \{ (-{\rm i}\nu-\rho) [H(g^{-1}k)]\} dg
=\int_G f(g)\exp \{ (-{\rm i}\nu-\rho) [H(g^{-1}g_0^{-1}k)]\} dg .
\end{equation}

Let us analyze the expression $H(g^{-1}g_0^{-1}k)$. Since
\[
g_0^{-1}k=k_{g_0}hn,\qquad g^{-1}k_{g_0}=(k_{g_0})_gh'n',
\]
then
\[
g^{-1}g_0^{-1}k=(k_{g_0})_gh'h n''n=(k_{g_0})_g\exp
[H(g^{-1}k_{g_0})] \exp [H(g_0^{-1}k)]n''n.
\]
Therefore,
\[
H(g^{-1}g_0^{-1}k)= H(g^{-1}k_{g_0})+ H(g_0^{-1}k).
\]
Then
\[
\exp \{ (-{\rm i}\nu-\rho) [H(g^{-1}g_0^{-1}k)]\} = \exp \{ (-{\rm
i}\nu-\rho) [H(g^{-1}k_{g_0})]\} \exp \{ (-{\rm i}\nu-\rho)
[H(g_0^{-1}k)]\}.
\]
Substituting this expression into \eqref{283'}, we obtain that
under the transition
\[
f(g)\to \pi^H(g_0)f(g)\equiv f(g_0^{-1}g)
\]
the function $\tilde f (\tilde k, \nu)$ turns into the function
\begin{gather*}
\exp \{ (-{\rm i}\nu-\rho) [H(g_0^{-1}k)]\} \int_G f(g) \exp \{
(-{\rm i}\nu-\rho) [H(g^{-1}k_{g_0})]\} dg
\\
\qquad{} =\exp \{ (-{\rm i}\nu-\rho) [H(g_0^{-1}k)]\} \tilde f
(\tilde k_{g_0},\nu) =\pi_{-{\rm i}\nu-\rho} (g_0)\tilde f (\tilde
k, \nu).
\end{gather*}
This proves our assertion.

The transform \eqref{283} at f\/ixed $\nu$ is called a {\it
Poisson transform} on $G/K$ (see \cite{Hel-94}). The
function~$f(g)$ from \eqref{283} can be restored by means of the
function $\tilde f (\tilde k, \nu)$ as \cite{Hel-94}
\begin{equation}
\label{284} f(g)=|W|^{-1} \int_\mathcal{F} \int_{K/M} \tilde f
(\tilde k, \nu) \exp \{ ({\rm i}\nu-\rho) [H(g^{-1}k)]\}
|c(\nu)|^{-2}d\tilde k\, d\nu .
\end{equation}
Moreover, the Plancherel formula
\begin{equation}
\label{285} \int_{G/K} |f(\tilde g)|^2 d\tilde g=|W|^{-1}
\int_\mathcal{F} \int_{K/M} |\tilde f (\tilde k, \nu)|^2
|c(\nu)|^{-2} d\tilde k\, d\nu
\end{equation}
holds. In \eqref{284} and \eqref{285}, $|W|$ means an order of the
Weyl group $W$, $d\tilde g$ is a $G$-invariant measure on $G/K$,
$d\tilde k$ is a $K$-invariant measure on $K/M$, and $d\nu$ is the
Lebesgue measure on $\mathcal{F}$. The multiplier $c(\nu)$ in the
Plancherel measure $|c(\nu)|^{-2}d\nu$ is determined by the
formula
\begin{equation}
\label{286} c(\nu)= \int_{\bar N} \exp \{ (-{\rm i}\nu-\rho)
[H(\bar n)]\} d\bar n =\frac{I({\rm i}\nu)}{I(\rho)}.
\end{equation}
Here $H(\bar n)$ is determined by the Iwasawa decomposition $\bar
n=k\exp [H(\bar n)]n$, $k\in K$, $n\in N$, $H(\bar n)\in
\mathfrak{a}$, of the element $\bar n \in \bar N$, and $d\bar n$
is the normalized measure on $\bar N$ (see \cite{War}). For
$I(\nu)$ the following formula holds:
\begin{equation}
\label{287} I(\nu)= \prod_{\alpha\in \Sigma^+} B\left(
\frac{m(\alpha)}2, \frac{m(\alpha/2)}4 + \frac{\langle\nu,\alpha
\rangle}{ \langle\alpha,\alpha \rangle}\right) ,
\end{equation}
where $\Sigma^+$ is the set of positive restricted roots of the
pair $(\mathfrak{g},\mathfrak{a})$ (without taking into account
multiplicities), $m(\alpha)$ is a multiplicity of the root
$\alpha$, and $B(.,.)$ is the beta-function. The formulas
\eqref{286} and \eqref{287} determine the Plancherel measure in
\eqref{284} and \eqref{285}. A proof of formulas
\eqref{284}--\eqref{287} can be found in \cite{War}. Note that
formulas \eqref{283}--\eqref{285} demand  certain normalization of
the measures $dg$, $d\tilde k$ and $d\nu$. A violation of this
normalization leads to multiplication of expressions by constants.

The function $\tilde f(\tilde k,\nu)$ on $K/M\times \mathcal{F}$
satisf\/ies the additional condition (see \cite{89})
\[
\int_\mathcal{F} \int_{K/M} |\tilde f(\tilde k,\nu)|^2
|c(\nu)|^{-2}
 d\tilde k\, d\nu =
\int_\mathcal{F} \int_{K/M} |\tilde f(\tilde k,w\nu)|^2
|c(\nu)|^{-2}
 d\tilde k\, d\nu ,
\]
where $w$ is any element of the Weyl group $W$. Hence, the
relation \eqref{285} can be written in the form
\begin{equation}
\label{288} \int_{G/K} |f(\tilde g)|^2 d\tilde g=
\int_\mathcal{F^+} \int_{K/M} |\tilde f (\tilde k, \nu)|^2
|c(\nu)|^{-2}  d\tilde k\, d\nu.
\end{equation}

Formulas \eqref{283}, \eqref{284} and \eqref{288} can be continued
onto the spaces $L^2(G/K)$ and $L^2(K/M\times \mathcal{F}_+)$ (the
latter space is taken with respect to the measure $|c(\nu)|^{-2}
d\tilde k\, d\nu$). Then they give a~decomposition of the
quasi-regular representation $\pi^H$ of the group $G$ into
irreducible unitary constituents. According to this decomposition
we may state that {\it the representation $\pi^H$ of $G$
decomposes into a direct integral of the unitary representations
$\pi_\lambda\equiv \pi_{{\rm i}\nu-\rho}$, $\nu\in \mathcal{F}^+$,
of $G$ and each of these representations appears in the
decomposition only once:}
\[
\pi^H=\int_{\mathcal{F}^+} \oplus \pi_{{\rm i}\nu-\rho}
|c(\nu)|^{-2} d\nu.
\]

In order to separate in $L^2(G/K)$ those spaces, on which
irreducible representations $\pi_{{\rm i}\nu-\rho}$ are realized,
we have to take by means of the integral transform (see formula
\eqref{284})
\begin{equation}
\label{289} \int_{K/M} \tilde f (\tilde k, \nu) \exp \{ ({\rm
i}\nu-\rho) |H(g^{-1}k)]\} d\tilde k
\end{equation}
an image in $L^2(G/K)$ of the space $L^2(K/M)\equiv L^2_0(K)$, on
which $\pi_{{\rm i}\nu-\rho}$ is realized.

\subsection[Decomposition of quasi-regular representation $\pi^C$]{Decomposition of quasi-regular representation $\boldsymbol{\pi^C}$}
\label{Quasi-pi}

The problem of  decomposition of the quasi-regular representation
$\pi^C$ into irreducible representations of $G$ can be solved by
means of the usual Fourier transform on the commutative group~$A$.
Let $C^\infty_c(G/MN)$ be the space of inf\/initely
dif\/ferentiable functions on $G/MN$ with a compact support. This
space is everywhere dense in $L^2(G/MN)$. Functions $f\in
C^\infty_c(G/MN)$ are considered as functions $f(\tilde k,h)$ on
$(K/M)A$ (see subsection 11.7). For these functions the following
transform can be constructed:
\begin{equation}
\label{290} \tilde f (\tilde k,\nu)=\int_A f(\tilde k,h) \exp \{
({\rm i}\nu+\rho)(\log h)\} dh,\ \ \ \ \nu\in \mathcal{F}.
\end{equation}
It is easy to see that
\begin{equation}
\label{291} f(\tilde k,h)= \int_{\mathcal{F}}\tilde f (\tilde
k,\nu) \exp \{ (-{\rm i}\nu-\rho)(\log h)\} d\nu,
\end{equation}
and the Plancherel formula
\begin{equation}\label{292}
\int_{G/MN} |f(x)|^2dx\equiv \int_{K/M} d\tilde k \int_A |f(\tilde
k,h)|^2 e^{ 2\rho (\log h)} dh =\int_{K/M} d\tilde k
\int_{\mathcal{F}} |\tilde f (\tilde k,\nu)|^2 d\nu
\end{equation}
holds.

Formulas \eqref{290}--\eqref{292} can be continued to the spaces
$L^2(G/MN)$ and $L^2(K/M\times \mathcal{F})$. Using the formula
\eqref{290} it is easy to show that if a function $f(\tilde
k,h)\in L^2(G/MN)$ transforms under the representation $\pi^C$,
then the function $\tilde f (\tilde k,\nu)$ at each f\/ixed $\nu$
transforms under the irreducible unitary representation
$\pi_{-{\rm i}\nu-\rho}$ of the group $G$. Since $f(\tilde k,h)$,
as a function on $A$, does not satisfy additional conditions, then
$\tilde f (\tilde k,\nu)$, as a function on $\mathcal{F}$, also
does not satisfy any additional conditions. Formulas
\eqref{290}--\eqref{292} give a decomposition of the
representation $\pi^C$ of the group $G$ into irreducible
representations $\pi_{{\rm i}\nu-\rho}$ of $G$:
\[
\pi^C =\int_\mathcal{F} \oplus \pi_{{\rm i}\nu-\rho} d\nu.
\]
The representations $\pi_{{\rm i}\nu-\rho}$ and $\pi_{w({\rm
i}\nu)-\rho}$ are unitary equivalent for any $w\in W$ and other
equivalence relations in the set of the representations $\pi_{{\rm
i}\nu-\rho}$ do not exist. Hence, {\it the representation~$\pi^C$
decomposes into a direct integral of all irreducible class $1$
representations of the principal unitary series, and each of these
representations is contained in the decomposition $|W|$ times,
where $|W|$ is an order of the group $W$.}

In order to separate in $L^2(G/MN)$ a subspace of functions which
are transformed under the irreducible representation $\pi_{{\rm
i}\nu-\rho}$ it is necessary to make the integral transform
\begin{equation}
\label{293} \hat f (\tilde k, h;\nu)=\int_A
 f (\tilde k,hh')e^{-({\rm i}\nu-\rho)(\log h')} dh'.
\end{equation}
(Speaking in a mathematically strict way, this space can be
separated not in $L^2(G/MN)$, but in the space of inf\/initely
dif\/ferential functions on $G/MN$.) The functions \eqref{293} are
homogeneous in~$h$ of homogeneity degree $({\rm i}\nu-\rho)$. It
is easy to show that $\hat f (\tilde k, h;\nu)$ are functions on
the cone $G/MN\sim (K/M)A$ and are transformed under the
representation $\pi_{{\rm i}\nu-\rho}$.

Let us consider other spectral problems, in particular,
restrictions of representations $\pi_\lambda$ of the group $G$
onto the subgroups $K$, $\bar N$, $\bar N M$, $K_\varepsilon$, and
decompositions of these restrictions into irreducible
constituents.

\subsection[Restriction of representations $\pi_\lambda$ onto $K$]{Restriction of representations $\boldsymbol{\pi_\lambda}$ onto $\boldsymbol{K}$}
\label{Restr-K}

A  multiplicity of an irreducible unitary representation
$\omega$ of the subgroup $K$ in $\pi_\lambda
{\downarrow}_K$  coincides with
a multiplicity of the unit (identity) representation of the subgroup
$M$ in $\omega {\downarrow}_M$. This statement solves completely
the problem of restriction of $\pi_\lambda$ onto $K$. It is known
\cite{89} that if the group~$G$ is of real rank 1 (that is, $G$
coincides with one of the groups $SU(1,n)$, $SO_0(1,n)$,
$Sp(1,n)$, $F_{4(-20)}$), then the restriction $\pi_\lambda
\vert_K$ decomposes into a sum of all irreducible representations
of $K$ which are of class 1 with respect to $M$ and their
multiplicities in the decomposition are equal to 1.

\subsection[Restriction of representations $\pi_\lambda$ onto $\bar N$]{Restriction of representations $\boldsymbol{\pi_\lambda}$ onto $\boldsymbol{\bar N}$}
\label{Restr-barN}

The representation $\pi_\lambda$ can be realized as follows. We
take functions $f$ on $G$ satisfying the condition
\begin{equation}
\label{294} f (g)=f (xmnh)=\exp (-\lambda(\log h)) f(x),
\end{equation}
where $x$ are representatives of cosets of $G/MNA$. We construct a
Hilbert space of such functions with the norm
\begin{equation}
\label{295} \Vert f\Vert^2= \int_K |f(k)|^2 dk.
\end{equation}
The representation $\pi_\lambda$ is realized in this space by the
formula
\[
\pi_\lambda(g_0) f(g)=f(g_0^{-1}g),\qquad g_0\in G.
\]
Functions $f$ satisfying the condition \eqref{294} are determined
uniquely by their values on representatives $x$. Therefore,
$\pi_\lambda$ can be realized on functions $f(x)$. In order to
give an action of the operators $\pi_\lambda(g)$ on functions
$f(x)$ we take into account that $\pi_\lambda(g) f(x)=f(g^{-1}x)$
and decompose $g^{-1}x$ into a product of elements of $X$, $M$,
$N$ and $A$ (where $X$ denotes the set of representatives $x$):
\begin{equation}
\label{296} g^{-1}x=x_gm'n'h_g.
\end{equation}
Since $f(g^{-1}x)=f(x_gm'n'h_g)=\exp (-\lambda (\log h_g))f(x_g)$,
then
\begin{equation}
\label{297} \pi_\lambda(g)f(x)=\exp (-\lambda (\log h_g))f(x_g),
\end{equation}
where $h_g$ and $x_g$ are determined by \eqref{296}. When we realize  
the representation $\pi_\lambda$ on the space
of functions $f(x)$ it is desirable to have an expression for a
norm in this space expressed by means of an integral over $X$.

According to the Gelfand--Naimark--Bruhat decomposition of the
group $G$, almost each element $g\in G$ can be decomposed in the
form
\[
g=\bar n hmn= \bar n mn'h,\qquad \bar n \in \bar N,\quad m\in
M,\quad n'\in N,\quad h\in A.
\]
Hence, the subgroup $\bar N$ can be taken as a set $X$ of
representatives $x$ of cosets of $G/MNA$. An action formula for
operators $\pi_\lambda(g)$, $g\in G$, upon functions $f(\bar n)$
can be obtained in each concrete case by means of formulas
\eqref{296} and \eqref{297}. In this space we have the norm
\begin{equation}
\label{298} \Vert f\Vert^2=\int_{\bar N} \vert f(\bar n)\vert
^2d\bar n ,
\end{equation}
where $d\bar n$ is an invariant measure on $\bar N$, instead of
the norm \eqref{295}. The representation $\pi_{{\rm i}\nu-\rho}$,
$\nu\in \mathcal{F}$, of the principal unitary series, realized on
the Hilbert space $L^2(\bar N)$ (with the norm \eqref{298}), is
unitary and is unitary equivalent to the corresponding
representation in the Hilbert space with the norm \eqref{295}.

Restrict the representation $\pi_\lambda$ of the group $G$,
realized on the space $L^2(\bar N)$, upon the subgroup $\bar N$.
It is easy to see that the operators $\pi_\lambda(\bar n_0)$,
$\bar n_0 \in \bar N$, act on $L^2(\bar N)$ as
\[
\pi_\lambda(\bar n_0) f(\bar n)=f(\bar n_0^{-1}\bar n).
\]
Thus, a restriction of the representation $\pi_\lambda$ upon the
subgroup $\bar N$ is unitary equivalent to the regular
representation of this subgroup. We conclude that a decomposition
of regular representation of $\bar N$ gives a decomposition of the
representation ${\pi_\lambda}{\downarrow_{\bar N}}$ into
irreducible constituents.

In particular, if $G=SO_0(1,n)$, then $\bar N$ is a commutative
group isomorphic to the group $T_{n-1}$ of real translations of
dimension $n-1$ in $\mathbb{R}^{n-1}$. If
$(x_1,x_2,\dots,x_{n-1})\in T_{n-1}$, then an irreducible unitary
representation of $T_{n-1}$ can be written in the form
\[
x=(x_1,x_2,\dots,x_{n-1})\to e^{{\rm i}(x\cdot p)},\qquad
p=(p_1,p_2,\dots,p_{n-1}), \qquad p_i\in \mathbb{R}.
\]
Here $x\cdot p=x_1p_1+x_2p_2+\cdots +x_{n-1}p_{n-1}$. The regular
representation of the group $\bar N=T_{n-1}$ decomposes into a
direct integral of all these irreducible unitary representations
and each of these representations is contained in the
decomposition once. The restriction of the representation
$\pi_\lambda$ of $G$ upon the subgroup $\bar N=T_{n-1}$ decomposes
into the same direct integral.

\subsection[Restriction of representations $\pi_\lambda$ onto $\bar N M$]{Restriction of representations $\boldsymbol{\pi_\lambda}$ onto $\boldsymbol{\bar N M}$}
\label{Restr-barNM}

We represent the group $G$ in the form $G=(\bar N M/M)AMN$. Almost
each element $g\in G$ decomposes uniquely as a product of elements
of $(\bar N M/M)$, $A$, $M$ and $N$. Therefore, the representation
$\pi_\lambda$ of $G$ can be realized on the space of functions
given on the set $(\bar N M/M)$. This set is homeomorphic to the
subgroup $\bar N$. Hence, a norm on this space can be given by the
formula \eqref{298}. According to formulas \eqref{296} and
\eqref{297}, a restriction of the representation~$\pi_\lambda$
upon $\bar N M$ acts upon the functions $f(x)$, $x\in (\bar N
M/M)$ by the formula
\[
\pi_\lambda(s)f(x)=f(s^{-1}x),\qquad  s\in \bar N M.
\]
Therefore, a restriction of the representation $\pi_\lambda$ of
the group $G$ upon the subgroup $\bar N M$ is unitary equivalent
to the quasi-regular representation of this subgroup, realized on
the homogeneous space $\bar N M/M$.

In particular, if $G=SO_0(1,n)$, then $\bar N M\sim ISO(n-1)$. The
quasi-regular representation of the group $ISO(n-1)$ on the space
$L^2(ISO(n-1)/SO(n-1))$ decomposes into a direct integral of all
unitary irreducible representations of $ISO(n-1)$ of class 1 with
respect to the subgroup $M\sim SO(n-1)$ and each of these
representations of $ISO(n-1)$ is contained in the decomposition
only once (see, for example, \cite{VKII}). These representations
are given by a real positive number.

\subsection[Restriction of representations $\pi_\lambda$ onto $K_\varepsilon$]{Restriction of representations $\boldsymbol{\pi_\lambda}$ onto $\boldsymbol{K_\varepsilon}$}
\label{Restr-Keps}

The generalized Iwasawa decomposition
\[
G=\bigcup_{i=1}^r K_\varepsilon w_i AN,
\]
where the equality is understood on a dense subspace of $G$, can
be written in the form
\begin{equation}
\label{299} G=\bigcup_{i=1}^r (K_\varepsilon/M) w_i MAN.
\end{equation}
Almost each element of $G$ decomposes as a product of elements of
$(K_\varepsilon/M) w_i$ ($i=1,2,\dots,r$), $M$, $A$, $N$. Then
elements of the set
\begin{equation}
\label{300} X=\bigcup_{i=1}^r (K_\varepsilon/M) w_i
\end{equation}
can be taken as representatives of cosets of $G/MAN$. Thus, the
representation $\pi_\lambda$ of the group~$G$ can be realized on a
space of functions given on $X$. The representation $\pi_{{\rm
i}\nu-\rho}$, $\nu\in \mathcal{F}$, realized on the Hilbert space
$L^2_0(K)$ of invariant (with respect to $M$) functions on $K$
with the norm \eqref{295}, is unitary equivalent to the
representation $\pi_{{\rm i}\nu-\rho}$ realized by formula
\eqref{297} on the Hilbert space $L^2_0(X)$ of functions on the
set \eqref{300} with the scalar product
\[
\langle f,f' \rangle=\sum_{i=1}^r \int_{K_\varepsilon}
f(k_\varepsilon w_i)\overline{f'(k_\varepsilon
w_i)}dk_\varepsilon,
\]
where $dk_\varepsilon$ is an invariant measure on $K_\varepsilon$.
It follows from \eqref{296} and \eqref{297} that a restriction of
the representation $\pi_\lambda$ of $G$ upon the subgroup
$K_\varepsilon$ acts on functions $f$ given on $X$ by the formula
\begin{equation}
\label{301} \pi_\lambda(k^0_\varepsilon)f(k_\varepsilon w_i)=
f((k^0_\varepsilon)^{-1}k_\varepsilon w_i),\qquad
k^0_\varepsilon\in K_\varepsilon.
\end{equation}

Thus, a restriction of the representation $\pi_\lambda$ of $G$
upon the subgroup $K_\varepsilon$ is unitary equivalent to the
orthogonal sum of $r$ copies of the quasi-regular representation
of $K_\varepsilon$ on the space of functions on $K_\varepsilon/M$.
Since $M$ is a compact subgroup in $K_\varepsilon$, then the
decomposition of the quasi-regular representation of
$K_\varepsilon$ into irreducible constituents can be easily
obtained from the decomposition into irreducible constituents of
the regular representation of $K_\varepsilon$ (see~\cite{90}).

In particular, if $G=SO_0(1,n)$, then $M=SO(n-1)$. One of the
possibilities for a subgroup~$K_\varepsilon$ is $SO_0(1,n-1)$. We
have $M\subset K_\varepsilon$ and $r=2$. A restriction of the
representation~$\pi_\lambda$ of the group $SO_0(1,n)$ upon the
subgroup $K_\varepsilon=SO_0(1,n-1)$ is unitary equivalent to the
orthogonal sum of two quasi-regular representations of
$K_\varepsilon$ on the space of functions on
$SO_0(1,n-1)/SO(n-1)$. This quasi-regular representation
decomposes into a direct integral of all representations of class
1 (with respect to $SO(n-1)$) from the principal unitary series of
$SO_0(1,n-1)$ and each of these representations is contained in
the decomposition only once (see, for example, \cite{VKII}).

\section[Expansion of functions on the cone $G/MN$]{Expansion of functions on the cone $\boldsymbol{G/MN}$}
\label{Exp-cone}

In order to obtain basis functions on the cone $G/MN$ for each
coordinate system, we can act as in the case $G=SO_0(1,4)$. For
this, for each coordinate system it is necessary to f\/ind
dif\/ferential operators, invariant with respect to the group $G$
and its subgroups characterizing the corresponding coordinate
systems, then to create a system of dif\/ferential equations for
their eigenfunctions, and to solve this system. As  shown below,
an expansion of functions on $G/MN$ is reduced to expansion of
functions, given on homogeneous spaces of subgroups of smaller
dimension. If the latter expansion is known, then we do not need
to create a system of dif\/ferential equations and to solve it. In
other words, we shall construct basis functions by using the
method of reduction of this problem to the problem for a smaller
dimension (without using invariant dif\/ferential operators).
However, the basis functions, which will be obtained, are
eigenfunctions of an appropriate system of invariant
dif\/ferential operators. Let us consider each coordinate system
separately.

\subsection[Expansion for $S$-system]{Expansion for $\boldsymbol{S}$-system}
\label{Exp-cone-S}

Formulas \eqref{290}--\eqref{292} reduce an expansion of functions
$f(x)\equiv f(\tilde k,h)$ on $G/MN$ to an expansion of functions,
given on $K/M$. Indeed, under the integral in \eqref{291} we have
a function~$\tilde f (\tilde k,\nu)$, which for each f\/ixed $\nu$
is transformed under irreducible representation $\pi_{{\rm
i}\nu-\rho}$ of the group~$G$ and belongs to $L^2_0(K)\equiv
L^2(K/M)$. The function $\tilde f (\tilde k,\nu)$ can be expanded
in matrix elements of irreducible representations of the compact
subgroup $K$, invariant on the right with respect to the subgroup
$M$. To obtain this expansion it is enough to apply the
Peter--Weyl theorem.

Multiplying these representation matrix elements by the
exponential functions $\exp [(-{\rm i}\nu-\rho)(\log h)]$ and by
$\exp [({\rm i}\nu-\rho)(\log h)]$ (at f\/ixed $\nu$) we obtain
basis functions for irreducible spaces for the group $G$.
Unif\/ication of these basis functions for all values of $\nu$
gives a (continuous) basis of the space $L^2(G/MN)$.

For example, if $G=SO_0(1,n)$, then $K=SO(n)$ and $M=SO(n-1)$.
Expansion of functions given on the cone $G/MN$ is reduced by
formula \eqref{291} to expansions on the sphere $SO(n)/SO(n-1)$. A
basis for expansion of the latter functions consists of matrix
elements of ``null'' column of irreducible representations of the
subgroup $SO(n)$ with highest weights $(m,0,0,\dots,0)$, that is
of representations of class 1 with respect to the subgroup
$SO(n-1)$ (see \cite{VKII}, Chapter 9). We denote these matrix
elements by $D^m_{\alpha,0}(k)$, where $\alpha$ enumerates basis
elements of the space, where the irreducible representation with
highest weight $(m,0,\dots,0)$ acts. Then the collection of
functions (see formula \eqref{291})
\[
D^m_{\alpha,0}(\tilde k)\exp [(-{\rm i}\nu-\rho)(\log h)], \qquad
m=0,1,2,\dots, \qquad -\infty <\nu<\infty ,
\]
with dif\/ferent $\alpha$ constitutes the basis of
$L^2(SO_0(1,n)/MN)$, where $MN\sim ISO(n-1)$.

\subsection[Expansion for $T$-system]{Expansion for $\boldsymbol{T}$-system}
\label{Exp-cone-T}
Let us change decompositions \eqref{290}--\eqref{292}. Namely, we
consider functions $f\in C_c^\infty (G/MN)$ as functions $f(\bar
n,h)$ on $\bar NA$. Then instead of relations
\eqref{290}--\eqref{292} we have
\begin{gather}
\label{302} \tilde f(\bar n,\nu)=\int_A f(\bar n,h) e^{({\rm
i}\nu+\rho) (\log h)} dh,
\\
\label{303} f(\bar n,h)=\int_\mathcal{F} \tilde f(\bar n,\nu)
e^{(-{\rm i}\nu-\rho) (\log h)} d\nu,
\\
 \label{304}
\int_{G/MN} |f(x)|^2 dx \equiv \int_{\bar N}d\bar n \int_A |f(\bar
n,h)|^2 e^{2\rho(\log h)} dh =\int_{\bar N} d\bar n
\int_\mathcal{F} |\tilde f(\bar n,\nu)|^2 d\nu.
\end{gather}
Thus, anexpansion of functions $f\in C_c^\infty (G/MN)$ in basis
functions is reduced to an expansion in basis functions, given on
$\bar N$. For each f\/ixed $\nu$, the functions $\tilde f(\bar
n,\nu)$ are transformed under the left regular representation of
the subgroup $\bar N$. They can be expanded in basis functions by
means of decomposition of the regular representation of $\bar N$
into irreducible constituents.

Let, for example, $G=SO_0(1,n)$. Then the subgroup $\bar N$ is
isomorphic to $(n-1)$-dimensional group $T_{n-1}$ of translations.
The exponential functions
\begin{gather}
\label{305} \exp {\rm i}(x\cdot p),\qquad
x=(x_1,x_2,\dots,x_{n-1})\in T_{n-1},
\\
p=(p_1,p_2,\dots,p_{n-1}),\qquad -\infty <p_i<\infty,\nonumber
\end{gather}
where $x\cdot p=x_1p_1+x_2p_2+\cdots +x_{n-1}p_{n-1}$, constitute
a continuous basis for functions given on~$\bar N$. Thus, the
functions $f(x)\equiv f(\bar n,h)$ can be expanded in the
functions
\begin{equation}
\label{306} \exp {\rm i}(x\cdot p)\exp [(\exp (-{\rm
i}\nu-\rho)(\log h)],\qquad -\infty <p_i<\infty,\qquad -\infty
<\nu<\infty.
\end{equation}
Using formulas \eqref{302}--\eqref{304} and formulas for expansion
of functions $f(\bar n)$ in the basis \eqref{305}, it is easy to
write down formulas for expansions of functions $f(\bar n,h)$ in
the basis \eqref{306} and the corresponding Plancherel formula.

\subsection[Expansion for $O$-system]{Expansion for $\boldsymbol{O}$-system}
\label{Exp-cone-O}

In this coordinate system the cone $G/MN$ is parametrized by means
of the set $(\bar N M/M)A$ (see formula \eqref{276}). We consider
functions $f\in C^\infty_c(G/MN)$ as functions $f(y,h)$ on $(\bar
N M/M)\times A$. The expansions
\begin{gather}
\label{307} \tilde f(y,\nu)=\int_A f(y,h) e^{({\rm
i}\nu+\rho)(\log h)} dh,
\\
\label{308} f(y,h)=\int_\mathcal{F} \tilde f(y,\nu) e^{(-{\rm
i}\nu-\rho)(\log h)} d\nu,
\\
\label{309} \int_{G/MN} |f(x)|^2 dx=\int_{\bar N M/M} dy \int_A
|f(y,h)|^2 e^{2\rho (\log h)} dh =\int_{\bar N M/M}
dy\int_\mathcal{F} |\tilde f(y,\nu)|^2 d\nu
\end{gather}
hold. As we see,  expansion of functions $f\in C^\infty_c(G/MN)$
in basis functions is reduced to  expansion of functions given on
$\bar NM/M$. At each f\/ixed $\nu$ the functions $\tilde f(y,\nu)$
are transformed under the quasi-regular representation of the
subgroup $\bar NM$. By means of  decomposition of this
quasi-regular representation into irreducible representations of
the subgroup $\bar NM$, the functions $\tilde f(y,\nu)$ can be
expanded in matrix elements of irreducible representations of
$\bar NM$ right invariant with respect to the subgroup $M$.

If $G=SO_0(1,n)$, then the subgroup $\bar NM$ is isomorphic to the
group $ISO(n-1)$. The quasi-regular representation of $ISO(n-1)$
on the space of functions given on the homogeneous space
$ISO(n-1)/SO(n-1)$ decomposes into a direct integral of
irreducible unitary representations of $ISO(n-1)$ which are of
class 1 with respect to the subgroup $SO(n-1)$. Matrix elements of
these representations of $ISO(n-1)$ in $SO(n-1)$-basis are
expressed in terms of Bessel functions and are given in
\cite{VKII}. In expansion of functions given on $ISO(n-1)/SO(n-1)$
only matrix elements of the ``null'' column take part. If
$D^R_{\alpha,0}(r)$, $r\in ISO(n-1)$, $0<R<\infty$, are these
matrix elements, then the functions
\[
D^R_{\alpha,0}(r) e^{(-{\rm i}\nu-\rho)(\log h)},\qquad
0<R<\infty,\qquad -\infty< \nu<\infty,
\]
constitute a collection of basis functions on $SO_0(1,n)/SO(n)$ in
the $O$-coordinate system.

\subsection[Expansion for $H$-system]{Expansion for $\boldsymbol{H}$-system}
\label{Exp-cone-H}

In this coordinate system, the cone $G/MN$ is parametrized by the
set
\[
X=\bigcup_{i=1}^r (K_\varepsilon/M)w_i,\qquad w_i\in
W_\varepsilon\backslash W.
\]
We represent a function $f\in C^\infty_c(G/MN)$ as a function
$f(y,h)$ on $X\times A$. Then the transforms
\begin{gather}
\label{310} \tilde f(y,\nu)=\int_A f(y,h) e^{({\rm
i}\nu+\rho)(\log h)} dh,
\\
\label{311} f(y,h)=\int_\mathcal{F} \tilde f(y,\nu) e^{(-{\rm
i}\nu-\rho)(\log h)} d\nu,
\\
\label{312} \int_{G/MN} |f(x)|^2 dx=\int_{X} dy \int_A |f(y,h)|^2
e^{2\rho (\log h)} dh =\int_{X} dy\int_\mathcal{F} |\tilde
f(y,\nu)|^2 d\nu
\end{gather}
hold. Thus, an expansion of functions $f\in C^\infty_c(G/MN)$ in
basis functions in $H$-coordinate system is reduced to an
expansion of functions given on $K_\varepsilon/M$. At each f\/ixed
$\nu$ and $w_i$ the functions $\tilde f(y,\nu)$ are transformed
under the quasi-regular representation of the subgroup
$K_\varepsilon$. By means of  decomposition of this quasi-regular
representation into irreducible constituents, the functions
$\tilde f(y,\nu)$ can be expanded in matrix elements of
irreducible representations of $K_\varepsilon$, invariant on the
right with respect to the subgroup $M$. Moreover, matrix elements
of representations of $K_\varepsilon$ can correspond to
dif\/ferent chains of subgroups of $K_\varepsilon$. Each basis
function is non-vanishing only on one set $(K_\varepsilon/M)w_i$.

If $G=SO_0(1,n)$, then the subgroup $SO_0(1,n-1)$, containing the
subgroup $M=SO(n-1)$, can be taken as the subgroup
$K_\varepsilon$. In this case $r=2$. The elements $w_1\equiv 1$
and $w_2$ correspond to the upper and lower parts of the
hyperboloid $H^{n-1}= SO_0(1,n)/SO(n-1)$. Thus,
\[
X=(K_\varepsilon/M)\bigcup (K_\varepsilon/M)w.
\]
The problem of construction of the whole collection of basis
functions on the cone $G/MN$ in this case is reduced to
construction of basis functions for the space $L^2(H^{n-1})$ of
functions on the hyperboloid $SO_0(1,n-1)/SO(n-1)$. On this
hyperboloid dif\/ferent coordinate systems can be chosen. A
detailed consideration of this case see in \cite{VKII}, Chapters 9
and 10.

\section[Expansion of functions on the hyperboloid $G/K$]{Expansion of functions on the hyperboloid $\boldsymbol{G/K}$}
\label{Exp-hyp}

The f\/irst way of construction of collections of basis functions
on the hyperboloid $G/K$, corresponding to dif\/ferent coordinate
systems, is to use reasoning described for the case when
$G=SO_0(1,4)$ in Section~7. For $SO_0(1,4)$, for each coordinate
system we have found a complete system of dif\/ferential
operators, invariant with respect to the group $G$ and its
subgroups, which characterize the corresponding coordinate system.
Then we have constituted the corresponding system of
dif\/ferential equations and have solved it.

The second way for solving the problem of construction of
collections of basis functions consists in using the relations
\eqref{283}--\eqref{285}. Since it is not possible to construct a
system of invariant dif\/ferential operators for the case of a
generic group $G$, we use the second way. For each concrete group
$G$, the collection of basis functions, which will be found, a
priori is a solution of the corresponding system of invariant
dif\/ferential equations. Let us consider each coordinate system
separately.

\subsection[Expansion for $S$-system]{Expansion for $\boldsymbol{S}$-system}
\label{Exp-hyp-S}

According to the relation \eqref{284}, a function $f(\tilde g)\in
C^\infty_c(G/K)$ can be represented as
\begin{equation}
\label{313} f(\tilde g)=|W|^{-1} \int_\mathcal{F}\left[ \int_{K/M}
\tilde f(\tilde k,\nu) \exp (\{ ({\rm i}\nu-\rho) [H(g^{-1}k)]\}
d\tilde k\right] |c(\nu)|^{-2} d\nu.
\end{equation}
Let us consider the intrinsic integral on the right-hand side of
this relation. The function $\tilde f(\tilde k,\nu)$, as a
function on $K$ (constant on cosets with respect to $M$), can be
expanded in matrix elements $D^\omega_{\alpha\alpha'}(k)$ of
irreducible representations $\omega$ of the group $K$. Only the
matrix elements, invariant on the right with respect to the
subgroup $M$, enter to the expansion. This condition implies
a~certain conditions for $\alpha'$. Thus, by the Peter--Weyl
theorem we have
\begin{equation}
\label{314} \tilde f(\tilde k,\nu)=\sum_{\omega,\alpha,\alpha'}
A^{\omega\nu}_{\alpha\alpha'} D^\omega_{\alpha\alpha'}(\tilde k)
\end{equation}
and
\begin{gather}
\label{315} A^{\omega\nu}_{\alpha\alpha'}=\int_{K/M} \tilde
f(\tilde k,\nu) \overline{ D^\omega_{\alpha\alpha'}(\tilde k)}
d\tilde k,
\\
\label{316} \sum_{\omega,\alpha,\alpha'}
|A^{\omega\nu}_{\alpha\alpha'}|^2 =\int_{K/M} | \tilde f(\tilde
k,\nu)|^2 d\tilde k.
\end{gather}
We substitute the expression \eqref{314} for $\tilde f(\tilde
k,\nu)$ into the intrinsic integral on the right-hand side of the
relation \eqref{313} (that is, into the integral \eqref{289}) and
permute summation and integration. As a result, we obtain
\begin{equation}
\label{317} \mathcal{I}(g)\equiv \sum_{\omega,\alpha,\alpha'}
A^{\omega\nu}_{\alpha\alpha'} \int_{K/M}
D^\omega_{\alpha\alpha'}(\tilde k) \exp \{ ({\rm i}\nu-\rho)
[H(g^{-1}k)]\} d\tilde k.
\end{equation}
Since in formula \eqref{313} we are interested in functions $f$
given on $G/K$ we can suppose that $g=k'h$, where $k'$ are
representatives of cosets in $K/M$ and $h\in A^+$. Thus, we put
$g=k'h$. Making in \eqref{317} the substitution $k\to k'k$ and
taking into account the relation $d\tilde k=d(k'\tilde k)$, we
f\/ind that
\begin{gather*}
\mathcal{I}(g)\equiv \mathcal{I}(k'h)
=\sum_{\omega,\alpha,\alpha'} A^{\omega\nu}_{\alpha\alpha'}
\int_{K/M} D^\omega_{\alpha\alpha'}(k'\tilde k)\exp \{ ({\rm
i}\nu-\rho) [H(h^{-1}k)]\} d\tilde k
\\
\phantom{\mathcal{I}(g)\equiv \mathcal{I}(k'h)}{}
=\sum_{\omega,\alpha,\alpha'} \sum_{\alpha''}
A^{\omega\nu}_{\alpha\alpha'} D^\omega_{\alpha\alpha''}( k')
\int_{K/M} D^\omega_{\alpha''\alpha'}(\tilde k)\exp \{ ({\rm
i}\nu-\rho) [H(h^{-1}k)]\} d\tilde k.
\end{gather*}
The integral
\[
\int_{K/M} D^\omega_{\alpha''\alpha'}(\tilde k)\exp \{ ({\rm
i}\nu-\rho) [H(h^{-1}k)]\} d\tilde k \equiv
\mathcal{D}^\nu_{\omega(\alpha''\alpha'),0} (h)
\]
is a matrix element of the ``null'' column of the operator
$\pi_{{\rm i}\nu-\rho}(h)$ in the $K$-basis, and
\begin{equation}
\label{318} \mathcal{I}(g)\equiv \mathcal{I}(k'h)=
\sum_{\omega,\alpha,\alpha'} A^{\omega\nu}_{\alpha\alpha'}\left[
\sum_{\alpha''} D^\omega_{\alpha\alpha''}( k')
\mathcal{D}^\nu_{\omega(\alpha''\alpha'),0} (h)\right].
\end{equation}
Substituting this expression for $\mathcal{I}(g)$ into \eqref{313}
one obtains the relation
\begin{equation}
\label{319} f(\tilde g)\equiv f(\tilde k h)=
|W|^{-1}\int_\mathcal{F} \left[ \sum_{\omega,\alpha,\alpha'}
A^{\omega\nu}_{\alpha\alpha'}
F^{\omega\nu}_{\alpha\alpha'}(k',h)\right] |c(\nu)|^{-2}d\nu,
\end{equation}
where
\begin{equation}
\label{320} F^{\omega\nu}_{\alpha\alpha'}(k',h)= \sum_{\alpha''}
D^\omega_{\alpha\alpha''}( k')
\mathcal{D}^\nu_{\omega(\alpha''\alpha'),0} (h).
\end{equation}
From \eqref{315} and \eqref{283} we f\/ind the transform inverse
to the transform \eqref{319}:
\begin{gather}
A^{\omega\nu}_{\alpha\alpha'}=\int_{K/M} d{\tilde k}' \int_G dg\,
f(g) \exp \{ (-{\rm i}\nu-\rho) [H(g^{-1}k')]\}
\overline{D^\omega_{\alpha\alpha'}( {\tilde k}')}\nonumber\\
\phantom{A^{\omega\nu}_{\alpha\alpha'}}{}=\int_G
f(g)\overline{F^{\omega\nu}_{\alpha\alpha'}(k',h)} dg,\label{321}
\end{gather}
where $g=khk'$, $k,k'\in K$, $h\in A$. It follows from \eqref{316}
and \eqref{288} that the Plancherel formula
\begin{equation}
\label{322} \int_{G/K} |f(\tilde g)|^2 d\tilde g =
\int_{\mathcal{F}^+} \sum_{\omega,\alpha,\alpha'}
|A^{\omega\nu}_{\alpha\alpha'}|^2 |c(\nu)|^{-2}d\nu
\end{equation}
holds. Thus, functions \eqref{320} constitute a complete
collection of basis functions for the hyperboloid $G/K$.

If $G=SO_0(1,n)$, then $K=SO(n)$ and $M=SO(n-1)$. Irreducible
representations of $K$, which are of class 1 with respect to
$SO(n-1)$, are given by one non-negative integer $m$, and the
matrix elements $D^\omega_{\alpha\alpha'}(\tilde k)$ from
\eqref{320} and \eqref{315} take the form $D^m_{\alpha,0}(k)$,
$k\in K$. We suppose that these matrix elements are taken in the
basis corresponding to the chain of subgroups
\[
SO(n)\supset M\equiv SO(n-1)\supset SO(n-2)\supset \cdots .
\]
Thus, functions \eqref{320} are products of matrix elements of
``null'' column of a representation of the subgroup $SO(n)$ and
matrix elements  of ``null'' column of the operator $\pi_{{\rm
i}\nu-\rho}(h)$:
\[
 F^{m\nu}_{\alpha,0}(\tilde k,h)=D^m_{\alpha,0}(\tilde k)
\mathcal{D}^\nu_{m,0}(h)
\]
(see \cite{VKII} for an explicit form of these functions).

\subsection[Expansion for $T$-system]{Expansion for $\boldsymbol{T}$-system}
\label{Exp-hyp-T}

Let us write down the formulas \eqref{283}--\eqref{285} in such
form that the representation $\pi_{{\rm i}\nu-\rho}$ is realized
on functions given on $\bar N$ (not on functions, given on $K/M$).
According to \eqref{296} and~\eqref{297}, the operators $\pi_{{\rm
i}\nu-\rho}(g)$, $g\in G$, act on the space $L^2(\bar N)$ with the
norm \eqref{298} by the formula
\begin{equation}
\label{323} \pi_{\lambda}(g)f(\bar n)=\exp \{
-\lambda[H'(g^{-1}\bar n)]\} f({\bar n}_g),
\end{equation}
where $H'(g^{-1}\bar n)$ and ${\bar n}_g$ are determined by
\begin{gather*}
g^{-1}\bar n={\bar n}_gm \exp [H'(g^{-1}\bar n)] n,
\\
{\bar n}_g\in \bar N,\qquad m\in M,\qquad n\in N,\qquad
H'(g^{-1}\bar n)\in \mathfrak{a}.
\end{gather*}
Due to these formulas, instead of the transform \eqref{283} we
have the transform
\begin{equation}
\label{324} \tilde f ({\bar n}, \nu)=\int_{G/K} f(\tilde g) \exp
\{ (-{\rm i}\nu-\rho)[H'(g^{-1}\bar n)]\} d\tilde g.
\end{equation}
Integration over $G/K$ here coincides with integration on the
right-hand side of \eqref{262}, where $N$ is replaces by $\bar N$.
If the functions $f(\tilde g)$ are transformed under the
quasi-regular representation $\pi^H$ of the group $G$, then the
functions $\tilde f ({\bar n}, \nu)$ are transformed under the
representations $\pi_{-{\rm i}\nu-\rho}$ (by the formula
\eqref{323}). Instead of formulas \eqref{284} and \eqref{288} we
have
\begin{gather}
\label{325} f(\tilde g)=|W|^{-1} \int_\mathcal{F} \int_{\bar N}
\tilde f ({\bar n}, \nu) \exp \{ ({\rm i}\nu-\rho)[H'(g^{-1}\bar
n)]\} |c(\nu)|^{-2} d\bar n\, d\nu,
\\
\label{326} \int_{G/K} |f(\tilde g)|^2dg= \int_{\mathcal{F}^+}
\int_{\bar N} |\tilde f ({\bar n}, \nu)|^2  |c(\nu)|^{-2} d\bar
n\, d\nu.
\end{gather}

Now we expand $\tilde f ({\bar n}, \nu)$, as a function on $\bar
N$, in matrix elements of irreducible unitary representations of
the group $\bar N$:
\begin{equation}
\label{327} \tilde f ({\bar n}, \nu)=\int_{\hat N}
\sum_{\alpha,\alpha'}
A^{\omega\nu}_{\alpha\alpha'}D^\omega_{\alpha\alpha'}(\bar
n)d\mu(\omega),
\end{equation}
where $\hat N$ denotes a set of irreducible unitary
representations $\omega$ of the group $\bar N$. The
coef\/f\/icients $A^{\omega\nu}_{\alpha\alpha'}$ are determined by
the formula
\begin{equation}
\label{328} A^{\omega\nu}_{\alpha\alpha'}=\int_{\bar N} \tilde f
({\bar n}, \nu) \overline{D^\omega_{\alpha\alpha'}(\bar n)}d\bar n
\end{equation}
and the Plancherel formula
\begin{equation}
\label{329} \int_{\bar N} |\tilde f ({\bar n}, \nu)|^2d\bar n =
\int_{\hat N} \sum_{\alpha,\alpha'}
|A^{\omega\nu}_{\alpha\alpha'}|^2 d\mu(\omega)
\end{equation}
holds.

We substitute the expression \eqref{327} for $\tilde f ({\bar n},
\nu)$ into the intrinsic integral (the integral over~$\bar N$) of
the right-hand side of \eqref{325}. If  permutation of the
integrations is possible in the obtained expression (we denote
this expression by $\mathcal{I}(g)$), then
\[
\mathcal{I}(g)= \int_{\hat N} \sum_{\alpha,\alpha'}
A^{\omega\nu}_{\alpha\alpha'}d\mu(\omega) \int_{\bar N}
D^\omega_{\alpha\alpha'}(\bar n)\exp \{ ({\rm i}\nu-\rho)
[H'(g^{-1}\bar n ]\} d\bar n.
\]
Taking into account the $T$-parametrization of $G$, we assume that
elements $g$ have the form ${\bar n}'h$, ${\bar n}'\in \bar N$,
$h\in A$. Therefore, replacing $\bar n$ by ${\bar n}'{\bar n}$ we
f\/ind
\[
\mathcal{I}(g)\equiv\mathcal{I}({\bar n}'h)= \int_{\hat N}
\sum_{\alpha,\alpha'}
A^{\omega\nu}_{\alpha\alpha'}d\mu(\omega)\sum_{\alpha''}
D^\omega_{\alpha\alpha''}({\bar n}') \int_{\bar N}
D^\omega_{\alpha''\alpha'}(\bar n)\exp \{ ({\rm i}\nu-\rho)
[H'(h^{-1}\bar n ]\} d\bar n.
\]
If the integral
\[
\int_{\bar N} D^\omega_{\alpha''\alpha'}(\bar n)\exp \{ ({\rm
i}\nu-\rho) [H'(h^{-1}\bar n ]\} d\bar n
\]
exists, we denote it as
$\mathcal{D}^\nu_{\omega(\alpha''\alpha')}(h)$. Then it follows
from \eqref{325} that
\[
f(\tilde g)\equiv f({\bar n}'h)=|W|^{-1} \int_\mathcal{F}
\int_{\hat N} \sum_{\alpha,\alpha'}
A^{\omega\nu}_{\alpha\alpha'}F^{\omega\nu}_{\alpha\alpha'} ({\bar
n}',h)d\mu(\omega) |c(\nu)|^{-2}d\nu,
\]
where
\begin{equation}
\label{330} F^{\omega\nu}_{\alpha\alpha'}({\bar n}',h)
=\sum_{\alpha''} D^\omega_{\alpha\alpha''}({\bar n}')
\mathcal{D}^\nu_{\omega(\alpha''\alpha')}(h).
\end{equation}
Using formulas \eqref{324} and \eqref{328} one obtains the inverse
transform
\begin{equation}
\label{331} A^{\omega\nu}_{\alpha\alpha'}=\int_{\bar N} d\bar n
\int_{G/K} d\tilde g \, f(\tilde g) \exp \{ (-{\rm i}\nu-\rho)
[H'(g^{-1}\bar n ]\} \overline{D^\omega_{\alpha\alpha'}({\bar
n})}.
\end{equation}
A Plancherel formula for these transforms follows from \eqref{326}
and \eqref{329}. Thus, the functions \eqref{330} are basis
functions on $G/K$ in the $T$-coordinate system.

\subsection[Expansion for $O$-system]{Expansion for $\boldsymbol{O}$-system}
\label{Exp-hyp-O}

Let us write down the formulas \eqref{283}--\eqref{285} in such
form that the representation $\pi_{{\rm i}\nu-\rho}$ is realized
on functions given on $\bar N M/M$ (not on $K/M$). According to
\eqref{296} and \eqref{297}, the operators $\pi_{{\rm
i}\nu-\rho}(g)$, $g\in G$, act on the space $L^2(\bar N M/M)$ by
the formula
\begin{equation}
\label{332} \pi_{\lambda}(g)f(x)=\exp \{ -\lambda[H''(g^{-1}x)]\}
f(x_g),\qquad x\in (\bar N M/M),
\end{equation}
where $H''(g^{-1}x)$ and $x_g$ are determined by
\begin{gather*}
g^{-1}x=x_gm \exp [H''(g^{-1}x)] n,
\\
x_g\in (\bar NM/M),\qquad m\in M,\qquad n\in N,\qquad
H''(g^{-1}x)\in \mathfrak{a}.
\end{gather*}

Due to these formulas, instead of the transform \eqref{283} we
consider the transform
\begin{equation}
\label{333} \tilde f (x, \nu)=\int_{G/K} f(\tilde g) \exp \{
(-{\rm i}\nu-\rho)[H''(g^{-1}x)]\} d\tilde g.
\end{equation}
Integration on $G/K$ here coincides with integration on the
right-hand side of \eqref{265}, where $N$ is replaces by $\bar N$.
If the functions $f(\tilde g)$ are transformed under the
quasi-regular representation $\pi^H$ of the group $G$, then the
functions $\tilde f (x, \nu)$ are transformed under the
representation $\pi_{-{\rm i}\nu-\rho}$ (by the formula
\eqref{332}). Instead of formulas \eqref{284} and \eqref{288} we
have
\begin{gather}
\label{334} f(\tilde g)=|W|^{-1} \int_\mathcal{F} \int_{\bar NM/M}
\tilde f (x, \nu) \exp \{ ({\rm i}\nu-\rho)[H''(g^{-1}x)]\}
|c(\nu)|^{-2} dx\, d\nu,
\\
\label{335} \int_{G/K} |f(\tilde g)|^2d\tilde g=
\int_{\mathcal{F}^+} \int_{\bar NM/M} |\tilde f (x, \nu)|^2
|c(\nu)|^{-2}dx\, d\nu.
\end{gather}

Now our reasoning is as in the previous case. We expand $\tilde f
(x, \nu)$, as a function on $\bar NM$, constant on cosets with
respect to the subgroup $M$, in matrix elements of irreducible
unitary representations of the group $\bar NM$ invariant on the
right with respect to the subgroup $M$:
\begin{equation}
\label{336} \tilde f (x, \nu)=\int_{\hat {NM}_0}
\sum_{\alpha,\alpha'}
A^{\omega\nu}_{\alpha\alpha'}D^\omega_{\alpha\alpha'}(x)d\mu(\omega),
\end{equation}
where $\hat {NM}_0$ denotes a set of nonequivalent irreducible
unitary representations of the group $\bar NM$ of class 1 with
respect to the subgroup $M$. Invariance of matrix elements
$D^\omega_{\alpha\alpha'}(x)$ on the right with respect to $M$
imposes some conditions on values of $\alpha'$. The
coef\/f\/icients $A^{\omega\nu}_{\alpha\alpha'}$ are determined by
the formula
\begin{equation}
\label{337} A^{\omega\nu}_{\alpha\alpha'}=\int_{\bar NM/M} \tilde
f (x, \nu) \overline{D^\omega_{\alpha\alpha'}(x)}dx
\end{equation}
and the Plancherel formula
\begin{equation}
\label{338} \int_{\bar NM/M} |\tilde f (x, \nu)|^2dx = \int_{\hat
{NM}_0} \sum_{\alpha,\alpha'} |A^{\omega\nu}_{\alpha\alpha'}|^2
d\mu(\omega)
\end{equation}
holds.

We substitute the expression \eqref{336} for the function $\tilde
f (x, \nu)$ into the intrinsic integral (the integral over $x$) of
the right-hand side of \eqref{334}. If  permutation of the
integrations is possible in the obtained expression (we denote
this expression by $\mathcal{I}(g)$), the
\[
\mathcal{I}(g)= \int_{\hat {NM}_0} \sum_{\alpha,\alpha'}
A^{\omega\nu}_{\alpha\alpha'}d\mu(\omega) \int_{\bar NM/M}
D^\omega_{\alpha\alpha'}(x)\exp \{ ({\rm i}\nu-\rho) [H''(g^{-1}x
]\} dx.
\]
Due to the $O$-parametrization of $G$, elements $g$ have the form
$yh$, where $y$ are representatives of cosets of $\bar NM/M$ and
$h\in A$. Replacing $x$ by $yx$ and using the relation $dx=d(yx)$
we f\/ind
\begin{gather*}
\mathcal{I}(g){\equiv}\mathcal{I}(yh){=} \int_{\hat {NM}_0}\!
\sum_{\alpha,\alpha'}
A^{\omega\nu}_{\alpha\alpha'}d\mu(\omega)\sum_{\alpha''}
D^\omega_{\alpha\alpha''}(y) \!\int_{\bar NM/M}\!\!\!
D^\omega_{\alpha''\alpha'}(x)\exp \{ ({\rm i}\nu{-}\rho)
[H''(h^{-1}x)]\} dx.
\end{gather*}
If the integral
\[
\int_{\bar NM/M} D^\omega_{\alpha''\alpha'}(x)\exp \{ ({\rm
i}\nu-\rho) [H''(h^{-1}x) ]\} dx\equiv
\mathcal{D}^\nu_{\omega(\alpha''\alpha')}(h)
\]
exists, then it follows from \eqref{334} that
\[
f(\tilde g)=|W|^{-1} \int_\mathcal{F} \int_{\hat {NM}_0}
\sum_{\alpha,\alpha'}
A^{\omega\nu}_{\alpha\alpha'}F^{\omega\nu}_{\alpha\alpha'}
(y,h)d\mu(\omega) |c(\nu)|^{-2}d\nu,
\]
where
\begin{equation}
\label{339} F^{\omega\nu}_{\alpha\alpha'}(y,h) =\sum_{\alpha''}
D^\omega_{\alpha\alpha''}(y)
\mathcal{D}^\nu_{\omega(\alpha''\alpha')}(h).
\end{equation}
Using formulas \eqref{333} and \eqref{337} one obtains the inverse
transform
\[
A^{\omega\nu}_{\alpha\alpha'}=\int_{\bar NM/M} dx \int_{G/K}
d\tilde g \, f(\tilde g) \exp \{ (-{\rm i}\nu-\rho)
[H''(g^{-1}x]\} \overline{D^\omega_{\alpha\alpha'}(x)}.
\]
A Plancherel formula for these transforms follows from this
formula and from \eqref{335}. Thus, the functions \eqref{339} are
basis functions on $G/K$ in the $O$-coordinate system.

\subsection[Expansion for $H$-system]{Expansion for $\boldsymbol{H}$-system}
\label{Exp-hyp-H}

Let us give to formulas \eqref{283}--\eqref{285} such  form for
which the representations $\pi_{{\rm i}\nu-\rho}$ are realized on
functions on the set
\[
X=\bigcup_{i=1}^r (K_\varepsilon/M)w_i
\]
instead of functions given on $K/M$. According to \eqref{296} and
\eqref{297} the operators $\pi_\lambda (g)$, $g\in G$, act on the
space $L^2(X)$ by the formula
\[
\pi_\lambda (g) f(x)=\exp \{ -\lambda [ H''' (g^{-1}x)]\} f(x_g),
\qquad x\in X,
\]
where $H'''$ and $x_g$ are determined by the relation
\[
g^{-1}x=x_gm\exp [ H''' (g^{-1}x)] n,\qquad x_g\in X,\quad m\in
M,\quad n\in N,\quad H''' (g^{-1}x) \in \mathfrak{a}.
\]
Therefore, instead of the transform \eqref{283} we consider the
transform
\begin{equation}
\label{340} \tilde f (x,\nu)=\int_{G/K} f(\tilde g)\exp \{ (-{\rm
i}\nu-\rho) [ H''' (g^{-1}x)] d\tilde g.
\end{equation}
If the function $f(\tilde g)$ is transformed under the
quasi-regular representation $\pi^H$, then the function $\tilde f
(x,\nu)$ is transformed under the representation $\pi_{-{\rm
i}\nu-\rho}$. Instead of formulas \eqref{284} and \eqref{288} we
have
\begin{gather}
f(\tilde g)=|W|^{-1} \int_\mathcal{F} \int_{X} \tilde f (x, \nu)
\exp \{ ({\rm i}\nu-\rho)[H'''(g^{-1}x)]\} |c(\nu)|^{-2} dx\, d\nu
\nonumber\\
\label{341}    \phantom{f(\tilde g)}{} \equiv |W|^{-1}\sum_{i=1}^r
\int_\mathcal{F} \int_{K_\varepsilon} \tilde f (k_\varepsilon w_i,
\nu) \exp \{ ({\rm i}\nu-\rho)[H'''(g^{-1}k_\varepsilon w_i)]\}
|c(\nu)|^{-2}  dk_\varepsilon\, d\nu ,
\\
\label{342} \int_{G/K} |f(\tilde g)|^2d\tilde g= \sum_{i=1}^r
\int_{\mathcal{F}^+} \int_{K_\varepsilon} |\tilde f (k_\varepsilon
w_i, \nu)|^2  |c(\nu)|^{-2} dk_\varepsilon\, d\nu ,
\end{gather}
where $dk_\varepsilon$ is an invariant measure on $K_\varepsilon$.

For each f\/ixed $\nu$ and $w_i$ we expand the function $\tilde f
(k_\varepsilon w_i, \nu)$ (as a function on the
group~$K_\varepsilon$, constant on cosets with respect to $M$) in
matrix elements of irreducible unitary representations~$\omega$ of
the group $K_\varepsilon$ of class 1 with respect to the
subgroup~$M$:
\begin{equation}
\label{343} \tilde f (k_\varepsilon w_i,
\nu)=\int_{\hat{K_\varepsilon}} \sum_{\alpha,\alpha'} A^{\omega\nu
i}_{\alpha,\alpha'} D^\omega_{\alpha\alpha'}
(k_\varepsilon)d\mu(\omega),
\end{equation}
where $\hat{K_\varepsilon}$ means the set of irreducible unitary
representations of $K_\varepsilon$. The coef\/f\/icients
$A^{\omega\nu i}_{\alpha\alpha'}$ are determined by the formula
\begin{equation}
\label{344}
 A^{\omega\nu i}_{\alpha\alpha'}=\int_{K_\varepsilon}
\tilde f (k_\varepsilon w_i, \nu)
\overline{D^\omega_{\alpha\alpha'} (k_\varepsilon)}
dk_\varepsilon.
\end{equation}
The Plancherel formula
\begin{equation}
\label{345} \int_{K_\varepsilon} |\tilde f (k_\varepsilon w_i,
\nu)|^2 dk_\varepsilon = \int_{\hat K_\varepsilon}
\sum_{\alpha,\alpha'} | A^{\omega\nu i}_{\alpha\alpha'}|^2
d\mu(\omega)
\end{equation}
holds. We substitute the expression \eqref{343} for the function
$\tilde f (k_\varepsilon w_i, \nu)$ into the intrinsic integral in
\eqref{341} (we denote this integral by $\mathcal{I}(g)$). If
permutation of the integrals is allowed in the resulting
expression for $\mathcal{I}(g)$, then
\begin{equation}
\label{346} \mathcal{I}(g)=\int_{\hat
K_\varepsilon}\sum_{\alpha,\alpha'} A^{\omega\nu
i}_{\alpha\alpha'} d\mu(\omega) \int_{K_\varepsilon}
D^\omega_{\alpha\alpha'} (k_\varepsilon) \exp \{ ({\rm i}\nu-\rho)
[H'''(g^{-1}k_\varepsilon w_i)]\} dk_\varepsilon.
\end{equation}
Suppose that elements $g$ here have the form $k'_\varepsilon
w_ih$, $h\in A$. Making in \eqref{346} the substitution
$k_\varepsilon \to k'_\varepsilon k_\varepsilon$ and using the
relation $d k_\varepsilon =d( k'_\varepsilon k_\varepsilon)$, we
f\/ind that
\begin{gather*}
\mathcal{I}(g)\equiv \mathcal{I}(k'_\varepsilon w_jh)= \int_{\hat
K_\varepsilon} \sum_{\alpha,\alpha'} A^{\omega\nu
i}_{\alpha\alpha'} d\mu(\omega) \sum_{\alpha''}
D^\omega_{\alpha\alpha''}(k_\varepsilon')  \\
\phantom{\mathcal{I}(g)\equiv\mathcal{I}(k'_\varepsilon w_jh)=}{}
\times \int_{K_\varepsilon} D^\omega_{\alpha''\alpha'}
(k_\varepsilon) \exp \{ ({\rm i}\nu-\rho)[H'''(h^{-1} w_j^{-1}
k_\varepsilon w_i)]\} dk_\varepsilon.
\end{gather*}
If the latter integral on the right hand side exists, we denote it
as $\mathcal{D}^{\nu i}_{\omega(\alpha'',\alpha')}(w_jh)$. Then
the formula \eqref{341} gives
\[
f(\tilde g)\equiv f(k_\varepsilon'w_jh)= |W|^{-1}\sum_{i=1}^r
\int_\mathcal{F} \int_{\hat K_\varepsilon} \sum_{\alpha,\alpha'}
A^{\omega\nu i}_{\alpha\alpha'} F^{\omega\nu i}_{\alpha\alpha'}
(k_\varepsilon', w_jh) |c(\nu)|^{-2}d\mu(\omega)\, d\nu ,
\]
where
\begin{equation}
\label{347} F^{\omega\nu i}_{\alpha\alpha'} (k_\varepsilon',
w_jh)= \sum_{\alpha''} D^\omega_{\alpha\alpha''}(k_\varepsilon')
\mathcal{D}^{\nu i}_{\omega(\alpha'',\alpha')}(w_jh).
\end{equation}
The inverse transform is of the form
\[
A^{\omega\nu i}_{\alpha\alpha'}=\int_{K_\varepsilon}
dk_\varepsilon \int_{G/K} d\tilde g\, f(\tilde g) \exp \{ (-{\rm
i}\nu-\rho)[H'''(g^{-1}
 k_\varepsilon w_i)]\} \overline{
D^\omega_{\alpha\alpha'}(k_\varepsilon)}.
\]
The Plancherel formula for these transforms follows from
\eqref{342} and \eqref{345}. Thus, the functions \eqref{347} are
basis functions on $G/K$ in the $H$-coordinate system.

In the expansions considered above representations of the
subgroups ${\overline N}$, ${\overline N}M$ and $K_\varepsilon$
were considered in discrete bases. The corresponding expansions
can be similarly obtained also for the case when these
representations are considered in ``continuous'' bases.

\subsection[Expansion of $K$-invariant functions on $G/K$]{Expansion of $\boldsymbol{K}$-invariant functions on $\boldsymbol{G/K}$}

\label{K-invar-func}
{\it A function $f$ on $G/K$ is called $K$-invariant if $f(k\tilde
g)=f(\tilde g)$, $\tilde g\in G/K$.} Since elements of $G/K$ are
determined by representatives $g=kh$, $k\in K$, $h\in A$, then
$K$-invariant functions $f$ on $G/K$ can be considered as
functions on $A$:
\[
f(\tilde g)=f(kh)=F(h).
\]
If $M'$ is a normalizer of the subgroup $A$ in $K$, then for $m\in
M'$ we have $mAm^{-1}=A$ and
\[
F(mhm^{-1})=F(h).
\]
In other words, the function $F$ is uniquely determined by values
on $\overline{A^+}$.

We consider the Fourier transform \eqref{283} of $K$-invariant
functions $f$ on $G/K$:
\[
\tilde f (\tilde k, \nu)=\int_G f(g) \exp \{ (-{\rm
i}\nu-\rho)[H(g^{-1}k)]\} dg,\qquad \nu\in \mathcal{F}.
\]
Let us show that $\tilde f (\tilde k, \nu)$ is independent of
$\tilde k$. For $k_0\in K$ we have
\[
\tilde f (k_0\tilde k, \nu)=\int_G f(g) \exp \{ (-{\rm
i}\nu-\rho)[H(g^{-1}k_0k)]\} dg.
\]
Since $H(g^{-1}k_0k)=H((k_0^{-1}g)^{-1}k)$ and $d(k_0g)=dg$, then
\begin{gather*}
\tilde f (k_0\tilde k, \nu)=\int_G f(k_0g) \exp \{ (-{\rm
i}\nu-\rho)[H(g^{-1}k)]\} d(k_0g)
\\
\phantom{\tilde f (k_0\tilde k, \nu)}{} =\int_G f(g) \exp \{
(-{\rm i}\nu-\rho)[H(g^{-1}k)]\} dg =\tilde f (\tilde k, \nu).
\end{gather*}
This means that $\tilde f (\tilde k, \nu)$ is independent of
$\tilde k$, that is $\tilde f (\tilde k, \nu)=\tilde F(\nu)$.

Thus, for $K$-invariant functions $f$ on $G/K$ we have the
expansion
\begin{equation}
\label{348} f(\tilde g)=|W|^{-1} \int_\mathcal{F} \int_{K/M}
\tilde F(\nu) \exp \{ ({\rm i}\nu-\rho)[H(g^{-1}k)]\}
|c(\nu)|^{-2} d\tilde k\, d\nu ,
\end{equation}
where
\begin{equation}
\label{349} \tilde F(\nu)= \int_G f(\tilde g) \exp \{ (-{\rm
i}\nu-\rho)[H(g^{-1}k)]\} dg.
\end{equation}

The function
\begin{equation}
\label{350} \varphi_\nu(g)= \int_{K/M} \exp \{ ({\rm
i}\nu-\rho)[H(g^{-1}k)]\} d\tilde k
\end{equation}
is called a {\it zonal spherical function} of the group $G$
corresponding to the representation $\pi_{{\rm i}\nu-\rho}$. This
function is in fact the matrix element
\begin{equation}
\label{351}
 d^{{\rm i}\nu-\rho}_{11}(g)= \langle 1,
 \pi_{{\rm i}\nu-\rho}(g)1 \rangle
\end{equation}
of the representation $\pi_{{\rm i}\nu-\rho}$ of the group $G$. It
is easy to check that
\[
\varphi_\nu(kgk')= \varphi_\nu(g),\qquad k,k'\in K,
\]
that is, $\varphi_\nu(g)$ {\it is a $K$-invariant function on
$G/K$}.

It follows from \eqref{348} and \eqref{350} that
\begin{equation}
\label{352} f(\tilde g)= |W|^{-1} \int_\mathcal{F} \tilde F(\nu)
\varphi_\nu(g) |c(\nu)|^{-2}d\nu .
\end{equation}

According to the equality \eqref{258}, the measure $dg$ in
\eqref{349} can be decomposed into the product of the measures on
$K$ and $A$. Since $f(\tilde g)=F(h)$, then applying formula
\eqref{350} we have
\[
\tilde F(\nu)=c |W|^{-1} \int_\mathfrak{a} F(\exp H)
\varphi_\nu(\exp H) \left\vert \prod_{\alpha >0} \sinh \alpha
(H)\right\vert dH.
\]
Using the formula \eqref{258} again we obtain
\begin{equation}
\label{353} \tilde F(\nu)=\int_G f(\tilde g)
 \varphi_\nu(g) dg.
\end{equation}

The Plancherel formula \eqref{285} now takes the form
\begin{equation}
\label{354} \int_{G/K} |f(\tilde g)|^2dg= |W|^{-1}
\int_\mathcal{F} |\tilde F(\nu)|^2 |c(\nu)|^{-2}d\nu .
\end{equation}

The transform \eqref{353} is called a {\it spherical transform} of
$K$-invariant functions $f$ on $G/K$. The formula \eqref{352}
gives an inverse transform, which is an expansion of the function
$f$ in zonal spherical functions of the group $G$. The formula
\eqref{354} shows that the spherical transform is isometric.

In the spherical transform only the zonal spherical functions
$\varphi_\nu(g)$ of the representations $\pi_{{\rm i}\nu-\rho}$
from the principal unitary series participate. The formula
\eqref{350} determines zonal spherical functions for all
representations $\pi_{{\rm i}\nu-\rho}$ of the principal
nonunitary series, that is, for representations with arbitrary
complex linear forms $\nu$ on $\mathfrak{a}$.

\pdfbookmark[1]{References}{ref}
\LastPageEnding

\end{document}